\documentclass[11pt,a4paper]{article}
\usepackage{atlasphysics}
\usepackage{etex}

\usepackage{a4wide}

\usepackage{feynmp}
\usepackage{jheppub}
\usepackage{subfigure}

\usepackage{preprintcover}

\usepackage[maxfloats=75]{morefloats}
\usepackage{ifthen}
\usepackage{tikz}
\usetikzlibrary{calc}
\usepackage{bchart}

\usepackage{booktabs}
\usepackage{dcolumn}
\newcolumntype{d}[1]{D{.}{.}{#1}}

\usepackage{hyperref}
\usepackage{tabb}
\usepackage{upgreek}
\usepackage{relsize}
\usepackage{rotating}

\usepackage{color}
\usepackage{graphicx}
\usepackage{xspace}
\usepackage{multirow}
\usepackage{hhline}
\usepackage{bm}

\newcommand\TO     {\SPACE{\rightarrow}}

\newcommand\pb     {\mbox{\,pb}}

\newcommand\Hz     {\ifmmode{\,\mathrm{Hz}}\else
                      \,\textrm{Hz}\fi}

\newcommand\sigmu        {\upmu}

\newcommand\PYTHIA       {{\sc pythia}}

\newcommand\mue          {\mu e}

\renewcommand\ll         {\ell\ell}

\newcommand\all          {{\rm all}}

\newcommand\T            {\textrm{\footnotesize\sc t}}

\newcommand\miss         {\rm miss}

\newcommand\p            {P}

\definecolor{myWW}{rgb}{0,0,255}
\definecolor{myVV}{rgb}{305,69,94}
\definecolor{myMisid}{rgb}{180,81,92}
\definecolor{myDY}{rgb}{120,89,72}
\definecolor{myTop}{rgb}{60,70,92}
\definecolor{myr}{rgb}{0.584, 0, 0.102}
\definecolor{myb}{rgb}{0.004, 0.145, 0.431}
\newcommand\myb          {\color{myb}}
\newcommand\myr          {\color{myr}}

\newcommand{\slice}[5]{
  \pgfmathparse{0.5*#1+0.5*#2}
  \let\midangle\pgfmathresult

  \ifthenelse{\equal{#2}{0}}{}{
    \draw[fill=#5] (0,0) -- (#1:1) arc (#1:#2:1) -- cycle;
  }

  \node[label=\midangle:#4] at (\midangle:0.95) {};

  \ifthenelse{\equal{#3}{1}}{}{
    \ifthenelse{\equal{#3}{3}}{}{
      \pgfmathparse{min((#2-#1-10)/110*(-0.3),0)}
      \let\temp\pgfmathresult
      \pgfmathparse{max(\temp,-0.2) + 0.8}
      \let\innerpos\pgfmathresult
      \node at (\midangle:\innerpos){
        \ifthenelse{\equal{blue}{#5}}{\color{white} #3}{
          \ifthenelse{\equal{magenta}{#5}}{\color{white} #3}{
            \ifthenelse{\equal{cyan}{#5}}{\color{white} #3}{
              \ifthenelse{\equal{black}{#5}}{\color{white} #3}{
                #3
              }
            }
          }
        }
      };
    }
  }
}

\newcommand\no           {\!\!}
\newcommand\np           {\no\no}
\newcommand\nq           {\np\np}

\newcommand\z            {\phantom{0}}

\newcommand\dbline{\noalign{\vskip 0.10truecm\hrule\vskip 0.05truecm\hrule\vskip 0.10truecm}}
\newcommand\sgline{\noalign{\vskip 0.10truecm\hrule\vskip 0.10truecm}}
\newcommand\clineskip{\noalign{\vskip 0.10truecm}}

\graphicspath{{Figures/}}

\def\TO     {\,{\rightarrow}\,}

\def\Phi    {\protect\raisebox{1px}{\ensuremath{\phi}}}

\def\pb     {\mbox{\,pb}}
\def\ifb    {\mbox{\,fb$^{-1}$}}

\def\Hz     {\ifmmode{\,\mathrm{Hz}}\else
                      \,\textrm{Hz}\fi}
\def\TeV    {\ifmmode{\,\mathrm{Te\kern -0.1em V}}\else
                      \,\textrm{Te\kern -0.1em V}\fi}
\def\GeV    {\ifmmode{\,\mathrm{Ge\kern -0.1em V}}\else
                      \,\textrm{Ge\kern -0.1em V}\fi}

\def\PYTHIA       {\sc pythia}

\def\mW           {M_W}
\def\mZ           {M_Z}

\def\mue          {{\mu e}}

\def\ll           {{\ell\ell}}

\def\Jpsi         {{J/\psi}}

\def\T            {{\textrm{T}}}

\def\miss         {{\textrm{miss}}}

\def\MET          {{E_{\T}^{\miss}}}

\newcommand{\MTLead}{\ensuremath{m_{\rm T}^{\rm lead}}}

\def\dbline{\noalign{\vskip 0.10truecm\hrule\vskip 0.05truecm\hrule\vskip 0.10truecm}}
\def\sgline{\noalign{\vskip 0.10truecm\hrule\vskip 0.10truecm}}

\let\oldmarginpar\marginpar
\renewcommand\marginpar[1]{
  \-\oldmarginpar[\setlength{\marginparsep}{0.05in}\raggedright\footnotesize #1]
                 {\setlength{\marginparsep}{0.25in}\raggedright\footnotesize #1}
}
\newcommand{\margincomment}[1]{
  \ifthenelse{\equal{\turnoncomment}{True}}
             {
              \marginpar{\bf\boldmath\color{blue}\footnotesize#1}
             }
             {}
}

\newcommand{\hww}{\ensuremath{H{\rightarrow\,}WW^{\ast}}}
\newcommand{\mT}{\ensuremath{m_{\rm T}}}

\newcommand{\lumi}{20.3}
\newcommand{\lumieleven}{4.5}
\newcommand{\threelep}{\ensuremath{{\rm 3}{\ell}}}
\newcommand{\twolep}{\ensuremath{{\rm 2}{\ell}}}
\newcommand{\fourlep}{\ensuremath{{\rm 4}{\ell}}}

\PreprintCoverPaperTitle{Study of $(W/Z)H$ production and Higgs boson couplings using $\hww$ decays with the ATLAS detector}
\PreprintIdNumber{CERN-PH-EP-2015-104}
\PreprintCoverAbstract{
A search for  Higgs boson production in association with
a $W$ or $Z$ boson, in the \hww\ decay channel, is performed with a
data sample collected with the ATLAS detector at the LHC 
in proton--proton collisions 
at centre-of-mass energies $\sqrt{s}=7\tev$ and 8~\tev, corresponding
to integrated luminosities of
\lumieleven\ifb\ and \lumi\ifb,
respectively.
The $WH$  production mode is studied in two-lepton and three-lepton final states,
while two-lepton and four-lepton final states are used to search for the $ZH$
production mode. The observed significance, for the combined $WH$
and $ZH$ production, is 2.5 standard deviations while a significance of 0.9 standard
deviations is expected in the Standard Model Higgs boson hypothesis. The ratio of
the combined $WH$ and $ZH$ signal
yield to the Standard Model expectation, $\mu_{VH}$, is found to be
$\mu_{VH} =3.0^{+1.3}_{-1.1}{\, {(\rm
    stat.)}}^{+1.0}_{-0.7}{\,{(\rm sys.)}}$ for the Higgs boson mass of
125.36~\gev. The $WH$ and $ZH$ production modes are also combined with the
gluon fusion and
vector boson fusion  production modes studied in the $H \to WW^* \to \ell \nu \ell \nu$ decay channel,
resulting in an overall observed significance of 6.5 standard deviations
and $\mu_{\rm ggF+VBF+VH} = 1.16^{+0.16}_{-0.15} {\rm (stat.)}^{+0.18}_{-0.15} {\rm (sys.)}$.  
The results are interpreted in terms of scaling factors of the Higgs
boson couplings to vector bosons ($\kappa_V$) and fermions ($\kappa_F$); the combined results are: $|\kappa_V| = 1.06^{+0.10}_{-0.10}$, $|\kappa_F| = 0.85^{+0.26}_{-0.20}$.
}
\PreprintJournalName{JHEP}

\def\turnoncomment{True}

\begin{document}

\section{Introduction                                             \label{sec:introduction}        }
In the Standard Model (SM) of fundamental interactions, 
the Brout--Englert--Higgs~\cite{prl_13_321, prl_13_508, Guralnik:1964eu} 
mechanism induces the electroweak
symmetry breaking that provides mass to elementary particles. The
mechanism postulates the existence of an elementary scalar particle, the Higgs
boson.
The ATLAS and CMS collaborations at the CERN Large Hadron Collider (LHC) have observed the Higgs boson 
with a mass ($\mH$) of 
about 125~\gev~\cite{ATLAS-HiggsDiscovery,CMS-HiggsDiscovery}.
The measurements of the Higgs boson couplings to SM
particles, and its spin and CP quantum numbers, are essential tests of the SM~\cite{Aad:2013wqa,Aad:2013xqa,ATLAS-Hgg,ATLAS-HZZ,Chatrchyan:2013lba,CMS-couplings,CMS-SpinCP}. 
Higgs boson production in association with a $W$ or $Z$ (weak) boson, which are respectively denoted by $WH$ and $ZH$, and collectively referred to as  $VH$ associated
production in the following,
provides direct access to the Higgs boson couplings to weak bosons.
In particular, in the $WH$ mode with subsequent 
$\hww$ decay, the Higgs boson 
couples only to $W$ bosons, at both the production and decay vertices.

Searches for $VH$ production have been performed 
at both the Tevatron
and LHC colliders, in events with leptons, $b$-jets  and either missing transverse momentum
or two central jets. Evidence for $VH$ production has been recently reported in the Tevatron combination~\cite{PhysRevD.88.052014} while no $VH$ production has been observed so far at the  LHC~\cite{PhysRevD.89.012003,Aad:2014xzb,Aad:2014eha,Aad:2014eva,Khachatryan:2014ira,Chatrchyan:2013mxa,Chatrchyan:1633401}.

In this paper,  a search for Higgs boson production in
association with a weak boson, followed by $\hww$ decay, 
is presented. The data were 
collected in 2011 and 2012 by the  ATLAS experiment at centre-of-mass energies of $\sqrt{s}=7\tev$ and 8~\tev, respectively.
In the SM, for $\mH=125$~\gev,  the cross sections of the $WH$ and $ZH$ associated production modes,
followed by the $\hww$ decay, are  0.12~\pb\ and 0.07~\pb\ at $\sqrt{s}=7\tev$ and 0.15~\pb\ and 0.09~\pb\ at $\sqrt{s}=8\tev$~\cite{Heinemeyer:2013tqa}, respectively.
Four topologies are considered, with two, three or four charged leptons in the final state (only electrons or muons are considered).
The analyses are optimised to search for both the $WH$ and $ZH$ production modes; 
a combined result for $VH$ is also presented.
The $VH$ results are then further combined with the 
$\hww \to \ell \nu \ell \nu$ analysis of gluon fusion (ggF) and vector boson
fusion (VBF) production,
for which the ATLAS Collaboration has reported the observation of the Higgs boson 
in the $\hww$ decay channel with a significance of 6.1 standard deviations~\cite{HWWllpaper}.

The combination of the ggF, VBF and $VH$ analyses, presented in this paper, is used to 
determine the couplings of the Higgs boson to
vector bosons and, indirectly, to fermions, providing further constraints on the Higgs boson couplings. 

\section{Analysis overview                                        \label{sec:analysis}            }
Higgs boson production in association with a $W$ or $Z$ boson, followed by $\hww$ decay, is sought using events with two, three or four charged leptons in the final state.
Leptonic decays of $\tau$ leptons from $\hww\TO\tau\nu\tau\nu$ are considered as signal, while no specific selection is performed for events with hadronically decaying $\tau$ leptons in the final state. In the present analysis events from $VH(H\rightarrow\tau\tau$) are considered as background.
\begin{figure}[htp]
\begin{center}
\subfigure[]{\includegraphics[trim=6cm 20cm 6cm 4cm, width = 0.45\textwidth]{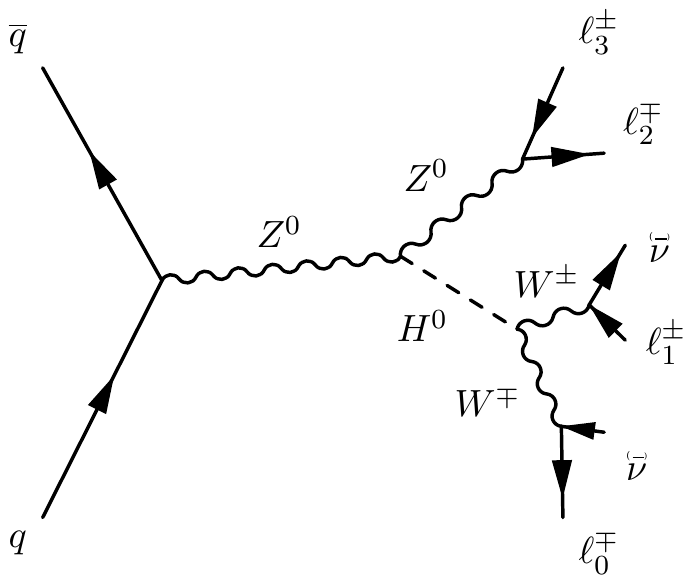}}
\subfigure[]{\includegraphics[trim=6cm 20cm 6cm 4cm, width = 0.45\textwidth]{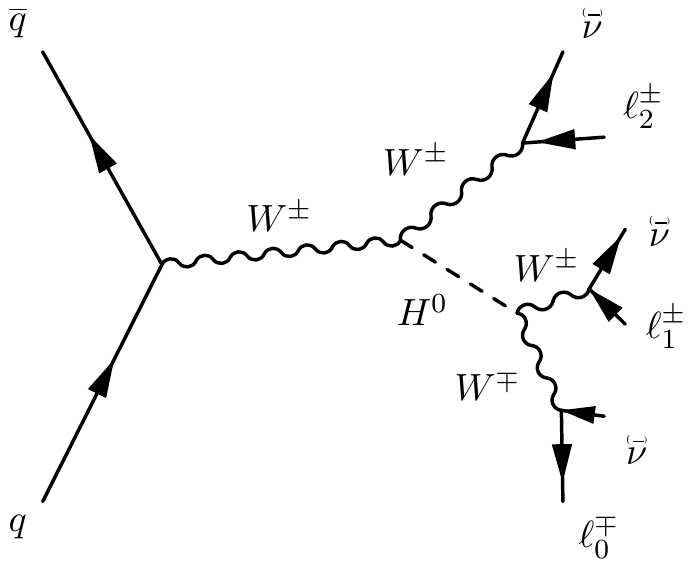}}
\subfigure[]{\includegraphics[trim=6cm 20cm 6cm 4cm, width = 0.45\textwidth]{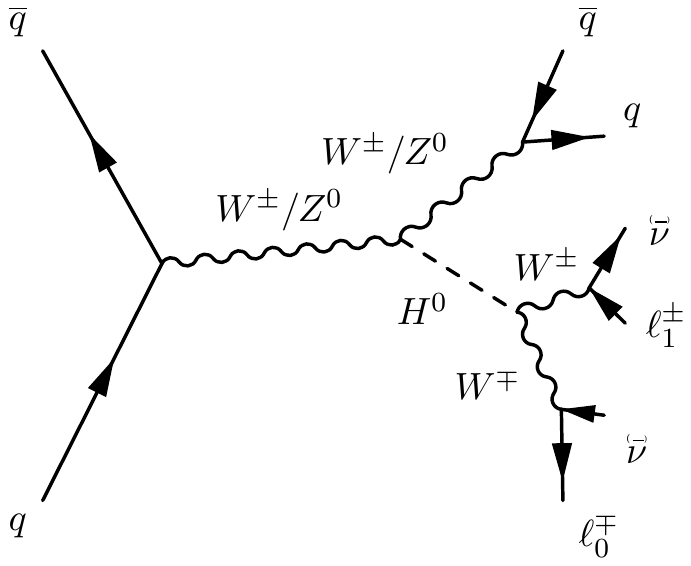}}
\subfigure[]{\includegraphics[trim=6cm 20cm 6cm 4cm, width = 0.45\textwidth]{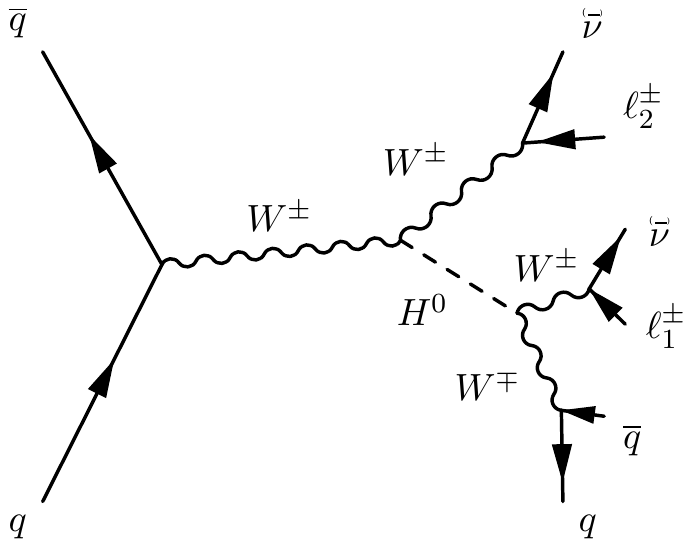}}
\caption{
         Tree-level Feynman diagrams of the $VH(\hww)$ topologies studied in this analysis:
         (a) \fourlep\ channel (b) \threelep\ channel (c) opposite-sign \twolep\ channel and (d) same-sign \twolep\ channel.
         For charged lepton external lines, the directions of arrows refer to the superscripted sign.
         Relevant arrows are assigned to the associated neutrino external lines.
         }
\label{fig:diagram_vh}
\end{center}
\end{figure}
The analysis is designed to select events which are  kinematically consistent with the $VH(\hww)$ process, in order  
to enhance the signal-to-background ratio. Figure~\ref{fig:diagram_vh} illustrates the relevant tree-level Feynman diagrams of the studied processes,
in which a Higgs boson is produced in association with a weak boson.

Four channels are analysed, defined as follows:  
\begin{itemize}
\item[(a)] {\bf \fourlep\ channel} (figure~\ref{fig:diagram_vh}(a)): 
The leading contribution consists of a process in which a virtual $Z$ boson radiates a Higgs boson, which in turn decays to a $W$ boson pair. The decays of the weak bosons produce four charged leptons and two neutrinos in the final state. The lepton pair with an invariant mass closest to the  $Z$ boson mass is labelled as ($\ell_2$, $\ell_3$), while the remaining leptons are labelled as $\ell_0$ and $\ell_1$ and are 
assumed to originate in the $\hww$ decay.
The main backgrounds to this channel are non-resonant $ZZ^{\ast}$ and $ZWW^{\ast}$ production.
\item[(b)] {\bf \threelep\ channel} (figure~\ref{fig:diagram_vh}(b)): The leading contribution consists of a process in which a virtual $W$ boson radiates a Higgs boson, and the Higgs boson decays to a $W$ boson pair. All the weak bosons decay leptonically producing three charged leptons and  three neutrinos in the final state. The lepton with unique charge is labelled as $\ell_0$, the lepton closest to $\ell_0$ in angle
 is labelled as $\ell_1$, and the remaining lepton is labelled as $\ell_2$. Leptons $\ell_0$ and $\ell_1$ are assumed to originate from the $\hww$ decay.
The most prominent background to this channel is $WZ/W\gamma^{\ast}$ production; non-resonant $WWW^{\ast}$ production is also a significant background having the same final state as the signal.
Other important backgrounds are
$ZZ^{\ast}$, $Z\gamma$, $Z$+jets, $\ttbar$ and $Wt$ production, 
as they pass the signal selection if a lepton is undetected or because of a misidentified or non-prompt lepton from a jet.
\item[(c)] {\bf Opposite-sign \twolep\ channel} (figure~\ref{fig:diagram_vh}(c)): The leading contribution consists of a process in which the weak boson $V$, which radiates the Higgs boson, decays hadronically and produces two energetic jets, while  $W$ bosons from the $\hww$ decay produce two oppositely charged leptons, labelled as  $\ell_0$ and $\ell_1$, and two neutrinos. The $WH$ process is expected to account for 70\% of the signal yield, while the $ZH$ process accounts for the remaining 30\%.
After requiring two leptons of different flavour,
the leading backgrounds for this channel are $\ttbar$ and $Wt$ processes. 
Other major components are 
$\Ztau$ and $WW$ production with two associated jets. 
Final states including $W$+jets and multijets may produce misidentified leptons, contaminating the signal region.  Other background sources include  $WZ/W\gamma^{\ast}$ production and other Higgs boson production and decay modes, especially  ggF production.
\item[(d)] {\bf Same-sign \twolep\ channel} (figure~\ref{fig:diagram_vh}(d)):
 The leading contribution consists of a process in which a $W$ boson radiates the Higgs boson, and then decays leptonically.  The radiated Higgs boson decays to 
two $W$ bosons, one decaying hadronically and the other, with the same charge as the first lepton, decaying leptonically.
The final state therefore contains
 two leptons with same charge, labelled as $\ell_1$ and $\ell_2$, two neutrinos  and two energetic jets.
Significant backgrounds in this channel are  $WZ/W\gamma^{\ast}$, $W\gamma$ and $W$+jets production; $WW$, $Z$+jets and top-quark processes also contribute to this final state.
Due to the overwhelming background from $\ttbar$ production, the selection is not optimised for events in which the lepton from the Higgs boson decay and the lepton from the associated $W$ boson have opposite charges.
\end{itemize}

\noindent All the channels described above are mutually exclusive due to the respective number of leptons with transverse momentum, $\pt$, greater than 15~\gev.
To maximise the analysis sensitivity to the $VH(\hww)$ process in each of these decay modes, the data samples for each topology, except for the opposite-sign 
\twolep\ channel, are further subdivided into several signal regions (SRs). Additional kinematic regions, with orthogonal selection criteria, designated as control regions (CRs), are used to normalise the major backgrounds in each SR by extracting normalisation factors.

The final results are extracted from a fit that simultaneously considers all SRs and CRs.
The \fourlep\ channel is split into two samples according to the number of same-flavour opposite-sign (SFOS) lepton pairs, namely  4$\ell$-2SFOS and 4$\ell$-1SFOS.  The sample containing two SFOS pairs  suffers from 
a higher background contamination than the sample with one SFOS pair.
The \threelep\ analysis requires at least one opposite-charge lepton pair, therefore the \threelep\ system must have total charge of $\pm1$. 
This analysis separates events with three same-flavour (SF) leptons, one SFOS lepton pair and zero SFOS lepton pairs, which have different signal-to-background ratios.
For the 3$\ell$-3SF and 3$\ell$-1SFOS channels a multivariate analysis is performed. The same-sign \twolep\ sample is divided into two sub-channels with one or two selected jets in the final state, namely 2$\ell$-SS1jet and 2$\ell$-SS2jet. 
The channel with two leptons of different flavour and opposite sign is denoted by 2$\ell$-DFOS in the following sections.

\section{The ATLAS detector                              \label{sec:atlas}               }
ATLAS~\cite{Aad:2008zzm}
is a multi-purpose particle physics detector with a forward-backward
symmetric cylindrical geometry\footnote{
ATLAS uses a right-handed coordinate system with its origin at the nominal interaction
point (IP) in the centre of the detector and the $z$-axis along the beam pipe. The $x$-axis
points from the IP to the centre of the LHC ring, and the $y$ axis points upward.
Cylindrical coordinates $(r,\phi)$ are used in the transverse plane, $\phi$ being the
azimuthal angle around the beam pipe. The pseudorapidity is defined in terms of the polar
angle $\theta$ as $\eta=-\ln\tan(\theta/2)$.} and close to 4$\pi$ coverage in solid
angle. 
It consists of an inner tracking detector (ID) surrounded by a thin 2 T
superconducting solenoid, electromagnetic and hadronic calorimeters, and
a muon spectrometer (MS) incorporating three large superconducting toroid magnets,
each with eight coils.

The ID covers the pseudorapidity
range $|\eta | <$ 2.5 and
consists of multiple layers of silicon pixel and microstrip
detectors, and a straw-tube transition radiation tracker.
The calorimeter system covers the pseudorapidity range $|\eta|< 4.9$. Within the region
$|\eta|< 3.2$, electromagnetic calorimetry is provided by barrel and endcap high-granularity
lead/liquid-argon (LAr) calorimeters.
An additional thin LAr presampler
covering $|\eta| < 1.8$ is used to correct for energy loss in the material upstream of the calorimeters.
Hadronic calorimetry is provided by a steel/scintillator-tile calorimeter,
covering
$|\eta| < 1.7$, and
two copper/LAr hadronic endcap calorimeters. The solid angle coverage is completed with
forward copper/LAr and tungsten/LAr calorimeter modules optimised for
electromagnetic and hadronic measurements, respectively.
The MS consists of separate trigger and high-precision tracking chambers
that measure the deflection of muons in the magnetic field generated by
superconducting air-core toroids. The precision chamber system covers the region
$|\eta| < 2.7$ with three stations of monitored drift-tube layers,
except for the forward region where the innermost station is
equipped with cathode strip chambers. The muon trigger system covers the
range $|\eta| < 2.4$ with resistive plate chambers in the barrel and
thin gap chambers in the endcap regions.
A three-level trigger system is used. The first-level trigger is
hardware-based,
using a subset of the detector information, and reduces the event rate to less than 
75 kHz.
This is followed by two software-based trigger levels, which together reduce the event rate to
about 400 Hz.

\section{Data samples                           \label{sec:DataMC}               }
The data were recorded using inclusive single-lepton and dilepton triggers.
Overall quality criteria were applied in order to suppress non-collision backgrounds such as cosmic-ray muons, beam-related backgrounds, or noise in the calorimeters.
The datasets used in the 8~\tev\ and 7~\tev\ analyses 
correspond to an integrated luminosity of \lumi\ifb\ and \lumieleven\ifb\, respectively. 
The analysis of the $\twolep$-SS channel was performed only
on the 8~\tev\ data sample, due to the low sensitivity of this channel.
The 8~\tev\ data 
 were taken at a  higher  instantaneous luminosity (${\cal L}\simeq 7\times10^{33}$ cm$^{-2}$s$^{-1}$) than that for 
the 7~\tev\ data (${\cal L}\simeq 3\times10^{33}$ cm$^{-2}$s$^{-1}$)
and with a higher number ($\simeq$ 21 versus $\simeq$ 9) of overlapping proton--proton collisions, producing higher
out-of-time and in-time pile-up~\cite{pileup}. 
The increased pile-up rate, rather than the increased centre-of-mass energy,  is the main
reason for the differences between 8~\tev\ and 7~\tev\ analysis selections. 

Table~\ref{tab:cross} lists the Monte Carlo (MC) generators used to
model the signal and background processes.
For the Higgs production processes the production cross section multiplied by the branching fraction of the \hww\ decay is shown,
while for the background processes the production cross section, including effects of cuts applied at the event generation, is presented.
\begin{sidewaystable}[htbp]
  \centering
            \vspace{0.3cm}
             \scalebox{0.75}{
               \begin{tabular}{l||ccc}
                \hline
               Process & Generator & $\sigma (\times {\rm Br})$ [pb] & Cross-section \\
                       &           &                               & normalisation \\
             \hline
             {\bf Higgs boson} &&\\
             ~~$VH$ ($\hww$)              & {\sc Pythia} ~\cite{Sjostrand:2006za,Sjostrand:2007gs} v8.165, v6.428 & 0.24, 0.20 & NNLO QCD +  NLO EW\\
             ~~$VH$ ( $H\rightarrow\tau\tau$) & {\sc Pythia}  v8.165, v6.428 & 0.07, 0.06 & NNLO QCD +  NLO EW\\
             ~~$gg\rightarrow H $ ($\hww$)                  & {\sc Powheg-Box}~\cite{powheg1,powheg2,powheg3,Bagnaschi:2011tu} v1.0 (r1655)+ {\sc Pythia} v8.165, v6.428 & 4.1, 3.3 & NNLO+NNLL QCD +  NLO EW\\
             ~~VBF ($\hww$)                  & {\sc Powheg-Box} ~\cite{Nason:2009ai} v1.0 (r1655)+ {\sc Pythia} v8.165, v6.428 & 0.34, 0.26 & NNLO QCD + NLO EW\\
             ~~\ttbar H  ($\hww$)            & {\sc Pythia} v8.165 &0.028, 0.019 &  NLO\\
              \hline

            {\bf Single boson}                                  &                                                            &\\
            ~~$Z/\gamma^{*}(\rightarrow\ell\ell)$+jets ($m_{\ell\ell} > 10 \gev$)    & {\sc Alpgen}~\cite{Alpgen} v2.14  +  {\sc Herwig}~\cite{Herwig} v6.52  & 16540, 12930 & NNLO\\
            ~~HF $Z/\gamma^{*}(\rightarrow\ell\ell)$+jets ($m_{\ell\ell} > 30 \gev$) & {\sc Alpgen} v2.14  + {\sc Herwig} v6.52                             & 126, 57       & NNLO\\
            ~~VBF $Z/\gamma^{*}(\rightarrow\ell\ell) ~ (m_{\ell\ell} > 7 \gev)$ & {\sc Sherpa}~\cite{Gleisberg:2008ta} v1.4.1 & 5.3, 2.8 & LO\\
            {\bf Top-quark} &&& \\
            ~~$\ttbar$      & {\sc Powheg-Box}\cite{Frixione:2007nw} v1.0 (r2129)+{\sc Pythia} v6.428  & 250, 180 & NNLO+NNLL\\
                            & {\sc MC@NLO}~\cite{Frixione:2002ik} v4.03 & & \\
            ~~$\ttbar W/Z$  & {\sc MadGraph}~\cite{Alwall:2011uj} v5.1.5.2, v5.1.3.28 +{\sc Pythia} v6.428 & 0.35, 0.25 & LO\\
            ~~$tqb$       & {\sc AcerMC}~\cite{Kersevan:2004yg} v3.8 +{\sc Pythia} v6.428 & 88, 65 & NNLL\\
            ~~$tb$, $tW$        & {\sc Powheg-Box}
                                  \cite{Alioli:2009je,Re:2010bp} v1.0 (r2092)+ {\sc
                                  Pythia} v6.428  &28, 20 & NNLL\\
            ~~$tZ$        & {\sc MadGraph} v5.1.5.2, v5.1.5.11 +{\sc Pythia} v6.428 &0.035, 0.025 & LO\\
            {\bf Dibosons} &&& \\
            ~~$WZ/W\gamma^{\ast}(\rightarrow\ell\ell\ell\nu) (m_{\ell\ell} > 7 \gev)$ & {\sc Powheg-Box}\cite{Melia:2011tj} v1.0 (r1508)+{\sc Pythia} v8.165, v6.428 & 12.7, 10.7 & NLO\\
            ~~$WZ/W\gamma^{\ast}(\rightarrow\ell\ell\ell\nu) ($min. $m_{\ell\ell} < 7 \gev)$ & {\sc Sherpa} v1.4.1 & 12.2, 10.5 & NLO\\
            ~~other~$WZ$ &  {\sc Powheg-Box}\cite{Melia:2011tj} v1.0 (r1508) + {\sc Pythia} v8.165 & 21.2, 17.2 & NLO\\
            ~~$q\bar{q}/qg\rightarrow
                 Z^{(\ast)}Z^{(\ast)}(\rightarrow\ell\ell\ell\ell,
                 \ell\ell\nu\nu)$ ($m_{\ell\ell} > 4 \gev$) & {\sc
                                                              Powheg-Box}\cite{Melia:2011tj}
                                                              v1.0
                                                              (r1556)
                                                              +{\sc
                                                              Pythia}
                                                              v8.165,
                                                              v6.428 &
                                                                       1.24, 0.79 & NLO\\

            ~~$q\bar{q}/qg\rightarrow Z^{(\ast)}Z^{(\ast)}(\rightarrow\ell\ell\ell\ell, \ell\ell\nu\nu)$ (min. $m_{\ell\ell} < 4 \gev$) &
                                               {\sc Sherpa} v1.4.1 & 7.3, 5.9 & NLO\\
            ~~other~$q\bar{q}/qg\rightarrow ZZ$ & {\sc Powheg-Box}\cite{Melia:2011tj} v1.0 (r1556) + {\sc Pythia} v8.165 & 6.9, 5.7 & NLO\\
            ~~$gg\rightarrow Z^{(\ast)}Z^{(\ast)}$ & gg2ZZ~\cite{gg2ZZ} v3.1.2 + {\sc Herwig} v6.52 (8~\tev\ only) & 0.59 & LO\\                          
            ~~$q\bar{q}/qg\rightarrow WW$ & {\sc Powheg-Box}\cite{Melia:2011tj} v1.0 (r1556) + {\sc Pythia} v6.428  & 54, 45 & NLO\\
            & {\sc Sherpa} v1.4.1 (for 2$\ell$-DFOS 8 \tev\ only) &  54 & NLO\\
            ~~$gg\rightarrow WW$        & gg2WW~\cite{gg2WW} v3.1.2 + {\sc Herwig} v6.52  & 1.9, 1.1 & LO\\
            ~~VBS $WZ$, $ZZ(\rightarrow\ell\ell\ell\ell, \ell\ell\nu\nu)$ $(m_{\ell\ell} > 7 \gev)$, $WW$ & {\sc Sherpa} v1.4.1  & 1.2, 0.88 & LO\\
            ~~$W\gamma~(\pt^{\gamma} > 8 \gev)$                 & {\sc Alpgen} v2.14 +{\sc Herwig} v6.52  & 1140, 970 & NLO\\
            ~~$Z\gamma~(\pt^{\gamma} > 8 \gev)$                 & {\sc Sherpa} v1.4.3  & 960, 810 & NLO\\
            {\bf Tribosons} &&& \\
            ~~$WWW^{\ast}, ZWW^{\ast}, ZZZ^{\ast}, WW\gamma^{\ast}$ &  {\sc MadGraph} v5.1.3.33, v5.1.5.10 + {\sc Pythia} v6.428 & 0.44, 0.18 & NLO\\
                    \hline
                   \end{tabular}
                     } 
  \caption{MC generators used to model the signal and background  processes. 
          The Higgs boson samples are normalised using
          the production cross section and the decay branching
          fraction computed for $\mh = 125 \gev$. 
          The values reported for the $VH$ ($\hww$) process include the NNLO contribution 
          from the $gg\rightarrow ZH $ ($\hww$) process.
          For generators and cross sections, wherever two
          comma-separated values are given, the first value refers to
          $\sqrt{s}=8$~\tev\ and the second to $\sqrt{s}=7$~\tev. When a single value is given, it refers to $\sqrt{s}=8 \tev$.
          The corresponding cross section times branching fraction of the \hww\ decay, $\sigma \times {\rm Br}$, are shown for the Higgs production processes,
          while for background processes the production cross section,
          including the effect of the leptonic branching fraction, and the $m_{\ell \ell}$ and  $p_T^{\gamma}$ cuts, as specified in the ``Process'' column, is presented.   
          `HF' refers to heavy-flavour jet production, and `VBS' refers to vector boson scattering.
When a lower cut  on $m_{\ell\ell}$ is specified, it is applied to
all SFOS lepton pairs,  while when an upper cut is indicated
it is applied to the SFOS pair of lowest mass in the event.  For the
{\sc Sherpa1.1}  $Z^{(\ast)}Z^{(\ast)}$ sample a lower cut of 4 \gev\ is applied,
in addition,  to the SFOS lepton pair of higher mass. 
          Cross sections are computed to different levels of accuracy (LO, NLO, NNLO or next-to-next-to-leading-logarithm, NNLL), 
as specified by the last column.
          }\label{tab:cross}
       \end{sidewaystable}
The samples
were simulated and normalised  for a Higgs boson of mass $\mH = 125 \gev$.
The $VH$ samples were simulated with {\sc Pythia} and  normalised to the next-to-next-to-leading-order
(NNLO) QCD calculations~\cite{Dittmaier:2011ti,Dittmaier:2012vm,Heinemeyer:2013tqa,Ciccolini:2003jy,Brein:2003wg}
with additional next-to-leading-order (NLO) electroweak (EW)
corrections computed with  {\sc Hawk}~\cite{Denner:2011rn} and applied  as a function of the transverse momentum of the
associated vector boson.
The  $gg \rightarrow ZH$ samples were simulated with {\sc
  Powheg-Box1.0} interfaced with  
{\PYTHIA 8} and  normalised to the NNLO QCD  calculations~\cite{Dittmaier:2012vm}.
Associated Higgs boson production with a $\ttbar $ pair ($t\bar{t}H$) is
simulated with {\sc Pythia8} and normalised to the
NLO QCD estimation~\cite{Dittmaier:2011ti,Dittmaier:2012vm,Heinemeyer:2013tqa}. 

The matrix-element-level calculations are interfaced to generators that model the parton shower, 
the hadronisation and the underlying event,
using either {\sc Pythia6}, {\sc Pythia8}, {\sc Herwig}  with the underlying
event modelled by {\sc Jimmy}~\cite{jimmy}, or {\sc Sherpa}.  
The CT10 parton distribution function (PDF) set~\cite{Lai:2010vv} is used 
for the {\sc Powheg-Box} and {\sc Sherpa} samples while
the CTEQ6L1 PDF set~\cite{cteq6} is used for {\sc Alpgen} and  {\sc AcerMC} samples.
The $Z/\gamma^{\ast}$ sample is  reweighted to the MRSTMCal~\cite{mrst} PDF set. 
The simulated samples are described in detail  in ref.~\cite{HWWllpaper}   with a few exceptions that are reported in the following.

The $Z/\gamma^{\ast}$ processes associated with light-  and heavy-flavour (HF) jets are modelled by {\sc Alpgen}+{\sc Herwig} with
 merged leading-order (LO) calculations. 
The simulation includes processes with up to five additional partons in the matrix element, or three additional partons in processes with $b$- or $c$-quarks.
An overlap-removal procedure is
 applied to avoid  double counting of HF in the light-jet
 samples. The sum of the two samples is normalised to the NNLO calculation of {\sc Dynnlo}~\cite{Catani:2009sm,PhysRevLett.98.222002}.
The $\ttbar W/Z$ and $tZ$ backgrounds are simulated using {\sc
  Madgraph} at LO interfaced with {\sc Pythia6}.
The production of four leptons from a pair of virtual 
$Z$ or $\gamma$ bosons, indicated by $ZZ{^\ast}$ in the following, contributes to the background in the 3$\ell$ channel
when one low-$\pt$ lepton is not detected.
Since this background is more prominent when one lepton pair has a
very low mass, a dedicated sample which requires at least one SFOS
pair with $m_{\ell\ell} < 4 \gev$, generated with {\sc Sherpa} and
normalised to the  NLO QCD cross section from the parton-level MC program  {\sc MCFM}~\cite{Campbell:2010ff}, is included.
Production of triboson processes is a major source of background, in particular
$WWW^{\ast}$ in the 3$\ell$ channel and $ZWW^{\ast}$ in the 4$\ell$
channel. They are modelled by {\sc Madgraph} interfaced with {\sc
  Pythia6} and normalised to the NLO cross section from ref.~\cite{Binoth:2008kt}.
\clearpage
All samples are processed using the full ATLAS detector simulation~\cite{atlassim} based on {\sc Geant4}~\cite{Agostinelli:2002hh},
except for $WH$, $WZ/W\gamma^{\ast}$ with $m_{\ell\ell} > 7 \gev$,
$q\overline{q}/qg \rightarrow WW$, $WW\gamma^{*}$, $\ttbar $
\noindent and  single top, which are instead simulated with {\sc Atlfast}-II~\cite{ATLAS:1300517}, a parameterisation of
 the response of the electromagnetic and hadronic calorimeters, and with {\sc Geant4} for other detector components.
The events are reweighted to ensure that the distribution of pile-up observed in the data is correctly reproduced.

\section{Event reconstruction and selection                                          \label{sec:selection}           }
\subsection{Event reconstruction}
The primary vertex of each event is selected as the vertex with the largest value of $\sum (p_{\rm T})^2$, where the sum is
over all the tracks associated with that particular
vertex. Furthermore, it is required to have at least three tracks with $p_{\rm T} >400$ MeV.

Muons are reconstructed in the region  $|\eta|<2.5$  by combining tracks reconstructed in the MS and ID~\cite{Aad:2014rra}. This analysis uses muon candidates referred to as ``Chain 1, CB muons" in ref.~\cite{Aad:2014rra}.
Electrons are identified within the region $|\eta|<2.47$, except in the transition region between barrel and endcap calorimeters ($1.37 < |\eta| < 1.52$), through the association of an ID track to a
calorimeter cluster whose shower profile is consistent with an
electromagnetic shower~\cite{electron-performance-new}.
Electron identification uses information from both the calorimetric and tracking system.
In the 7~\tev\ analysis a cut-based approach is adopted while in the 8~\tev\ analysis a likelihood-based selection 
is also exploited as described in~\cite{ele-reco}.
Following that reference, in the \fourlep\ and \threelep\ channels, electrons with \pt\ $<$ 20~\gev\ are required to satisfy the ``very tight" likelihood requirement, while electrons with \pt\ $>$ 20~\gev\ are required to satisfy the ``loose"
likelihood requirement. In the \twolep\ channels, electrons with \pt\ $<$ 25~\gev\ are required to satisfy the ``very tight" likelihood requirement, while electrons with \pt\ $>$ 25~\gev\ are required to satisfy the
``medium" likelihood requirement.    

Both a track-based and a calorimeter-based isolation selection are
applied to leptons. The isolation criteria are chosen to maximise the sensitivity to the $VH(\hww)$ process at $\mH = 125 \gev$. The track-based isolation 
is built on the computation of the scalar sum of the $\pt$ of tracks associated with the primary vertex
and inside a
cone, constructed around the candidate lepton, of size $\Delta R=0.2~\footnote{$\Delta R~ = \sqrt{(\Delta \eta)^{2} + (\Delta \phi)^{2}}$}$ and excluding the track of the candidate lepton.
 The calorimeter-based isolation uses the scalar sum of the
 transverse energies measured within a cone of $\Delta R=0.2$, 
excluding the energy of the calorimeter cluster associated with the lepton itself.
For the 8~\tev\ data the electron calorimeter-based isolation algorithm uses 
topological clusters~\cite{ele-reco}, while for the 7~\tev\ data it uses 
calorimeter cells. Cell-based isolation is used for  
muons in the calorimeter in both the 8~\tev\ and 7~\tev\ analyses. 
The calorimeter and track isolation criteria 
differ between the 8~\tev\ and 7~\tev\ data samples and
are not the same for all  the channels.
The upper bound of the calorimeter-based isolation energy varies
from 7\% to 30\% of the lepton $\pt$, while the sum of the 
$\pt$  of the tracks in the cone is required to be smaller than   4\% to 12\% of the lepton 
$\pt$, where tighter cuts are applied at low $\pt$. Less stringent isolation criteria on  energy and $\pt$ are required for the 7~\tev\
data sample, due to the lower level of pile-up compared to the 8~\tev\ data sample.

Jets are reconstructed from three-dimensional topological clusters~\cite{ATLAS-CONF-2010-053} over the region $|\eta| < $ 4.5 using the anti-$k_t$
algorithm~\cite{Cacciari:2008gp} with radius parameter $R = $ 0.4.
Jets are required to have $\pt$ larger than 25~\gev\ except for the forward region, $|\eta| > 2.4$, in which 
the threshold is raised to 30~\gev.  
In order to suppress the contamination of jets from pile-up, the following
selection is applied: the sum of the \pt\ of 
all tracks within $\Delta R$ = 0.4 of the jet axis
and that of the subset of these associated with the
primary vertex is computed. The ratio of the latter to the former is
required to be larger than 0.5 (0.75) for the 8 (7)~\tev\ data samples, for all jets with \pt\ $<$ 50~\gev\ and $|\eta| < 2.4$. 

The MV1 $b$-jet identification algorithm is used to tag jets 
containing a
$b$-hadron ~\cite{ATLAS-CONF-2011-102}.
For $b$-jets with  $|\eta| <$ 2.5 and \pt\ $>$ 20 (25)~\gev\ in the 8
(7)~\tev\ data analysis,
the selection has an efficiency of 85\%, estimated using simulated \ttbar\ events.
 It corresponds to a rejection 
of a factor of 10 against jets originating from light quarks or gluons~\cite{CONF-2013-109,CONF-2014-046}.

When two leptons are reconstructed within a cone of $\Delta R = 0.1$, 
or a lepton and a jet are reconstructed within $\Delta R = 0.3$, 
they are considered to be the same physical 
object and one of the two is removed. 
In the rare occurrence  of an overlap between two leptons of the same flavour, the higher-\pt\ lepton is kept 
while the lower-\pt\ lepton  is discarded. 
The muon is retained in the presence of an overlap 
with an electron, the electron is retained in the presence of an overlap 
with a jet, and  the jet is retained in the presence of an overlap
with a muon. 

Two variables describing the missing transverse momentum are employed in this study: one is calorimeter-based and the other is track-based.
The former, which  benefits from the large rapidity coverage of the
calorimeter and its sensitivity to neutral particles, is referenced as
$\vecmet$ ~\cite{ATLAS-CONF-2012-101}. The $\vecmet$ magnitude, $\met$, is used in the analysis selection.
The quantity $\vecmet$ is calculated as the negative vector sum of the momenta of muons, electrons, $\tau$ leptons, photons, jets and clusters of calorimeter cells that are not  associated with these objects (the ``soft term'').
In the 8~\tev\ analysis, to suppress the pile-up effect, the ratio of the scalar \pt\ sum
of all soft term  tracks associated with the primary vertex 
 to the scalar \pt\ sum of all soft term tracks from all vertices is employed. 
This ratio is used to scale all soft-event contributions to $\met$~\cite{ATLAS-CONF-2014-019}.
The track-based missing transverse momentum measurement is used to reduce the effects of pile-up on the resolution of the calorimeter-based variant~\cite{ATLAS-CONF-2010-020}.
 It is calculated as the vector sum of the transverse momenta of tracks with $\pt > 500 \mev$  that originate from the primary vertex.
This quantity is called  $\vecptmiss$, and the analysis selections are applied to its magnitude, $\ptmiss$.
In order to include neutral components in the calculation of \ptmiss\ in final states with jets,
the sum of track momenta in jets is replaced by their energy measured in the calorimeter.
 
\label{sec:reconstruction} 

\subsection{Event selection}\label{sec:eventselection} 
Events are required to contain a primary vertex.
The four channels are further split into eight 
signal regions, designed to optimise the sensitivity to the $VH(\hww)$ process,
with a specific set of selections applied to define each signal region.
The selection criteria rely on the number of leptons 
and their properties such as charge, flavour, $\pt$, 
and on the number of jets and $b$-tagged jets and on the magnitude of the missing transverse momentum.
Leptons with $\pt{} > 15\gev$ are selected and their number is used to divide the analysis in the
various channels. Similarly the analysis channels are subdivided in categories according the number of selected jets. 
Of particular importance are the invariant masses and opening angles among the selected objects,
most notably those of  opposite-sign  lepton pairs.
The spin-0 property of the Higgs boson, in conjunction with the $V\mbox{-}A$ structure of the weak interaction,
results in a preference for a small opening angle of lepton pairs from the 
$H \rightarrow WW^{\ast} \rightarrow \ell\nu\ell\nu$ decays.
On the other hand, as described in section~\ref{sec:analysis}, 
major backgrounds 
often contain $Z$ boson production or $\ttbar$ production which give rise to opposite-sign lepton pairs
with a large opening angle.
In the $\twolep$-SS channel, the lepton originating from  the Higgs boson decay is selected by minimising 
the invariant mass of the lepton and jet(s); 
cuts are then applied to the opening angle between this lepton and the closest jet in the transverse plane.
The definitions of the signal regions used for each channel are summarised
in table~\ref{tab:event_selection} and further detailed in sections~\ref{subs:4l}--\ref{subs:2lss}. 

\begin{sidewaystable}[hp!]
\begin{center}
\scalebox{0.85}{
\begin{tabular}{r||cc|ccc|ccc}
\hline 
Channel &  \multicolumn{2}{c|}{\fourlep} & \multicolumn{3}{c|}{\threelep}& \multicolumn{3}{c}{\twolep}    \\
\hline
Category &  2SFOS & 1SFOS  & 3SF & 1SFOS & 0SFOS & DFOS & SS2jet & SS1jet \\
\hline 
Trigger & \multicolumn{2}{c|}{single-lepton triggers} & \multicolumn{3}{c|}{single-lepton triggers}& \multicolumn{3}{c}{single-lepton \& dilepton triggers} \\
\hline
Num.  of leptons & 4 & 4 & 3 & 3 & 3 & 2 & 2 & 2  \\
$p_{\rm T,leptons}$ [\gev] & $>25,20,15$ & $>25,20,15$ & $>15$ &$>15$ & $>15$ & $>22,15$ & $>22,15$ & $>22,15$ \\
Total lepton charge & 0 & 0 & $\pm 1$ & $\pm 1$ & $\pm 1$ & 0 & $\pm 2$ & $\pm 2$  \\
Num.  of SFOS pairs  & 2& 1 & 2 & 1 & 0 & 0 & 0 & 0   \\
Num.  of jets & $\le 1$ & $\le 1$ & $\le 1$ & $\le 1$ & $\le 1$ & $\ge 2$ & 2 & 1  \\
$p_{\rm T,jets}$ [\gev] &$>25~(30)$ &$>25~(30)$ &$>25~(30)$ &$>25~(30)$ &$>25~(30)$ &$>25~(30)$ &$>25~(30)$&$>25~(30)$\\
Num.  of $b$-tagged jets & 0 & 0 & 0 & 0 & 0 & 0 & 0 & 0 \\
\met\ [\gev] & $>20$ & $>20$ & $>30$ & $>30$ & --- & $>20$ & $>50$ & $>45$   \\
\ptmiss\ [\gev] & $>15$ & $>15$ & $>20$ & $>20$ & --- & --- & --- & ---  \\
$|m_{\ell\ell} - m_{Z}|$ [\gev] & $<10~(m_{\ell_{2}\ell_{3}})$ & $<10~(m_{\ell_{2}\ell_{3}})$ & $>25$ & $>25$ & --- & --- & $>15$ & $>15$ \\
Min. $m_{\ell\ell}$ [\gev] & $>10~(m_{\ell_{0}\ell_{1}})$ & $>10~(m_{\ell_{0}\ell_{1}})$&  $>12$ & $>12$ & $>6$ & $>10$ & $>12~(ee,\mu\mu)$ & $>12~(ee,\mu\mu)$ \\
                        & & & & & & & $>10~(e\mu)$ & $>10~(e\mu)$\\
Max. $m_{\ell\ell}$ [\gev] & $<65~(m_{\ell_{0}\ell_{1}})$ & $<65~(m_{\ell_{0}\ell_{1}})$ &$<200$ & $<200$ & $<200$ & $<50$ & --- & ---  \\
$m_{4\ell}$ [\gev] & $>140$ & --- & --- & --- & --- & --- & --- & ---  \\
$p_{\rm T,4\ell}$ [\gev] & $>30$& --- & --- & --- & --- & --- & --- & ---  \\
$m_{\tau\tau}$ [\gev] & --- & --- & --- & --- & --- & $<(m_{Z} - 25)$ & --- & ---  \\
$\Delta R_{\ell_{0}\ell_{1}}$ & --- & ---& $<2.0$ & $<2.0$ & --- & --- & --- & --- \\
$\Delta \phi_{\ell_{0}\ell_{1}}$ [rad] & $<2.5~(\Delta \phi_{\ell_{0}\ell_{1}}^{\rm boost})$ & $<2.5~(\Delta \phi_{\ell_{0}\ell_{1}}^{\rm boost})$ & --- & --- & --- & $<1.8$ & --- & --- \\
$\mT$ [\gev] & --- & --- & --- & --- & --- & $<125$ &  ---&$>105~(\MTLead)$  \\
Min. $m_{\ell_{i}j(j)}$ [\gev] & --- & --- & --- & --- & --- & --- & $<115$ & $<70$ \\
Min. $\phi_{\ell_{i}j}$ [rad] & --- & --- & --- & --- & --- & --- & $<1.5$ & $<1.5$ \\
$\Delta y_{jj}$ & --- & --- & --- & --- & --- & $<1.2$ & --- & --- \\
$|m_{jj} - 85|$ [\gev] & --- & --- & --- & --- & --- & $<15$ & --- & --- \\
\hline 
\end{tabular}
}
\caption{
Definition of each signal region in this analysis. $\MTLead$ is the transverse mass of the leading lepton and the $\vecmet$ (see section \ref{subs:2lss} for the definition of $\MTLead$). For $p_{\rm T,leptons}$ in the \fourlep\ channel the three
values listed above refer to the leading, sub-leading, and to the two remaining leptons, respectively. For $p_{\rm T,leptons}$ in the \twolep\ channel the two values listed above refer to the leading and  sub-leading leptons, respectively. For $p_{\rm T,jets}$ the value in parentheses refers to forward jets ($|\eta| > 2.4$). 
}
\label{tab:event_selection}
\end{center}
\end{sidewaystable}

In all the  $\fourlep$ and $\threelep$ signal regions, events are recorded using inclusive single-lepton triggers,
which are fully efficient for high lepton multiplicity signatures.
For the $2\ell$ channels in 8~\tev\ data taking, dilepton triggers are also used.
In all channels at least one lepton must match a candidate reconstructed at trigger level. 
This requires the leading lepton in an event to have $\pt$ greater than 24~\gev\ in the 8~\tev\ data sample, and greater than 18~\gev\ and 20~\gev\ for muons and electrons respectively in the 7~\tev\ data sample.
Single lepton trigger efficiencies are measured with respect to
offline reconstructed leptons using leptonic \Zboson\ decays. The
measured values are approximately  95\% for electrons, 90\% for muons in the endcap and 70\%
for muons in the barrel.

\subsubsection{Four-lepton channel}\label{subs:4l}
Events in the $\fourlep$ channel are required to have exactly four leptons.
The $\pt$ of leading and sub-leading leptons must be above 25~\gev\ and 20~\gev, respectively, 
and the \pt\ of each of the remaining two leptons must exceed 15~\gev.
The total charge of the four leptons is required to be zero.
Only events with at least one SFOS lepton pair are accepted, and events are assigned to the 4$\ell$-2SFOS and 4$\ell$-1SFOS SRs according to the number of such pairs. 

In order to select final states with neutrinos, \met{} and \ptmiss\ are required to be above 20~\gev\ and above 15~\gev, respectively. 
In order to reduce the $\ttbar Z$ background, events are vetoed if they contain more than one jet.
Top-quark production is further suppressed by vetoing events with any $b$-tagged jet with \pt~above 20~\gev.
The invariant mass of $\ell_2$ and $\ell_3$, $m_{\ell_2\ell_3}$, is required to satisfy 
$|m_{\ell_2\ell_3}- m_{Z}| < 10 \gev$ (where $m_{Z}$ is the mass of the $Z$ boson), 
and the invariant mass of $\ell_0$ and $\ell_1$, $m_{\ell_0\ell_1}$, is required to be between 10~\gev\ and 65~\gev.
This requirement on  $m_{\ell_0\ell_1}$  greatly reduces the contamination from $ZZ^{(\ast)}$ production in events with two pairs of SFOS leptons.

The sensitivity is improved by exploiting two additional variables.
The variable $\Delta \phi_{\ell_{0}\ell_{1}}^{\rm boost}$ denotes the difference in azimuthal angle
between the two leptons from the Higgs boson candidate in 
the frame where
the Higgs boson's \pt\ is zero. 
The Higgs boson transverse momentum is approximated with 
$\vecptmissH \sim - \vecptmissZ - \vecptmissjet$, or with
$\vecptmissH \sim - \vecptmissZ$ if no jet is present.
The angular separation $\Delta \phi_{\ell_{0}\ell_{1}}^{\rm boost}$ is
required to be below 2.5~rad.
The magnitude of the vector sum of the lepton transverse momenta, \pt$_\mathrm{4\ell}$, can discriminate against 
the main background, $ZZ^{(\ast)}$, which has no neutrinos.
A cut requiring \pt$_\mathrm{4\ell}>$ 30~\gev\ is introduced for the 4$\ell$-2SFOS SR.
In this signal region the invariant mass of the four leptons is required to be above 140~\gev\ to remove 
events from the $H\to ZZ^{\ast} \to 4\ell$ decay, which are the target of another analysis~\cite{Aad:2014eva}.
In the signal extraction through the fit explained in section~\ref{sec:interpretation}, the 4$\ell$-2SFOS  and 4$\ell$-1SFOS SRs enter as two separate signal regions.

\subsubsection{Three-lepton channel}\label{subs:3l}
For the $\threelep$ channel, exactly three leptons with  $\pt > 15\gev$  are required with a total charge of $\pm1$. 
After this requirement, contributions from 
background processes that include more than one misidentified lepton, 
such as  $W$+jet production and inclusive $b{\bar b}$ pair production, are negligible.
Events are then separated into the 3$\ell$-3SF, 3$\ell$-1SFOS and 3$\ell$-0SFOS SRs, 
requiring three SF leptons, one SFOS lepton pair and zero SFOS lepton pairs, respectively.

In order to reduce the background from $t{\bar t}$
production, events are vetoed if they contain more than one jet.
The background from top-quark production is further suppressed by vetoing events if they contain any $b$-tagged jet with \pt~$> 20\gev$.
In order to select final states with neutrinos, \met{} is required to be above 30~\gev\ and \ptmiss\ above 20~\gev\ in the 3$\ell$-3SF and 3$\ell$-1SFOS SRs. 
In the 3$\ell$-0SFOS SR, \met\ or \ptmiss\ selections are not imposed because the main backgrounds also contain neutrinos.
The invariant mass of all SFOS pairs in the 3$\ell$-3SF and 3$\ell$-1SFOS SRs is required to satisfy $|m_{\ell\ell}- m_{Z}| > 25 \gev$.
This requirement suppresses $WZ$ and $ZZ^{*}$ events, and increases the $Z$+jets rejection.

A lower bound is set on the smallest invariant mass of pairs of oppositely charged leptons at 12~\gev\  in the 3$\ell$-3SF and 3$\ell$-1SFOS SRs, and 
at 6~\gev\ in the 3$\ell$-0SFOS SR. 
In addition, an upper bound on the invariant mass of oppositely charged leptons is set at 200~\gev\ in the three signal regions. 
These selections reject  backgrounds from HF and reduce the number of combinatorial lepton pairs from the $WZ/W\gamma^{\ast}$ process. 
The latter could indeed give larger mass values with respect to the  $WH$ process since it can proceed through the t- and u-channels, in addition to the s-channel, which is also present in $WH$ production.

The angular separation $\Delta R_{\ell_0\ell_1}$ is required to be smaller than 2 in the 3$\ell$-3SF and 3$\ell$-1SFOS SRs.
This cut favours the Higgs boson decay topology relative to that of $WZ/W\gamma^{\ast}$ events.

In the 3$\ell$-3SF and 3$\ell$-1SFOS SRs, the shape of a  multivariate
discriminant based on a Boosted Decision Trees
(BDT)~\cite{Hoecker:1019880}, which produces a multivariate classifier (``BDT Score''), is used to achieve a further separation
between  signal and background.
The main purpose of the multivariate classifier is to distinguish between the 
signal and the dominant $WZ/W\gamma^{\ast}$  and $ZZ^{\ast}$ backgrounds, and the BDT is trained against these two background processes.
The BDT parameters are chosen in order
to ensure that there is no overtraining, i.e. that the BDT is robust
against statistical fluctuations in the training samples.
The BDT input discriminating variables  
which provide the best separation between signal and background are 
the \pt\ of each lepton, the magnitude of their vector sum,
the invariant masses of the two opposite-sign lepton pairs ($m_{\ell_0\ell_1}$,  $m_{\ell_0\ell_2}$),
$\Delta R_{\ell_0\ell_1}$, $\met$, and \ptmiss. 
In the fit, the shape of the distribution of the ``BDT Score'', divided into six bins,
is used  to extract the number of observed events in the 3$\ell$-3SF and 3$\ell$-1SFOS SRs,
while the shape of the distribution of $\Delta R_{\ell_0\ell_1}$, divided into four bins,
is used to extract the number of observed events in the 3$\ell$-0SFOS SR. 
In the other channels only the event yield in each signal and control
region is used without shape information. 

\subsubsection{Opposite-sign two-lepton channel}\label{subs:2los}
In the $\twolep$-DFOS channel, exactly two leptons with $\pt$ larger than 22~\gev\ and 15~\gev\ are required.
Only opposite-sign $e\mu$ final states are considered in order to reduce the background from $Z$+jets, $WZ$ and $ZZ$ events.
A cut on the invariant mass of the lepton pair, $m_{\ell_0\ell_1} > 10\gev$, is applied to reject combinatorial dilepton backgrounds.
In order to select final states with neutrinos, \met{} is required to be above 20~\gev.
These selections reduce the 
background processes that contain jets faking leptons.
The presence of at least two jets with $\pt > 25\gev$
is required.
The background from top-quark production is reduced 
by vetoing events if they contain any $b$-tagged jets with $\pt > 20 \gev$.
To reject the $Z$+jets production that leads to $e \mu$ final states through \Ztau ~decay,
a requirement of $m_{\tau\tau} < (m_{Z} -25\gev)$ is applied, where $m_{\tau\tau}$ is the  dilepton invariant mass reconstructed using the collinear approximation~\cite{HWWllpaper}, namely under the assumptions that the lepton pair originates from $\tau$ lepton decays, the neutrinos are the only source of $\MET$ and they are collinear with the charged leptons.

Upper bounds 
on the invariant mass of the lepton pair, $m_{\ell_0\ell_1} < 50 \gev$, and the azimuthal angular separation of the lepton pair, $\Delta\phi_{\ell_0\ell_1} < 1.8$ rad, 
are applied to enhance the Higgs boson signal relative to the $WW$, $\ttbar$ and $W$+jets backgrounds.
Requirements on the rapidity separation between the two leading jets, $\Delta y_{jj} < 1.2$, and the invariant mass of the two leading jets, 
$|m_{jj}-85\gev| < 15 \gev$, are introduced to select jets from the associated $W/Z$ bosons.
The central value of the $m_{jj}$ selection interval
is larger than the $W$ boson mass 
in order to retain the acceptance for $ZH$ production with $Z \rightarrow \qqbar$ decay.

The selection $\mT < 125$~\gev\ is applied, where $\mT$ 
is the transverse mass of the dilepton system and $\met$, defined as
  $\mT = \sqrt{(E_{\rm T}^{\ell\ell}+\met)^{2} - |{\bf \pt}^{\ell\ell}+{\bf \met}|^{2}}$, 
where $E_{\rm T}^{\ell\ell} = \sqrt{|{\bf \pt}^{\ell\ell}|^{2}+m_{\ell\ell}^{2}}$.
The selections on $\Delta y_{jj}$ and $m_{jj}$ make this channel orthogonal to the ggF-enriched $n_{j} \ge 2$ category in ref.~\cite{HWWllpaper}, while orthogonality with respect to the VBF category is ensured by explicitly vetoing the BDT signal region of the VBF analysis~\cite{HWWllpaper}.
In the fit the 2$\ell$-DFOS channel enters as a single signal region.

\subsubsection{Same-sign two-lepton channel}\label{subs:2lss}
In the $\twolep$-SS channel, exactly two leptons with the same charge are required.
Lower bounds on lepton \pt{} are set to 22~\gev\ and 15~\gev\ and
both the same-flavour and different-flavour combinations are considered.
A lower bound on $m_{\ell_1\ell_2}$ is applied at 12~\gev\ for same-flavour lepton pairs and at 
10~\gev\ for different-flavour lepton pairs.
Despite the same-charge requirement, a wrong-charge assignment may allow background contributions 
from $Z$ boson decays. Therefore 
a veto on same-flavour lepton pairs with $|m_{\ell_1\ell_2}- m_{Z}| < 15 \gev$
is introduced. 

The 2$\ell$-SS2jet and 2$\ell$-SS1jet SRs require the number of jets to be exactly two or 
exactly one, respectively. 
Events with $b$-tagged jets having $\pt > 20~\gev$ are discarded.
The \met\ is required to be larger than 50~\gev\ in the 2$\ell$-SS2jet SR and larger than 45~\gev\ in the 2$\ell$-SS1jet SR.
Additional cuts are applied to events in the 2$\ell$-SS2jet and 2$\ell$-SS1jet SRs, on the following variables (see table~\ref{tab:event_selection} for details): 
the minimum invariant mass of a lepton and the jet(s) in the event, 
$m_{\ell_i {\rm j}}^{\rm min}$ ($m_{\ell_i {\rm jj}}^{\rm min}$);
the smallest opening angle between the lepton which minimises the above variable and a jet, 
$\Delta\phi_{\ell_i {\rm j}}^{\rm min}$; the transverse mass of the leading lepton and the $\vecmet$, $\MTLead = \sqrt{2 \times p_{\rm T, lead} \times \met \times (1 - \cos(\phi_{\rm lead} - \phi_{\met}))}$,
where $p_{\rm T, lead}$ and $\phi_{\rm lead}$ are respectively the transverse momentum and $\phi$ angle of the leading lepton.
Lower values of $m_{\ell_i {\rm j}}^{\rm min}$ ($m_{\ell_i {\rm jj}}^{\rm min}$)
and of $\Delta\phi_{\ell_i {\rm j}}^{\rm min}$  
favour Higgs boson decays relative to the
major backgrounds. High values of  $\MTLead$ help in reducing $W$+jets background.
The \pt\ threshold for the sub-leading muon in the $\mu\mu$ channel is increased to 20~\gev\ in both the SRs to suppress misidentified muons from $W$+jets and multijet production.

In the fit, the 2$\ell$-SS2jet and 2$\ell$-SS1jet SRs
are further split into four signal regions according to the combination of lepton flavours in each event: 
$ee$, $e\mu$, $\mu e$ and $\mu \mu$, where $e\mu$ refers to the case in which the electron has 
leading $\pt$ while $\mu e$ refers to the case in which the muon has leading $\pt$.
This splitting is motivated by the expected differences in the background contributions, for example $W\gamma$,
which is expected to be zero in the $\mu\mu$ channel but not in the other channels.

\subsubsection{Signal acceptance}\label{subs:signal}
The number of expected $VH(\hww)$ events surviving the event selections
is presented for each channel in table~\ref{tab:signal}.
The total acceptance for $WH(H \rightarrow WW^{\ast} \rightarrow \ell\nu\ell\nu)$, 
$WH(H\rightarrow  WW^{\ast} \rightarrow \ell\nu q q)$ and
$ZH(H\rightarrow  WW^{\ast} \rightarrow \ell\nu\ell\nu)$ is 3.7\%, 0.3\% and 1.9\%, respectively.
The analysis acceptance for the $ZH(H\rightarrow  WW^{\ast} \rightarrow \ell\nu q q)$ process is negligible.
The acceptance is defined as the ratio of the number of events in the SRs to the number of events expected according to the branching fractions for the various processes.
Associated Higgs boson production followed by the decay  $H\rightarrow\tau\tau$  
cannot be completely isolated from the selected final states. 
Therefore the results presented in this paper, with the exception of
 Section \ref{sec:couplings} for consistency of the analysed model,  include this process as part of the background, 
with the production cross section ($\sigma_{VH}$) and ${\rm Br}(H\rightarrow\tau\tau)$ fixed to the SM value.

\clearpage
\begin{table}[h!]
\begin{center}
\scalebox{0.93}{
\begin{tabular}{r||cc|ccc|ccc}
 \multicolumn{9}{l}{(a) 8~\tev\ data sample}                       \\
\hline 
Channel &  \multicolumn{2}{c|}{\fourlep} & \multicolumn{3}{c|}{\threelep}& \multicolumn{3}{c}{\twolep}    \\
\hline
Category &  2SFOS & 1SFOS  & 3SF & 1SFOS & 0SFOS & DFOS & SS2jet & SS1jet \\
\hline 
$WH~(H\rightarrow WW^{\ast})$&---                  &---                 & 0.56   & 1.4  & 1.3   & 1.5   & 1.0   & 1.8  \\
$ZH~(H\rightarrow WW^{\ast})$& 0.21   & 0.24  & 0.17  & 0.18 & 0.15 & 0.67 & 0.02 & 0.19  \\
\hline
\multicolumn{1}{c||}
{$VH~(H\rightarrow WW^{\ast})$}& 0.21 & 0.24 & 0.73 & 1.6 & 1.4  & 2.2 & 1.0 & 2.0  \\
\cline{2-9}
(all categories)  & \multicolumn{8}{c}{9.4 }\\

\hline
 \multicolumn{9}{c}{}                       \\
 \multicolumn{9}{l}{(b) 7~\tev\ data sample} \\ 
\cline{1-7}
$WH~(H\rightarrow WW^{\ast})$&---                   &--                   & 0.12 & 0.29 & 0.26 & 0.21 \\
$ZH~(H\rightarrow WW^{\ast})$&  0.023 & 0.021 & 0.013 & 0.033 & 0.028 & 0.075 \\
\cline{1-7}
\multicolumn{1}{c||}
{$VH~(H\rightarrow WW^{\ast})$}& 0.023 & 0.021 & 0.13 & 0.32 & 0.29 & 0.29 \\
\cline{2-7}
(all categories) & \multicolumn{6}{c}{1.1 } & &\\
\cline{1-7}
\cline{1-7}
\end{tabular}
}
\caption{Number of expected $VH(\hww)$ events in the signal regions, for $\mH = 125\gev$, in the (a) 8~\tev\ and (b) 7~\tev\ data samples.}
\label{tab:signal}
\end{center}
\end{table}

\section{Background modelling                                      \label{sec:bkg}                 }
The background contamination in the signal regions results from various physics processes,
each modelled by one of the following methods:
\begin{itemize}
\item Pure MC prediction: rates and differential distributions (shapes) are extracted from simulation and normalised to the cross sections in table~\ref{tab:cross};
\item MC prediction normalised to data: rates are extracted from data in control regions but shapes are extracted from simulation;
\item Pure data-driven prediction: rates and shapes are extracted from data.
\end{itemize}

Misidentified-lepton backgrounds ($W$+jets, multijets) in  the
\twolep\ channels are estimated by using a purely data-driven method,
which utilises the rate at which a jet is misidentified as a lepton~\cite{HWWllpaper}.
Table~\ref{tab:CRs} summarises the method adopted for each process in each signal region. 
The labels ``MC'' and ``Data'' represent the pure MC prediction and the
pure data-driven estimation.
For backgrounds modelled by  
simulation with a normalisation factor (NF) computed using data, 
the names of relevant control regions are shown as defined in tables~\ref{tab:cr_three_four_lep} and \ref{tab:cr_twolep}.
The ratio of $\ttbar$ yields to $tW$ yields is found to be compatible between all the CRs and associated SRs, 
thus only one NF is computed per CR for the ``Top'' category.
The ggF and VBF productions of Higgs bosons are treated
as background as discussed 
in section~\ref{sec:interpretation}.

Definitions of control regions in the $\fourlep$ and $\threelep$ analyses are presented
in table~\ref{tab:cr_three_four_lep}, and those defined in the $\twolep$ analyses are shown in table~\ref{tab:cr_twolep}.  The CRs are made orthogonal to the corresponding SRs  by inverting some selections with respect to the SR definitions.
Such selections are in boldface font in the tables and are further explained in the following sections. 

\begin{table}[h!]            

\begin{center}   
\scalebox{0.95}{    
\begin{tabular}{l||c|c|c|c}
\hline
Channel  & \fourlep\  & \threelep\ & \multicolumn{2}{c}{ \twolep\ } \\
\hline
Category       & 2SFOS, 1SFOS & 3SF, 1SFOS, 0SFOS & DFOS & SS2jet, SS1jet \\
\hline
Process      &   &  &  &  \\
~~$VVV$               & MC       & MC                    & MC             & MC           \\
~~$WZ/W\gamma^{\ast}$  & ---       & 3$\ell$-$WZ$ CR, 3$\ell$-$Z$jets CR & MC             & 2$\ell$-$WZ$ CR     \\
~~$ZZ^{\ast}$          & 4$\ell$-$ZZ$ CR & 3$\ell$-$ZZ$ CR, 3$\ell$-$Z$jets CR & MC             & MC           \\
~~OS $WW$             & ---      & MC                    & MC             & 2$\ell$-$WW$ CR     \\
~~SS $WW$             & ---      & MC                    & ---            & MC           \\
~~$W\gamma$           & ---      & ---                   & ---            & 2$\ell$-$W\gamma$ CR \\
~~$Z\gamma$           & ---      & 3$\ell$-$Z\gamma$ CR          & MC            & MC          \\
~~$Z/\gamma^{\ast}$    & ---      & 3$\ell$-$Z$jets CR, 3$\ell$-$ZZ$ CR & 2$\ell$-$Z\tau\tau$ CR  & 2$\ell$-$Z$jets CR  \\
~~$W$+jets            & ---      & ---                   & Data           & Data         \\
~~Multijets           & ---      & ---                   & Data           & Data         \\
~~Top                 & MC       & 3$\ell$-Top CR             & 2$\ell$-OSTop CR    & 2$\ell$-SSTop CR \\
\hline
\end{tabular}
}
\caption{Summary of background modelling.``$VVV$'' represents the triboson processes $WWW^{\ast}$, $ZWW^{\ast}$, $ZZZ^{\ast}$ and $WW\gamma^{\ast}$.
``Top'' processes include $\ttbar$ and single-top production dominated by $tW$ with $W\rightarrow \ell \nu$ decay, as well as $\ttbar W/Z$. 
         Some backgrounds are normalised by rescaling the MC yields
         by the data-to-MC ratio measured in CRs. For these backgrounds the 
         names of the most important CRs are listed.
         The symbol ``---'' denotes a negligible contribution to the total background in the signal region.}
\label{tab:CRs}\end{center}
\end{table}

\begin{table}[h!]
\begin{center}
\scalebox{0.67}{
\begin{tabular}{l||c|ccccc}
\hline 
Channel & \multicolumn{1}{c|}{\fourlep} & \multicolumn{5}{c}{ \threelep\ } \\
\hline
CR                            & $ZZ$                  & $WZ$        & $ZZ$               & $Z$jets          & Top       & $Z\gamma$ \\
\hline
Number of leptons             &4                         &3         &3                           &3                 &3             &3 \\
Total lepton charge           &0                         &$\pm1$    &$\pm1$                      &$\pm1$            &$\pm1$        &$\pm1$ \\
Number of SFOS                &2                         &2 or 1    & 2 or 1                     &2 or 1            &2 or 1        &2 or 1 \\
                              &                          &          & ($ee\mu$ or $\mu\mu\mu$) &          &                         &($\mu\mue$ or $eee$) \\
Number of jets                &$\le1$                    &$\le1$    &$\le1$                      &$\le1$            &$\ge1$        &$\le1$ \\
Number of $b$-jets     &0                         &0         &0                           &0                 &{$\mathbf{\ge1}$}   &0 \\
\met\ (and/or) \ptmiss\ [\gev]     &---                        &$>30$ and $>20$ &{\bf$\mathbf{<30}$ or $\mathbf{<20}$}    &{\bf$\mathbf{<30}$ and $\mathbf{<20}$}  &$>30$ and $>20$   &{\bf$\mathbf{<30}$ or $\mathbf{<20}$} \\
$|m_{\ell\ell} - m_{Z}|$ [\gev] &$<10(m_{\ell_2\ell_3})$   &$\mathbf{<25}$&--- &$\mathbf{<25}$ &$>25$      &--- \\
$|m_{\ell\ell\ell} - m_{Z}|$ [\gev] &---                   &---       &$<15$                       &$>15$     &---        &$\mathbf{<15}$ \\
Min. $m_{\ell\ell}$ [\gev]     &$\mathbf{>65}(m_{\ell_0\ell_1})$&$>12$     &$>12$                       &$>12$     &$>12$      &$>12$ \\
Max. $m_{\ell\ell}$ [\gev]     &---                       &$<200$    &$<200$                      &$<200$    &---        &$<200$ \\
$\Delta R_{\ell_{0}\ell_{1}}$    &---                       &$<2.0$    &$<2.0$                      &$<2.0$    &---        &$<2.0$ \\
\hline 
\end{tabular}
}
\caption{ Definition of control regions in the \fourlep\ and \threelep\ analyses. Selections indicated in boldface font are designed to retain the CR orthogonal to the relevant SR.}
\label{tab:cr_three_four_lep}
\end{center}
\end{table}

\begin{table}[h!]
\begin{center}
\scalebox{0.62}{
\begin{tabular}{l||cc|ccccc}
\hline 
Channel &  \multicolumn{2}{c|}{DFOS \twolep\ }& \multicolumn{5}{c}{SS \twolep\ } \\
\hline
CR                            &  OSTop & $Z\rightarrow \tau\tau$ & W$\gamma$ & $WZ$ & $WW$ & SSTop & $Z$jets \\
\hline
Number of leptons             &2        &2          &2              &{\bf 3}          &2               &2                 &2 \\
                              &         &           &{\bf$\mathbf{\ge1}$ conversion $\mathbf{e}$} &         &         &                  &\\
Total lepton charge           &0        &0          &$\pm2$         &$\pm1$           &{\bf 0}         &{\bf 0}           &{\bf 0} \\
Number of SFOS                &0        &0          &---            &---              &---             &---               &--- \\
Number of jets                &$\ge2$     &$\ge2$       &2 or 1         &2 or 1           &2 or 1          &2 or 1            &2 or 1 \\
Number of $b$-jets            &0        &0          &0              &0                &0               &$\mathbf{\ge1}$       &0 \\ 
\met\ [\gev]              &$>20$    &$>20$      &$>45$ (1j)     &$>45$ (1j)       &$>85$ (1j)      &$>45$ (1j,$ee,\mu\mu$)&$>45$ (1j) \\
                          &&&&&                                                                    &$>60$ (1j,$e\mu$)    &$<85$ (1j,$e\mu$) \\
                              &         &           &$>50$ (2j)     &$>50$ (2j)       &$>80$ (2j)      &$>50$ (2j,$ee,\mu\mu$)&$>50$ (2j,$ee,\mu\mu$) \\
                              &         &           &               &                 &                &$>60$ (2j,$e\mu$)        &$<80$ (2j,$e\mu$) \\
$|m_{\ell\ell} - m_{Z}|$ [\gev] &---      &---      &--- &$<15~({\mathrm{OS}}~ee,\mu\mu)$&$>15~(ee,\mu\mu)$&$>15~(ee,\mu\mu)$
 &$\mathbf{<15}~(ee,\mu\mu)$  \\
Min. $m_{\ell\ell}$ [\gev]     &{\bf$\mathbf{>90}$ (8~\tev)}&$>10$     &$>12~(ee,\mu\mu)$&$>12~(ee,\mu\mu)$&$>12~(ee,\mu\mu)$ &$>12~(ee,\mu\mu)$ &$>12~(ee,\mu\mu)$ \\
                              &{\bf$\mathbf{>80}$ (7~\tev)}&&&&&& \\
                              &          &          &$>10~(e\mu)$    &$>10~(e\mu)$      &$>10~(e\mu)$     &$>12~(e\mu)$       &$>55~(e\mu)$ \\
Max. $m_{\ell\ell}$ [\gev]     &---       &$<70$     &$<50$           &---             &---              &---              &$<80~(e\mu)$ \\
$m_{\tau\tau}$ [\gev]                 &$<(m_{Z} - 25)$  &---        &---            &---              &---             &---               & \\
$\Delta \phi_{\ell_{0}\ell_{1}}$ [rad]&---      &$\mathbf{>2.8}$&$<2.5$          &---             &---              &---              & \\
$\mT$ [\gev]              &---      &---     &$>105$ (1j)    &$>105$ (1j)      &$>105$ (1j)     &$>105$ (1j)           &---\\
Min. $m_{\ell_{i}j}$ [\gev]     &---      &---        &$<70$          &$<70$            &$<70$           &$<70$             &$<70$ \\
Min. $m_{\ell_{i}jj}$ [\gev]    &---      &---        &$<115$         &$<115$           &$<115$          &$<115$            &$<115$ \\
Min. $\phi_{\ell_{i}j}$  [rad]       &---      &---       &$<1.5$         &$<1.5$           &---             &---               &--- \\
$\pt^{\ell\ell}$ [\gev]        &---      &---        &$>30$          &---              &---             &---               &--- \\
\hline
\end{tabular}
}
\caption{Definition of control regions in the \twolep\ analyses.
Selections indicated in boldface font are designed to keep the CR orthogonal to the relevant SR.}
\label{tab:cr_twolep}
\end{center}
\end{table}

\subsection{Background in the four-lepton channel}
The main backgrounds that contribute to the 4$\ell$-2SFOS and 4$\ell$-1SFOS 
SRs are diboson processes,
dominated by $ZZ^{\ast}$ with \met\ from $\Ztau$ decay, 
and triboson processes,
in particular $ZWW^{\ast}$, which has the same signature as the signal.
These processes respectively account for about 85\% and 15\% of the total background 
contamination.
To normalise $ZZ^{\ast}$ a dedicated CR, the 4$\ell$-$ZZ$ CR, is defined by inverting the requirement on the invariant mass of dileptons from the Higgs boson candidate.
All the other minor background processes, listed in table~\ref{tab:CRs}, are modelled by simulation. 

\subsection{Background in the three-lepton channel}
Three classes of backgrounds contribute to the 3$\ell$ channel.
The first class comprises diboson processes: $WZ/W\gamma^{\ast}$,
$ZZ^{\ast}$ with an undetected lepton mainly due to its low-$\pt$, and $Z\gamma$, in which the photon converts to electron--positron pairs.
The $ZZ^{\ast}$ contribution in this channel is mainly due to single-resonant $ZZ^{\ast}$ production 
where the three-lepton invariant mass is just below the $Z$ boson mass.
The second class includes triboson processes, mainly $WWW^{\ast}$.
The last class of backgrounds are processes with a misidentified lepton, mainly $Z$+jets and top-quark pair production.

In the 3$\ell$-3SF and 3$\ell$-1SFOS SRs, $WZ/W\gamma^{\ast}$ and $ZZ^{\ast}$ represent the leading background contributions accounting for about 80\%
of the total background yields, with 65\%  from $WZ/W\gamma^{\ast}$ and 15\% from $ZZ^{\ast}$.
Production of $Z\gamma$, $VVV$, $Z$+jets and top-quarks share the remaining background fraction equally.
The 3$\ell$-0SFOS SR  contains contributions of similar size from $WZ/W\gamma^{\ast}$, $VVV$ and top-quark production. In this SR 
the total background event yield is about eight times lower than in the 3$\ell$-3SF and 3$\ell$-1SFOS SRs.

A 3$\ell$-$WZ$ CR is defined by reversing the $Z$-veto requirement,  in order to select events with a $Z$ boson decay.
The 3$\ell$-$ZZ$ CR and 3$\ell$-$Z\gamma$ CR are defined by requiring low 
\met\ values 
to reflect the absence of final-state neutrinos in the background process 
under study.
For these control regions, the invariant mass of the three leptons must be consistent with the 
$Z$ boson mass. 
These regions are further distinguished according to the flavour combination of the three leptons, namely
$eee$ or $\mu\mue$ for the 3$\ell$-$Z\gamma$ CR, and 
$\mu\mu\mu$ or $ee\mu$ for the 3$\ell$-$ZZ$ CR.

The 3$\ell$-$Z$jets CR is defined by reversing the \met\ and the $Z$-veto selections.
The properties of misidentified electrons and muons are different;
therefore the 3$\ell$-$Z$jets CR is further split 
into the misidentified-electron component ($eee+\mu\mu e$
events) and misidentified-muon component ($\mu\mu\mu+ee\mu$ events) and
an 
NF is assigned to each component.
In the 7~\tev\ data sample, the predicted $Z$+jets event yield with misidentified muons in the $\threelep$ SRs is negligible.
Furthermore, the number of events in the $3\ell$-$Z$jets CR with such a lepton flavour combination is too small
to reliably extract the NF.
Therefore the estimation of the misidentified-muon component is taken directly from simulation.

The 3$\ell$-Top CR is defined by requiring at least one $b$-tagged jet.
The $Z$+jets contribution is difficult to isolate from other processes that include $Z$ bosons.
Thus the NF for this process is constrained not only by the
3$\ell$-$Z$jets CR, but also in part by the 3$\ell$-$WZ$ CR and the 
3$\ell$-$ZZ$ CR, as indicated in table~\ref{tab:CRs}.

\subsection{Background in the opposite-sign two-lepton channel}

The dominant background in this channel is top-quark production, which accounts for about 50\% of the total contamination.
The 2$\ell$-OSTop CR is defined by requiring a high invariant mass of the lepton pair in the final state.
As the $b$-jet rejection criteria are the same in the CR and SR,  
the systematic uncertainties related to $b$-tagging largely cancel between the two regions.
The second dominant background is $\Ztau$, which accounts for 20\% of the total
background in the SR.
A dedicated control region, 2$\ell$-$Z\tau\tau$ CR, is defined by requiring a large opening angle between the two leptons.
The $WW$ process constitutes the third largest background, accounting for 10\% of the total.
Due to a difficulty in separating this process from $\ttbar$ events over a wide kinematic region, 
no dedicated CR is defined, and this process is modelled purely by MC simulation.

The contribution of backgrounds with misidentified leptons, $W$+jets and multijet, accounts for 10\% of the total background.
The misidentified-lepton background rate has an uncertainty of  40\%. 
Due to this large uncertainty, the misidentified-lepton background contributes significantly to this channel.
The $WZ/W\gamma^{\ast}$ production and the ggF production followed by $\hww$ decay, each representing 5\%  of the total background, are modelled with MC simulation.

\subsection{Background in the same-sign two-lepton channel}

The  $WZ/W\gamma^{\ast}$ and $W$+jets processes each account for one third of the total background.
Some of the
$WZ/W\gamma^{\ast}$ events with three leptons enter the selection when
one of the leptons escapes detection.
To normalise this process, the 2$\ell$-$WZ$ CR is defined by selecting events with three leptons.
The contamination from $W$+jets events with one misidentified lepton
is estimated by using the same data-driven method used in the $\twolep$-DFOS channel.

The remaining background processes contribute at the 10\% level or less.
The normalisation of $W\gamma$
is based on the 2$\ell$-$W\gamma$ CR, defined 
by requiring at least one electron consistent with a conversion, including a requirement that the electron does not have a hit in the innermost pixel layer.
The background contribution to the \twolep-SS channel
due to lepton charge misidentification, in otherwise charge-symmetric
processes, is found to be relevant only for electrons and affects
top-quark production, opposite-sign $WW$ and $Z$+jets. 
It represents 10\% of the total background in the \twolep-SS1jet SR and
3\% of the total background in the \twolep-SS2jet SR.
The 2$\ell$-SSTop CR, 2$\ell$-$WW$ CR and 2$\ell$-$Z$jets CR are defined selecting opposite-sign leptons to normalise these contributions.
Moreover, in  2$\ell$-SSTop CR at least one $b$-tagged jet is selected.
Due to the small production rate, no control region is defined to normalise 
the $WW$ events  from vector boson scattering with same charge, whose rate is taken directly from simulation. 

\subsection{Normalisation factors and composition of control regions}

NFs are computed through the signal extraction fit explained in section~\ref{sec:interpretation}. 
Table~\ref{tab:CRs} lists 
the main background processes, along with the CRs that contribute to the determination of their NFs.
The NFs, which are specific to each signal region, are fitted taking into account only the total
number of expected and observed events in each CR, separately for the
8~\tev\ and 7~\tev\ data samples, and are summarised in table~\ref{tab:NFs}. 
The numbers of observed and expected events from simulation in the 
  8~\tev\ and 7~\tev\ data analysis are summarised 
in table~\ref{table:CR34}.

Background spectra and the expected composition of the CRs in 8~\tev\ collisions are shown in figures~\ref{Figure:CR-3lep}--\ref{Figure:CR-SS}.
The $VH(\hww)$ component is 
shown on top of the background to demonstrate that
there is no significant signal leakage in the CRs.
The exact amount of signal leakage in each CR is presented in table~\ref{table:CR34}.

In these tables and figures, each background process normalised using CRs is presented separately,
while backgrounds that are not normalised using CRs are grouped together as ``Others''. 
The ggF and VBF production of Higgs bosons, 
and the $VH(H \rightarrow \tau\tau)$ process 
are included in the ``Others'' category, assuming $\mH = 125\gev$ and 
the SM value for the 
cross sections and for the branching fraction.

\begin{table}[h!]            
\begin{center}   
\begin{tabular}{l||c|c|cc}
 \multicolumn{5}{l}{(a) 8 \tev\ data sample} \\
\hline
Channel & \fourlep\ & \threelep\ & \multicolumn{2}{c}{ \twolep\ } \\
\hline
Category       & 2SFOS, 1SFOS & 3SF, 1SFOS, 0SFOS & DFOS & SS2jet, SS1jet \\
\hline
\hline
Process      &   &  &  &  \\
~~$WZ/W\gamma^{\ast}$  & ---                  & $1.08^{+0.08}_{-0.06}$              & ---                  & $0.94 \pm 0.10$\\
&&&&\\
~~$ZZ^{\ast}$          & $1.03^{+0.11}_{-0.10}$ & $1.28^{+0.22}_{-0.20}$             & ---                  & ---          \\
~~OS $WW$             & ---                  & ---                               & ---                  & $0.80 \pm 0.33$ \\
~~$W\gamma$           & ---                  & ---                               & ---                  & $1.06 \pm 0.12$ \\
~~$Z\gamma$           & ---                  & $0.62^{+0.15}_{-0.14}$              & ---                  & ---          \\
&&&&\\
~~$Z/\gamma^{\ast}$    & ---                  & $0.80^{+0.68}_{-0.53}$ ($\mu$-misid) & $0.90^{+0.18}_{-0.16}$ & $0.86 \pm 0.30$  \\
&&&&\\
                      &                      & $0.33^{+0.12}_{-0.11}$ ($e$-misid)   && \\
&&&&\\
~~Top                 & ---                  & $1.36^{+0.34}_{-0.30}$              & $1.05^{+0.16}_{-0.14}$ & $1.04 \pm 0.08$  \\
&&&&\\
\hline
 \multicolumn{5}{l}{} \\
 \multicolumn{5}{l}{(b) 7 \tev\ data sample} \\ \cline{1-4}
Process      &   &  &  &  \\
~~$WZ/W\gamma^{\ast}$  & ---                  & $1.02^{+0.12}_{-0.11}$
                                 & ---           & \\
&&&&\\
~~$ZZ^{\ast}$          & $1.59^{+0.36}_{-0.31}$ & $1.78^{+0.51}_{-0.42}$            & ---           & \\
&&&&\\
~~OS $WW$             & ---                  & ---                             & ---           & \\
~~$W\gamma$           & ---                  & ---                             & ---           & \\
~~$Z\gamma$           & ---                  & $0.45^{+0.09}_{-0.09}$            & ---           & \\
&&&&\\
~~$Z/\gamma^{\ast}$    & ---                  & $0.68^{+0.16}_{-0.15}$ ($e$-misid) & $1.11^{+0.38}_{-0.34}$ & \\
&&&&\\
~~Top                 & ---                  & $1.25^{+0.66}_{-0.52}$            & $0.93^{+0.16}_{-0.14}$ & \\
&&&&\\
\cline{1-4}
\cline{1-4}
\end{tabular}

\caption{Summary of background normalisation factors in the (a) 8~\tev\ and (b) 7~\tev\ data samples.
         The uncertainties include both the statistical and systematic components (see section~\ref{sec:systematics}).
         ``---'' denotes that the background process, 
               when considered, is normalised by MC simulation.}
\label{tab:NFs}
\end{center}
\end{table}

\begin{table}[h!]
\begin{center}
\scalebox{0.58}{
\begin{tabular}{l|| D{,}{\pm}{-1} | D{,}{\pm}{-1} D{,}{\pm}{-1} D{,}{\pm}{-1} D{,}{\pm}{-1} D{,}{\pm}{-1} | D{,}{\pm}{-1} D{,}{\pm}{-1}}
\multicolumn{9}{l}{ (a) 8~\tev\ data sample}                       \\  
\hline 
Channel            & \multicolumn{1}{c|}{ \fourlep\ } &
                                                        \multicolumn{5}{c|}{ \threelep\ } & \multicolumn{2}{c}{ DFOS \twolep\ }     \\ 
\hline
CR                 & \multicolumn{1}{c|}{$ZZ$} & \multicolumn{1}{c}{$WZ$} & \multicolumn{1}{c}{$ZZ$} & \multicolumn{1}{c}{$Z$jets} & \multicolumn{1}{c}{Top} & \multicolumn{1}{c|}{$Z\gamma$} & \multicolumn{1}{c}{$Z\tau\tau$} & \multicolumn{1}{c}{OSTop}\\
 \hline
Observed events           & \multicolumn{1}{c|}{122} & \multicolumn{1}{c}{578} & \multicolumn{1}{c}{60} & \multicolumn{1}{c}{251} & \multicolumn{1}{c}{55} & \multicolumn{1}{c|}{156} & \multicolumn{1}{c}{328} & \multicolumn{1}{c}{1169}\\ 
MC prediction             & 121 , 16 & 576 , 63  & 60 , 10     & 249 , 46 & 55 , 12     & 155 , 31 & 326 , 55 & 1160 , 150\\ 
\multicolumn{1}{l||}{MC (no NFs)}    & 118 , 10 & 543 , 50  & 48 , 4  & 351 , 40 & 48 , 6  & 188 , 17 & 354 , 56 & 1120 , 140\\ 
\hline
Composition ($\%$)        &&&&&&&&\\
$~~WZ/W\gamma^{\ast}$     &   \multicolumn{1}{c|}{---} & 89.3 , 1.5
                                                                                          & 5.5 , 1.0   & 25.9 , 3.5   & 20 , 4  &  1.68 , 0.31   &\multicolumn{1}{c}{---}&\multicolumn{1}{c}{---}\\
$~~ZZ^{\ast}$             &   99.49 , 0.17               & 6.7 , 1.2    & 90.1 , 2.1   & 38 , 5   & 3.6 , 1.2   &  47 , 6  &\multicolumn{1}{c}{---}&\multicolumn{1}{c}{---}\\
$~~Z\gamma$               &   \multicolumn{1}{c|}{---} & 0.54 , 0.17
                                                                                          & 0.6 , 0.5    & 5.5  , 1.5   & 2.4 , 0.9   &  43 , 7  &\multicolumn{1}{c}{---}&\multicolumn{1}{c}{---}\\
$~~Z$+jets                &   \multicolumn{1}{c|}{---} & 1.1 , 0.5  & 2.1 , 1.5    & 29 , 7   & 5.50 , 3.34   &  8.3 , 3.4   &78.2 , 2.8 &0.7 , 0.4\\
$~~$Top                   &    0.019 , 0.012             & 0.66 , 0.18  & 0.27 , 0.13  & 0.081 , 0.034  & 64 , 6  &  0.13 , 0.06 &10.5 , 1.6 &71.3 , 3.3\\
$~~$Others                &   0.49 , 0.17              & 0.80 , 0.16  & 1.16 , 0.20    & 0.87 , 0.13  & 3.6 , 0.6   &  0.33 , 0.06 &11.2 , 1.9 &27.8 , 3.2\\
$~~VH~(H\rightarrow WW^{\ast})$    &   0.026 , 0.006     & 0.93 , 0.16  & 0.26 , 0.11  & 0.37 , 0.09  & 0.52 , 0.13 &  0.052 , 0.011 &0.100 , 0.018&0.21 , 0.04\\
\hline 
\end{tabular}
}
\end{center}

\begin{center}
\scalebox{0.58}{
\begin{tabular}{l|| D{,}{\pm}{-1} D{,}{\pm}{-1} D{,}{\pm}{-1} D{,}{\pm}{-1} D{,}{\pm}{-1} }
\hline
Channel       & \multicolumn{5}{c}{SS \twolep\ } \\
\hline
CR            & \multicolumn{1}{c}{$W\gamma$} & \multicolumn{1}{c}{$WZ$} & \multicolumn{1}{c}{$WW$} & \multicolumn{1}{c}{SSTop}& \multicolumn{1}{c}{$Z$jets} \\
\hline 
\hline
Observed events           & \multicolumn{1}{c}{228} & \multicolumn{1}{c}{331} & \multicolumn{1}{c}{769} & \multicolumn{1}{c}{5142} & \multicolumn{1}{c}{39731} \\
MC prediction             & 229 , 41  & 311 , 66   & 742 , 63    & 5080 , 350   & 41000 , 14000\\
\multicolumn{1}{l||}{MC (no NFs)}    & 218 , 35  & 335 , 68   & 787 , 58    & 4930 , 330   & 47000 , 16000\\
\hline
Composition (\%)       &&&&& \\
$~~W\gamma$                    &85.0 , 2.4  &\multicolumn{1}{c}{---}  &0.46 , 0.14  &0.049 , 0.018 &0.022 , 0.007 \\
  $~~WZ/W\gamma^{\ast}$    &1.02 , 0.27 &85 , 4            &2.34 , 0.24  &0.200 , 0.029 &0.38 , 0.09\\
$~~WW$                   &0.37 , 0.08 &0.028 , 0.014&23.9, 2.3   &1.43 , 0.21   &0.57 , 0.15 \\
$~~Z$+jets               &4.2,1.6     &7.0, 3.5                 &7.0 , 2.0  &2.2 , 0.7   &97.7 , 0.5 \\
$~~$Top                  &0.68,0.20   &1.50, 0.29   &62.7 , 2.8   &95.5 , 0.8  &0.86 , 0.21 \\
$~~$Others               &8.7 , 1.2   & 5.3, 1.2  &3.2 , 0.4  &0.63 , 0.11   &0.44 , 0.11 \\
$~~VH~(H\rightarrow WW^{\ast})$   &\multicolumn{1}{c}{---}   &0.77, 0.17  &0.32 , 0.04  &0.036 , 0.005 &0.0077, 0.0020 \\
\hline 
\end{tabular}
}
\end{center}

\begin{center}
\scalebox{0.58}{
\begin{tabular}{l|| D{,}{\pm}{-1} | D{,}{\pm}{-1} D{,}{\pm}{-1} D{,}{\pm}{-1} D{,}{\pm}{-1} D{,}{\pm}{-1} | D{,}{\pm}{-1} D{,}{\pm}{-1}}
 \multicolumn{9}{l}{(b) 7~\tev\ data sample}                       \\  
\hline 
Channel          & \multicolumn{1}{c|}{ \fourlep\ } &
                                                      \multicolumn{5}{c|}{ \threelep\ }  & \multicolumn{2}{c}{ DFOS \twolep\ }     \\
\hline
CR               & \multicolumn{1}{c|}{$ZZ$} & \multicolumn{1}{c}{$WZ$} & \multicolumn{1}{c}{$ZZ$} & \multicolumn{1}{c}{$Z$jets} & \multicolumn{1}{c}{Top} & \multicolumn{1}{c|}{$Z\gamma$} & \multicolumn{1}{c}{$Z\tau\tau$} & \multicolumn{1}{c}{OSTop}\\
\hline
 Observed events          & \multicolumn{1}{c|}{24} & \multicolumn{1}{c}{101} & \multicolumn{1}{c}{18} & \multicolumn{1}{c}{68} & \multicolumn{1}{c}{9} & \multicolumn{1}{c|}{123} & \multicolumn{1}{c}{55} & \multicolumn{1}{c}{137}\\ 
 MC prediction            & 24 , 8  & 101 , 16    & 18 , 5  & 67 , 15     & 8 , 4   & 123 , 26      & 55 , 15    & 137 , 20\\ 
 \multicolumn{1}{l||}{MC (no NFs)}   & 15 , 5  & 99 , 10  & 10.7 , 0.6  & 81 , 7 & 8.1 , 1.4   & 208 , 12      & 51 , 12    & 145 , 18\\ 
\hline
 Composition  ($\%$)    & &&&&&&&\\
 $~~WZ/W\gamma^{\ast}$   &  \multicolumn{1}{c|}{---} & 87.5 , 2.5  & 3.1 , 1.1   & 6.9 , 1.4  & 14 , 5 & 0.61 , 0.15    &\multicolumn{1}{c}{---}&\multicolumn{1}{c}{---}\\ 
 $~~ZZ^{\ast}$           &  99.71 , 0.12  & 7.4  , 2.1  & 92.7 , 2.3  & 26 , 6  & 4.2 , 2.5   &  32 , 7               &\multicolumn{1}{c}{---}&\multicolumn{1}{c}{---}\\
 $~~Z\gamma$             &  \multicolumn{1}{c|}{---} & 1.8 , 0.8   &
                                                                     0.5 , 0.4   & 48 , 7  & 6 , 4   & 59 , 7    &\multicolumn{1}{c}{---}&\multicolumn{1}{c}{---}\\
 $~~Z$+jets              &  \multicolumn{1}{c|}{---} & 1.5 , 0.8   & 3.0 , 1.4   & 19 , 5  & 0.4 , 2.2   & 8.2 , 2.1     &76 , 6  & 0.14 , 0.15   \\
 $~~$Top                 &  0.031 , 0.015 & 0.7 , 0.4 & 0.01, 0.20 & 0.07 , 0.13 & 71, 10      & 0.03 , 0.04   &13 , 5  & 75.2 , 3.2\\
 $~~$Others              &  0.23 , 0.11 & 0.56 , 0.11 & 0.44 , 0.11  & 0.115 , 0.021 & 4.2 , 1.4  &  0.05 , 0.17 &11 , 4  & 24.7 , 3.2\\
 $~~VH~(H\rightarrow WW^{\ast})$  &  0.02 , 0.31 & 0.53 , 0.08 & 0.106 , 0.030 & 0.044, 0.008 & 0.41 , 0.17  &  0.0154 , 0.0027  &0.048 , 0.017 & 0.135 , 0.030 \\
\hline
\end{tabular}
}
\caption{Number of observed and predicted events and background
  composition in the CRs for the $\fourlep$,  $\threelep$ and
  $\twolep$ channels in the (a) 8~\tev\ and (b) 7~\tev\ data samples.
         Normalisation factors are taken into account in the calculation of the composition. 
The uncertainties on event yields include both the statistical and
systematic components (see section~\ref{sec:systematics}).
}
\label{table:CR34}
\end{center}
\end{table}

\begin{figure}[htp]
\begin{center}
\subfigure[]{\includegraphics[width = 0.48\textwidth]{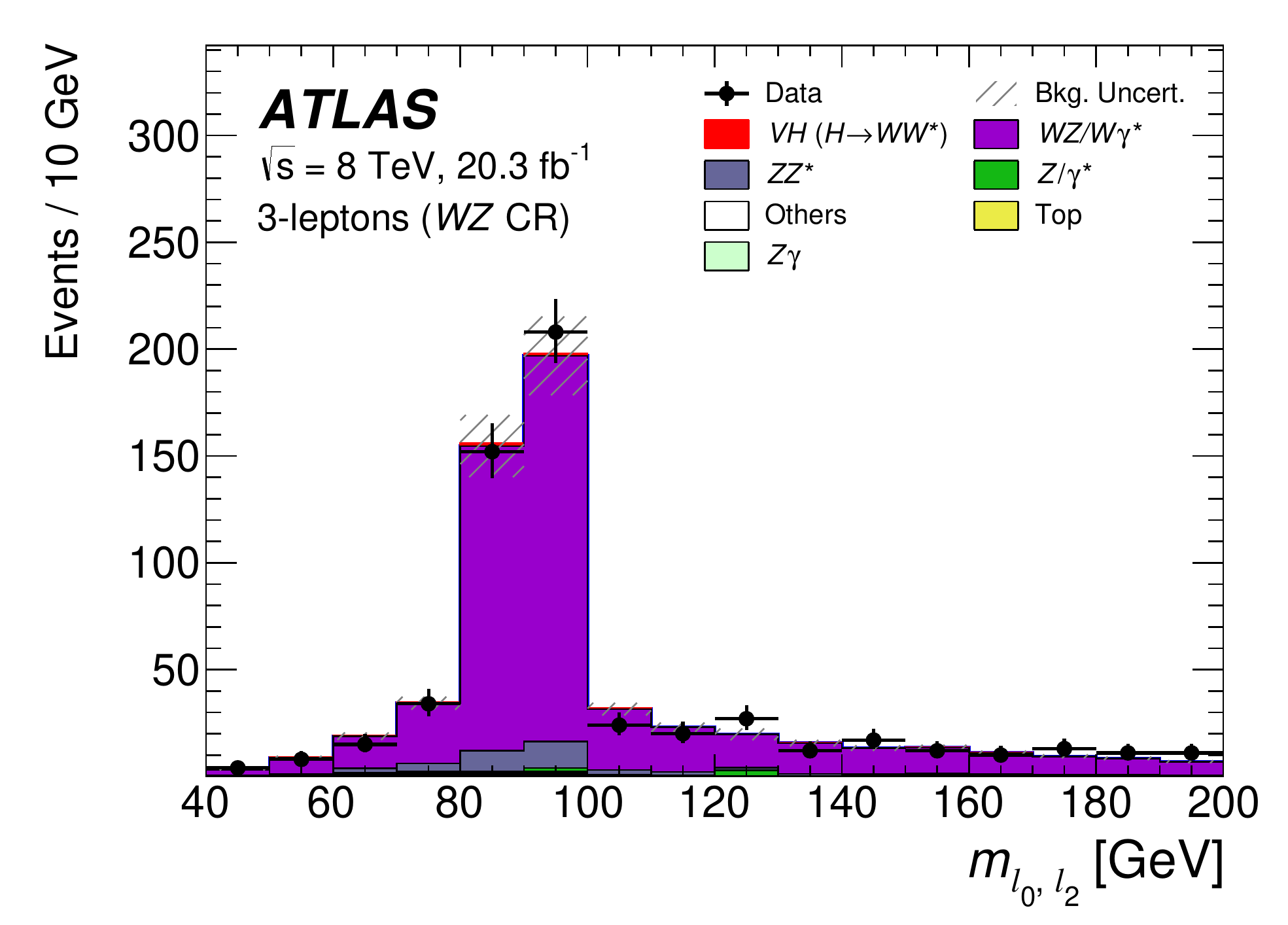}}
\subfigure[]{\includegraphics[width = 0.48\textwidth]{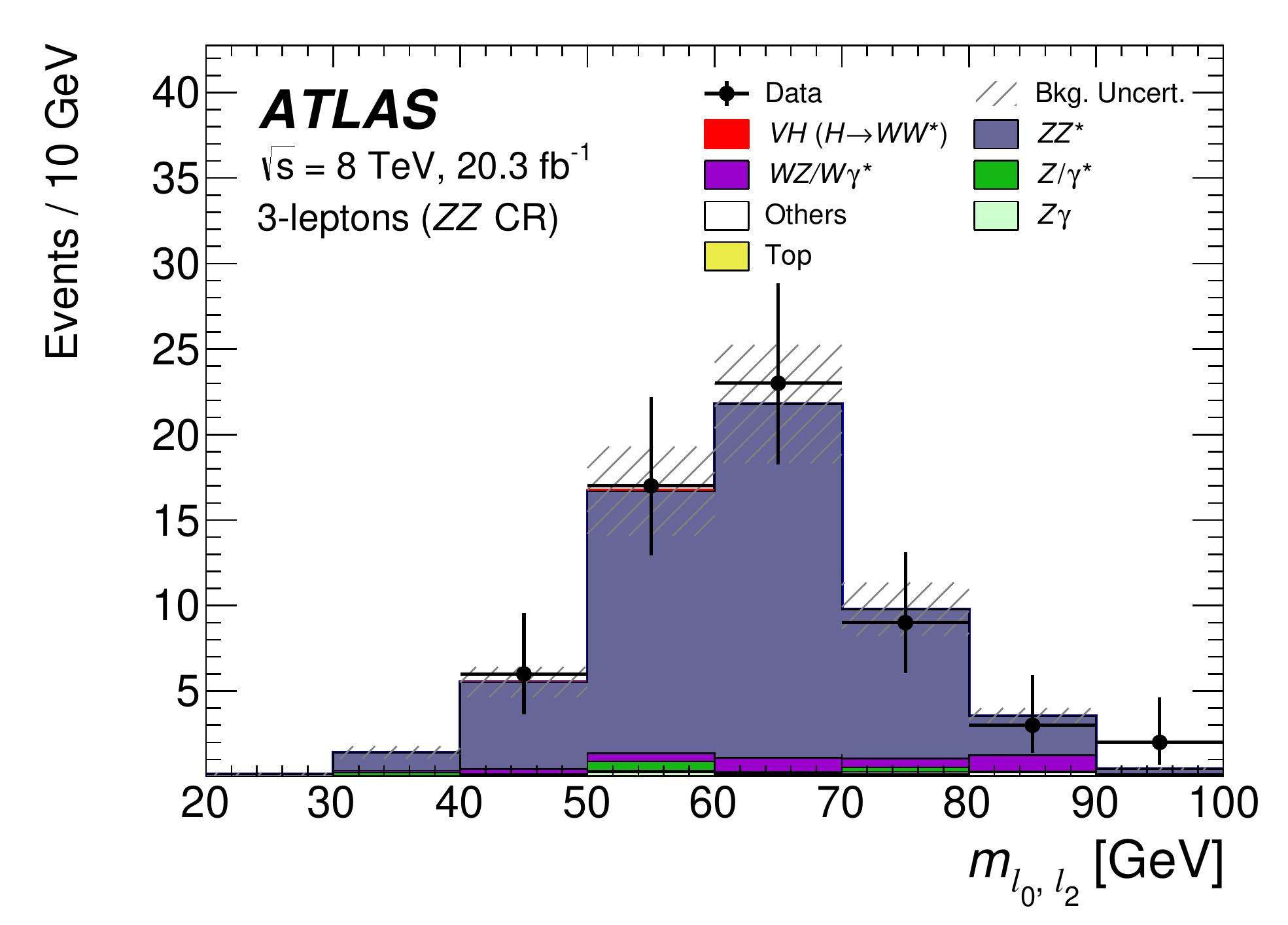}}
\subfigure[]{\includegraphics[width = 0.48\textwidth]{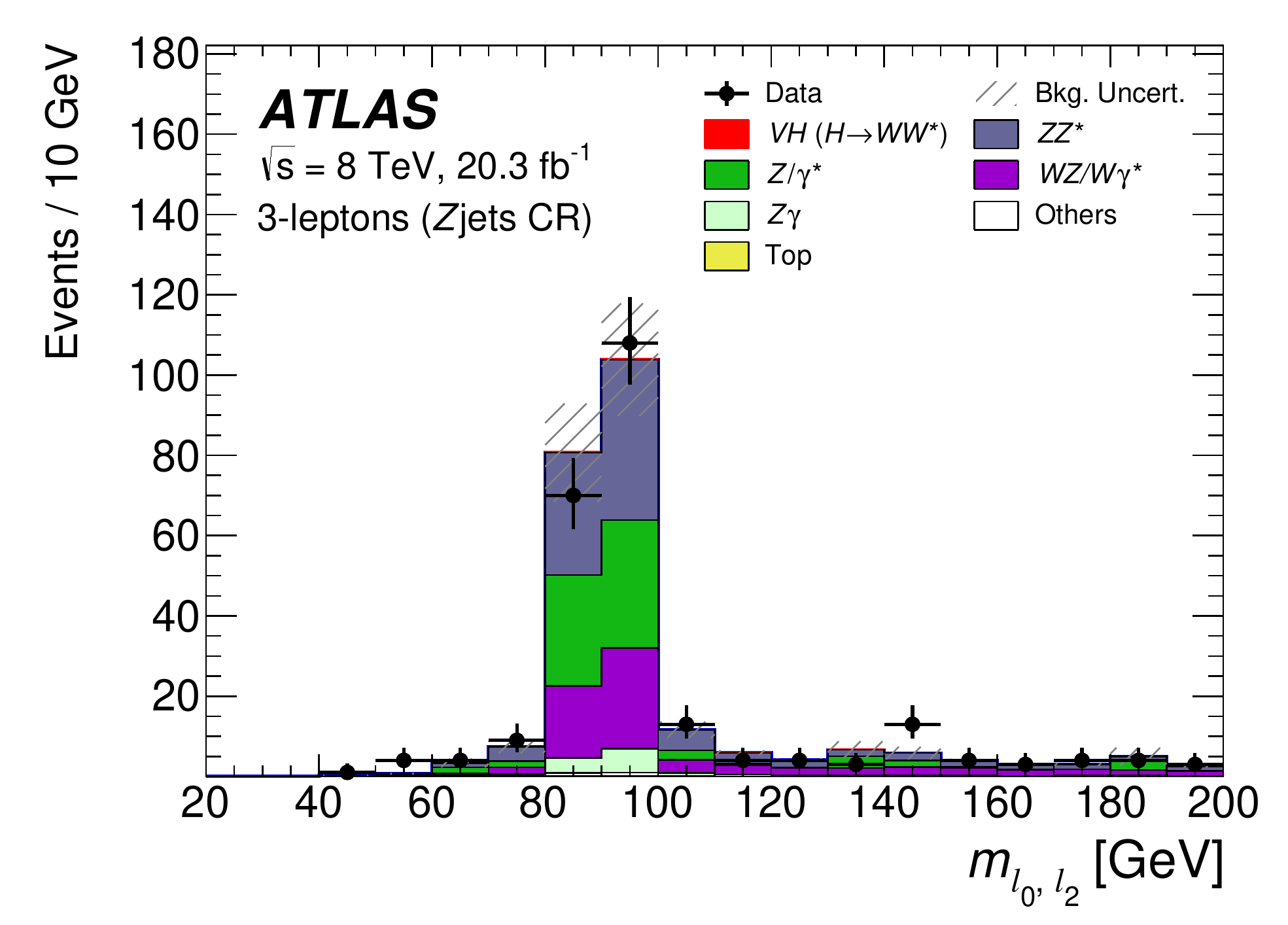}}
\subfigure[]{\includegraphics[width = 0.48\textwidth]{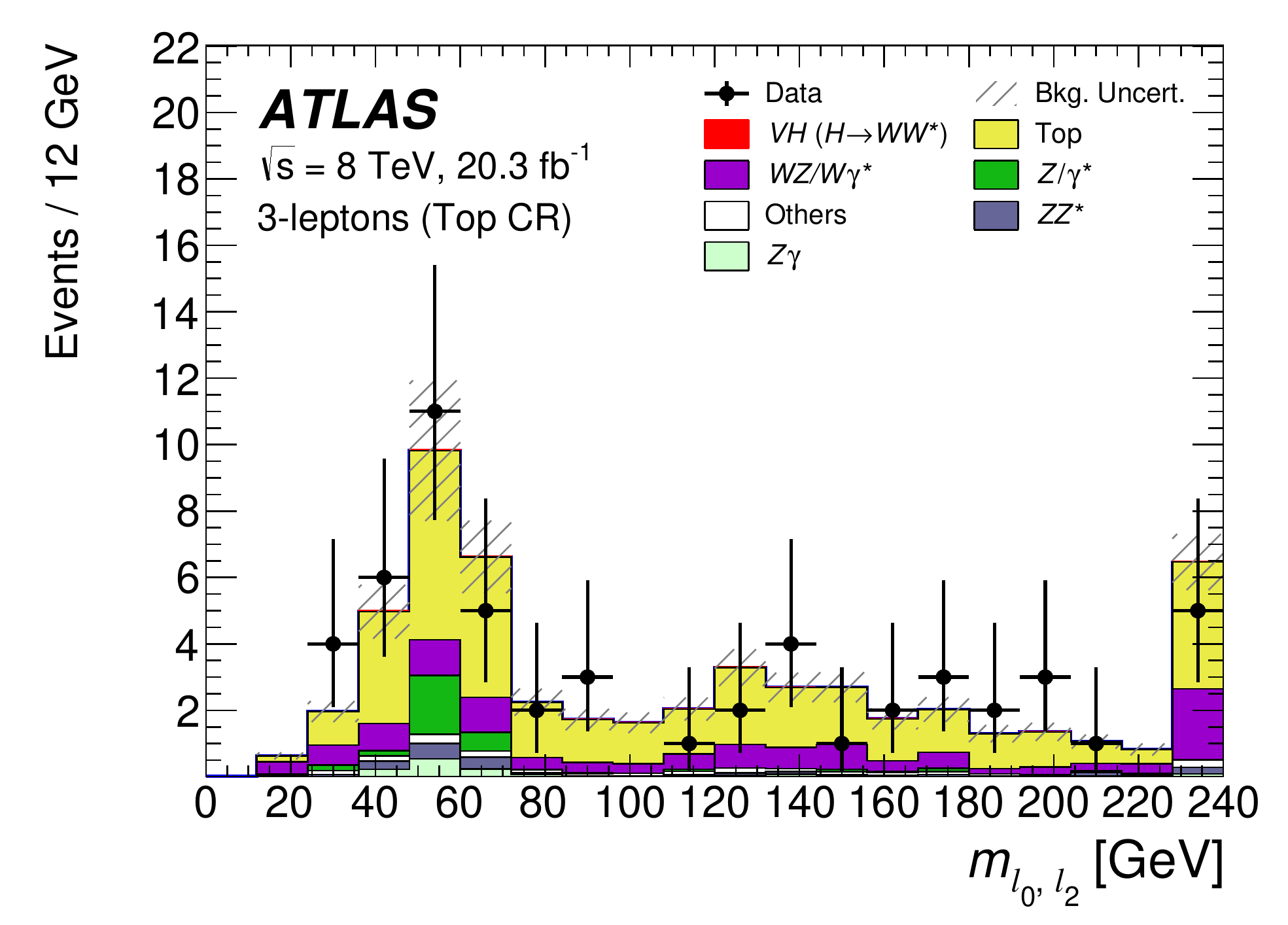}}
\subfigure[]{\includegraphics[width = 0.48\textwidth]{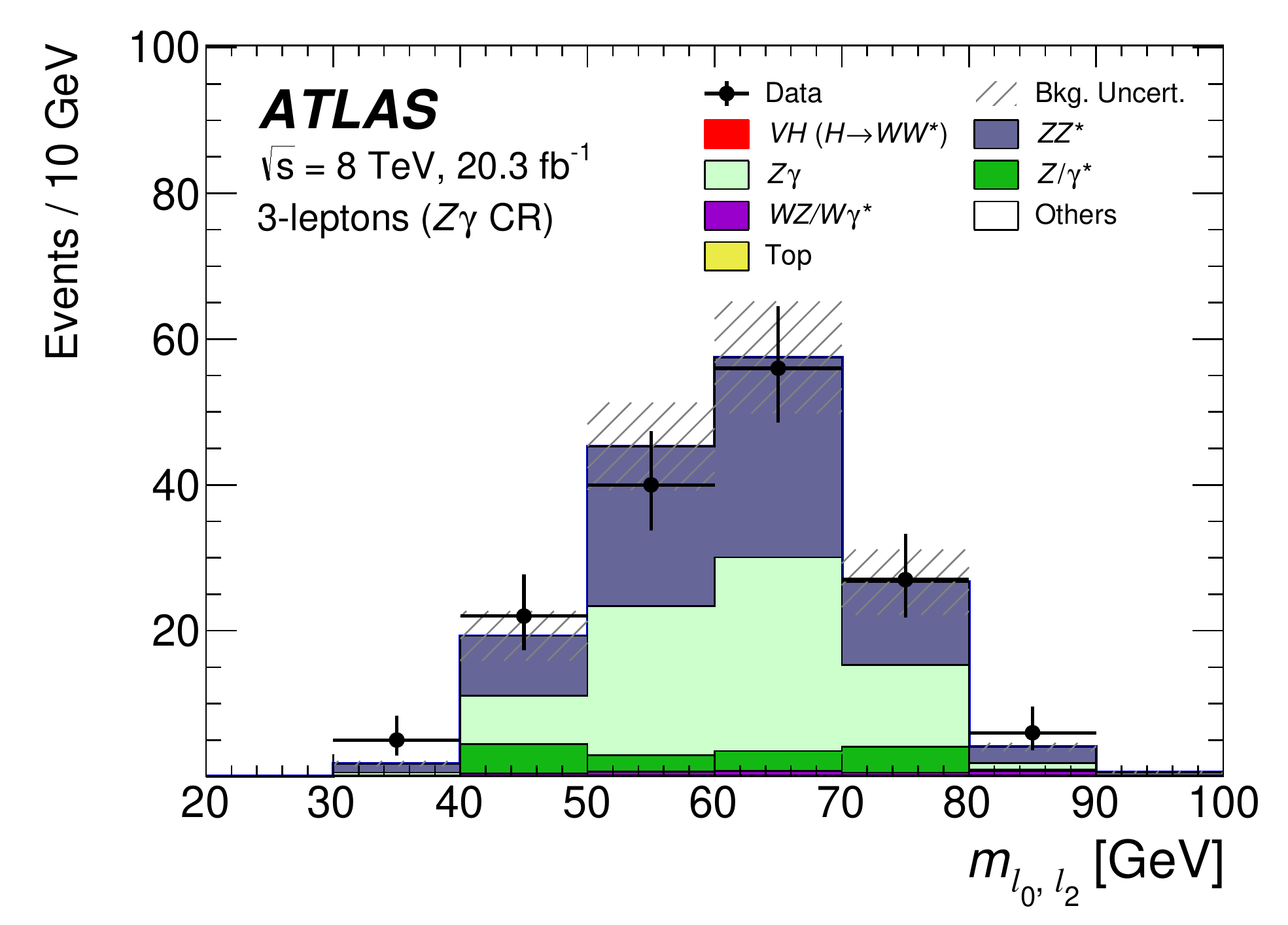}}
\caption{
Distributions of the invariant mass of leptons $\ell_0$ and $\ell_2$, defined in secton~\ref{sec:analysis}, in the five CRs defined in the 3$\ell$ channel:
(a) 3$\ell$-$WZ$ CR, (b) 3$\ell$-$ZZ$ CR, (c) 3$\ell$-$Z$jets CR, (d) 3$\ell$-Top CR and (e) 3$\ell$-$Z\gamma$ CR.
Data (points) are compared to the background plus $VH(\hww)$ (\mH=125~\gev) signal expectation 
(stacked filled histograms),
where the background contributions are normalised by applying the normalisation factors
shown in table~\ref{tab:NFs}.
The hatched area on the histogram represents total uncertainty, both statistical and systematic (see section~\ref{sec:systematics}), on the total background estimate.
The last bin includes overflows.
}
\label{Figure:CR-3lep}
\end{center}
\end{figure}

\begin{figure}[htp]
\begin{center}
\includegraphics[width = 0.48\textwidth]{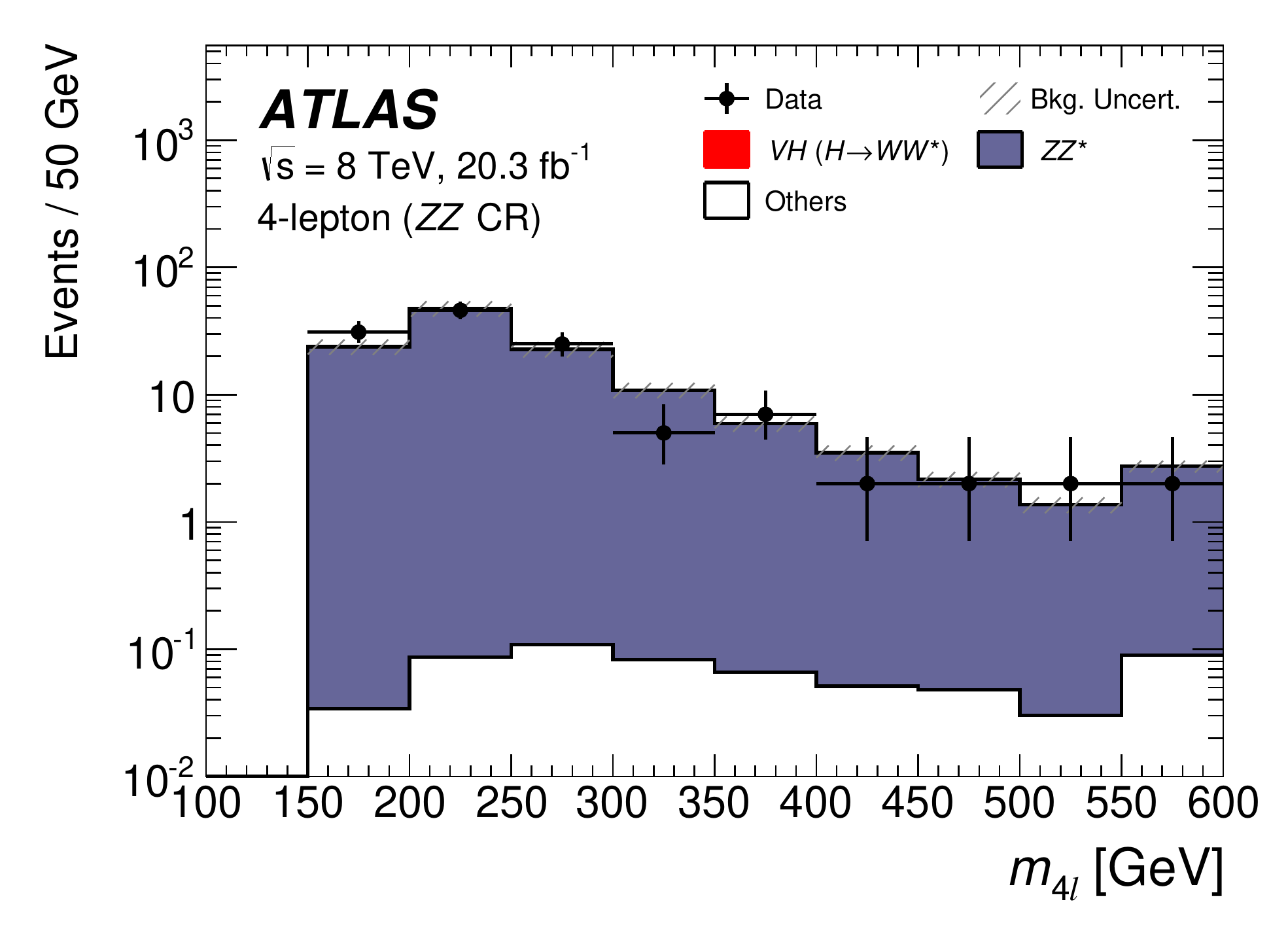}
\caption{
Distribution of the 4-lepton invariant mass $m_{4\ell}$ in the 4$\ell$-$ZZ$ CR control region.
Data (points) are compared to the background plus $VH(\hww)$ (\mH=125~\gev) signal expectation 
(stacked filled histograms),
where $ZZ^{\ast}$ events are normalised by applying the normalisation factor
shown in table~\ref{tab:NFs}. 
The hatched area on the histogram represents total uncertainty, both statistical and systematic (see section~\ref{sec:systematics}), on the total background estimate.
The last bin includes overflows.
}
\label{Figure:CR-4lep}
\end{center}
\end{figure}
\begin{figure}[htp]
\begin{center}
\subfigure[]{\includegraphics[width = 0.48\textwidth]{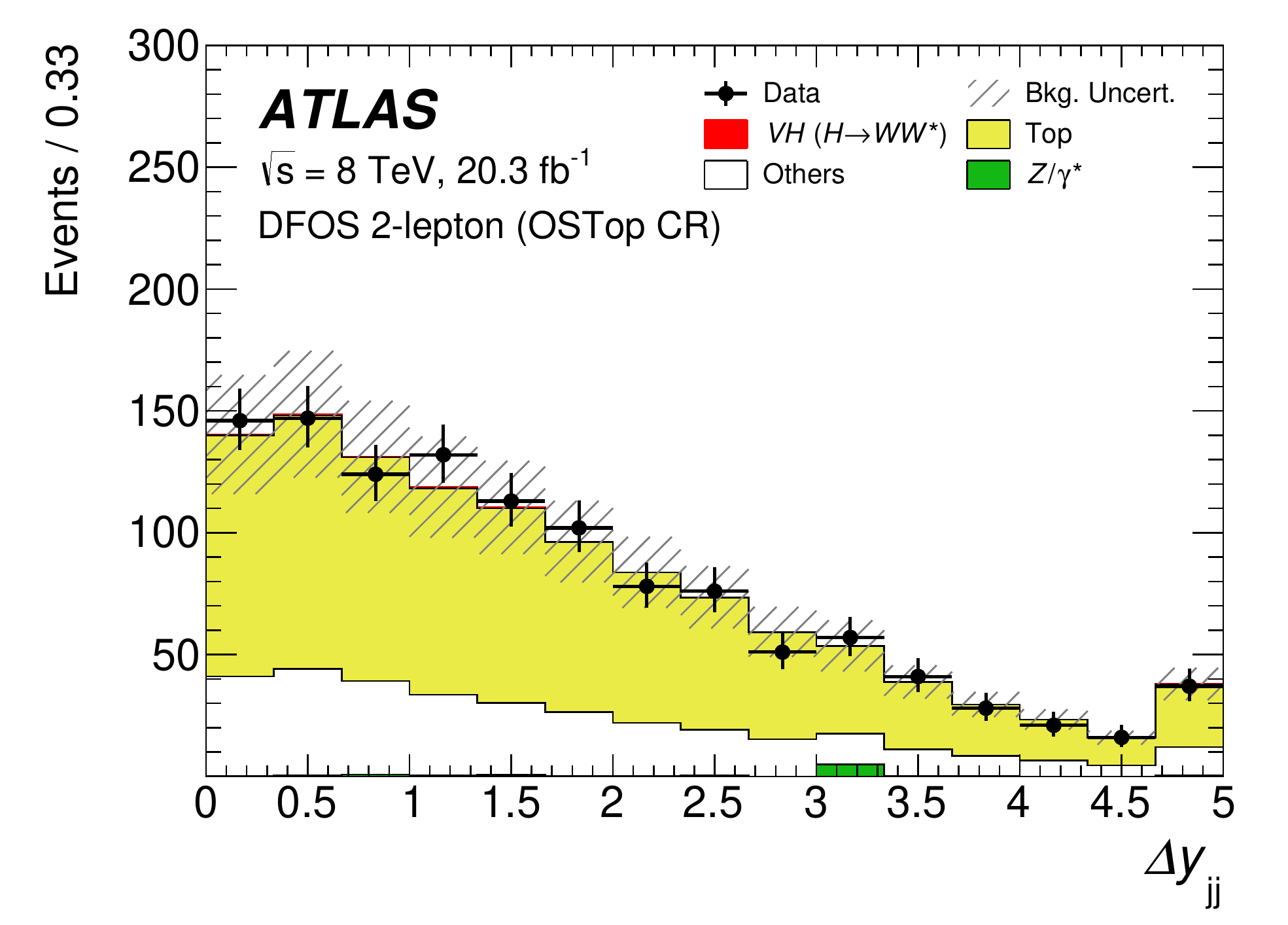}}
\subfigure[]{\includegraphics[width = 0.48\textwidth]{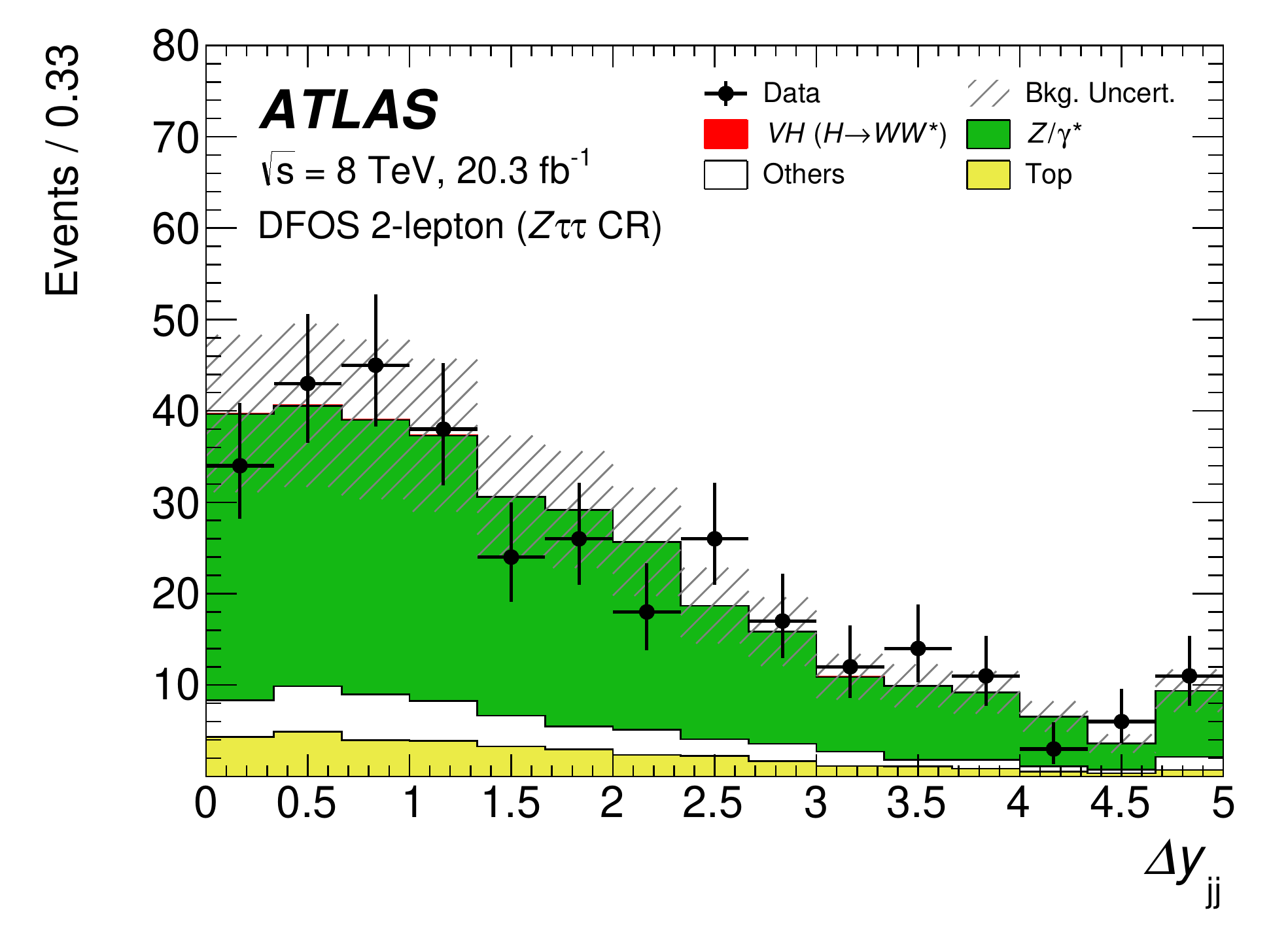}}
\caption{
Distributions of the difference in rapidity between the two leading jets $\Delta y_{jj}$ (a) in the 2$\ell$-OSTop CR and (b) in the 2$\ell$-$Z\tau\tau$ CR. 
Data (points) are compared to the background plus $VH(\hww)$ (\mH=125~\gev) signal expectation 
(stacked filled histograms),
where the background contributions are normalised by applying the normalisation factors
shown in table~\ref{tab:NFs}. 
The hatched area on the histogram represents total uncertainty, both statistical and systematic (see section~\ref{sec:systematics}), on the total background estimate.
The last bin includes overflows.
}
\label{Figure:CR-DFOS}
\end{center}
\end{figure}

\begin{figure}[htp]
\begin{center}
\subfigure[]{\includegraphics[width = 0.48\textwidth]{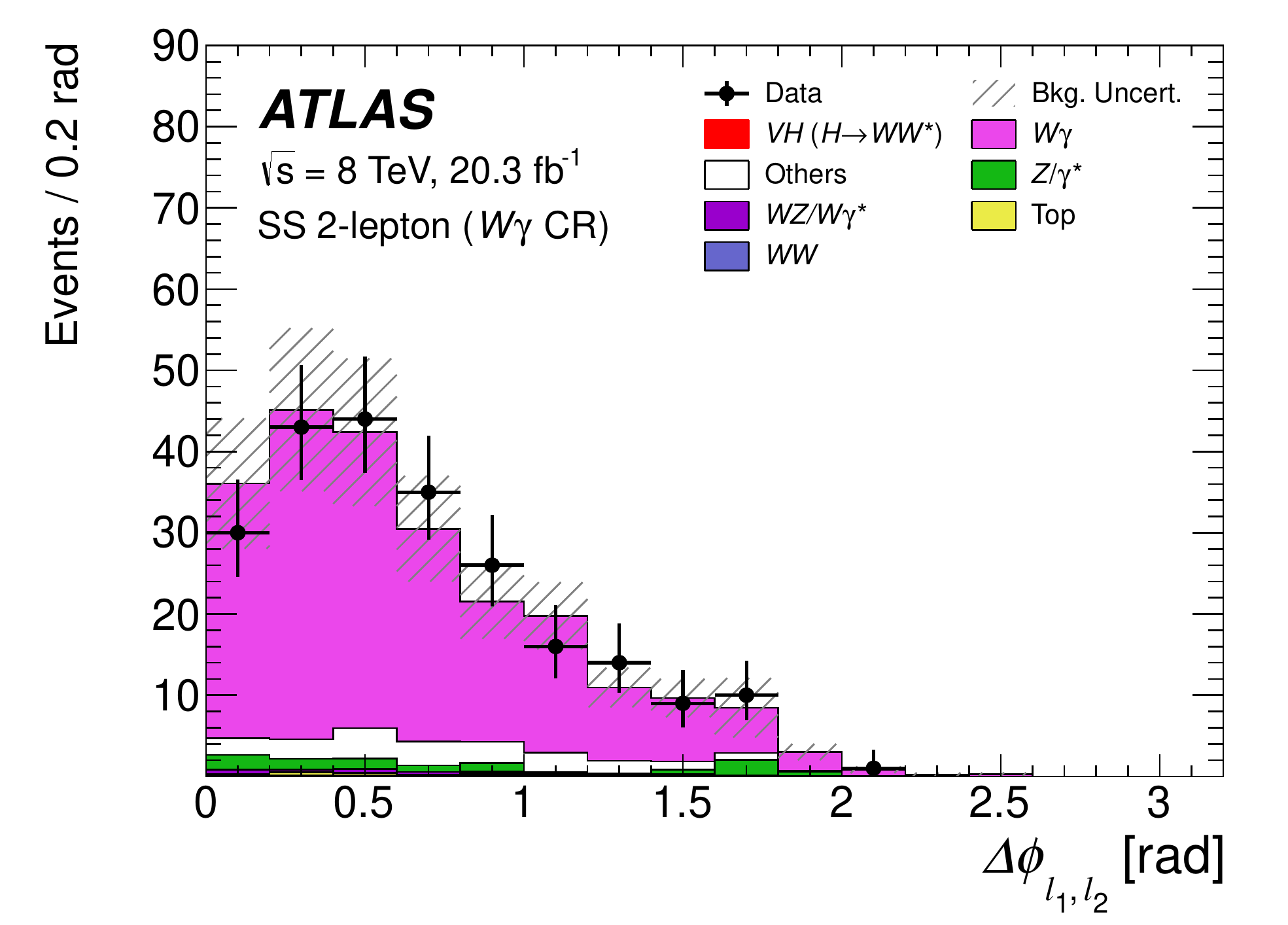}}
\subfigure[]{\includegraphics[width = 0.48\textwidth]{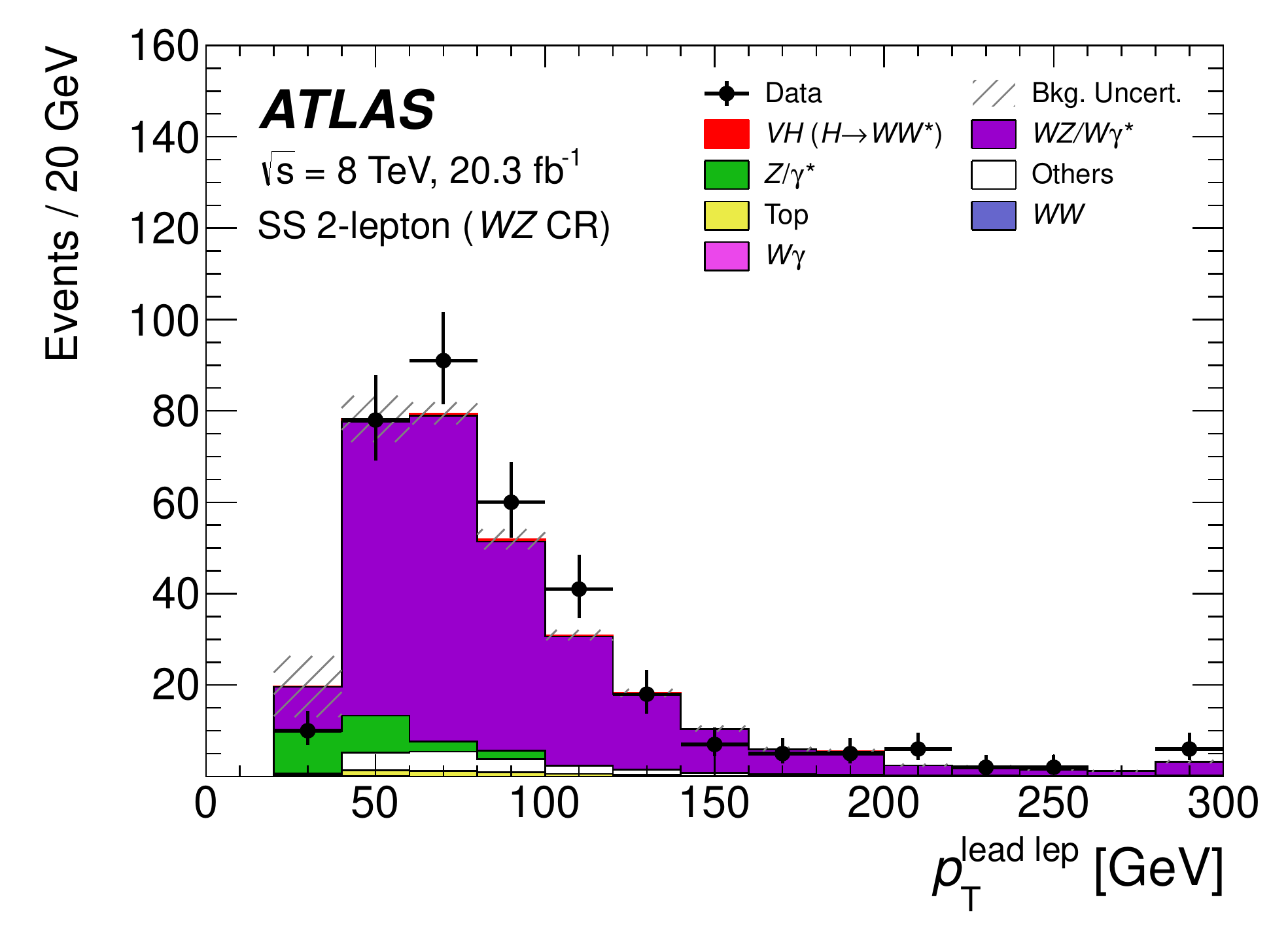}}
\subfigure[]{\includegraphics[width = 0.48\textwidth]{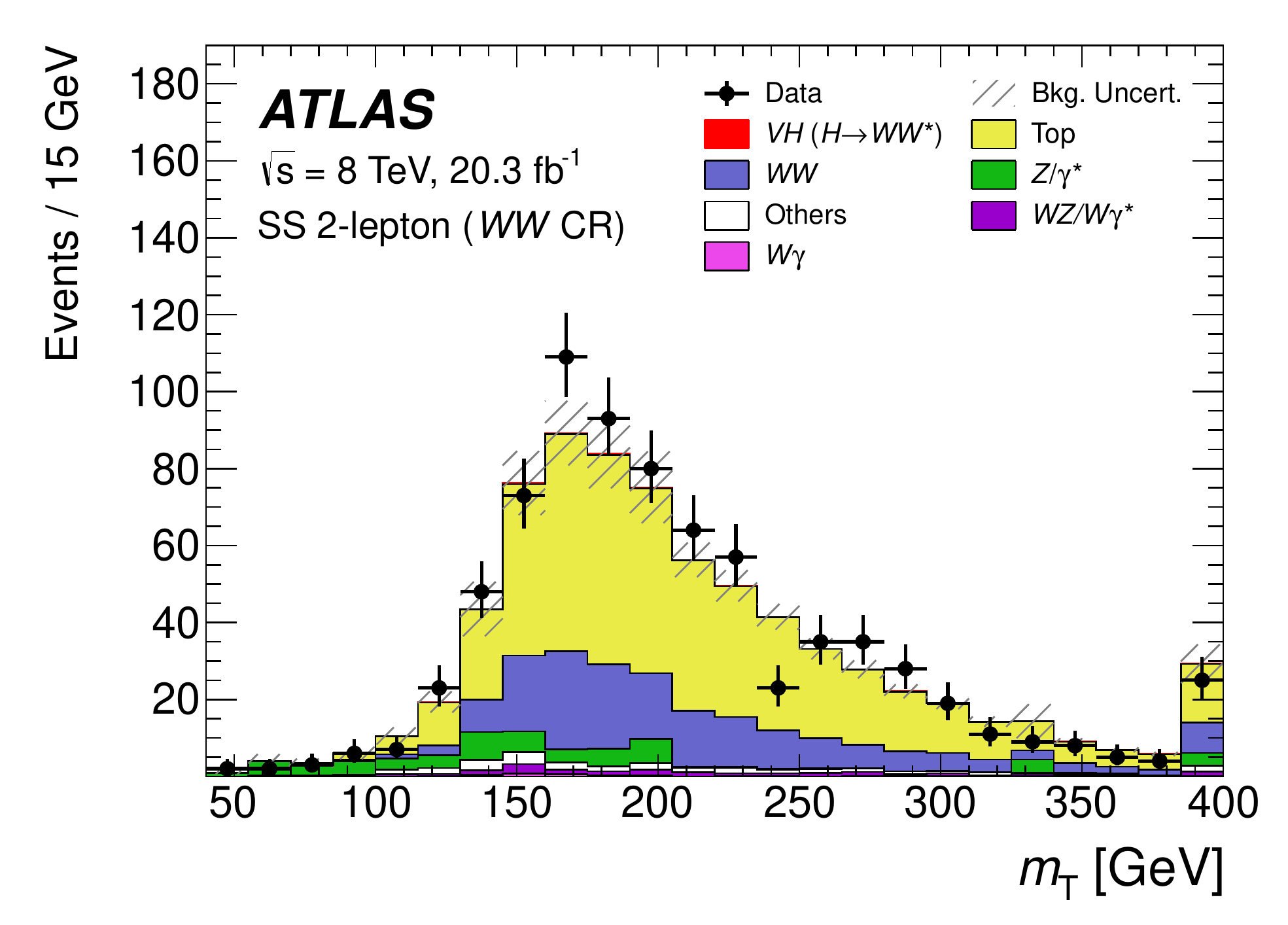}}
\subfigure[]{\includegraphics[width = 0.48\textwidth]{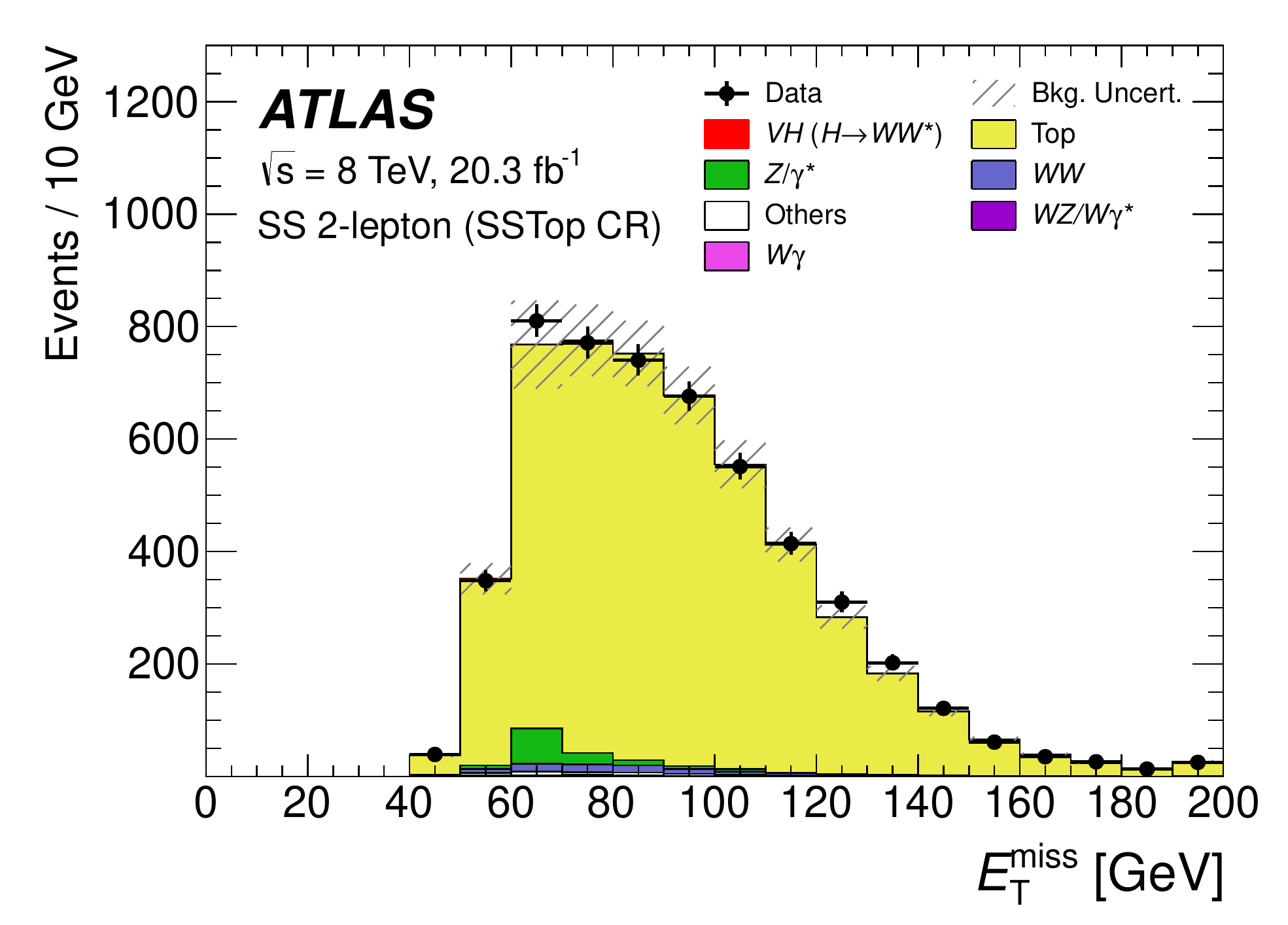}}
\subfigure[]{\includegraphics[width = 0.48\textwidth]{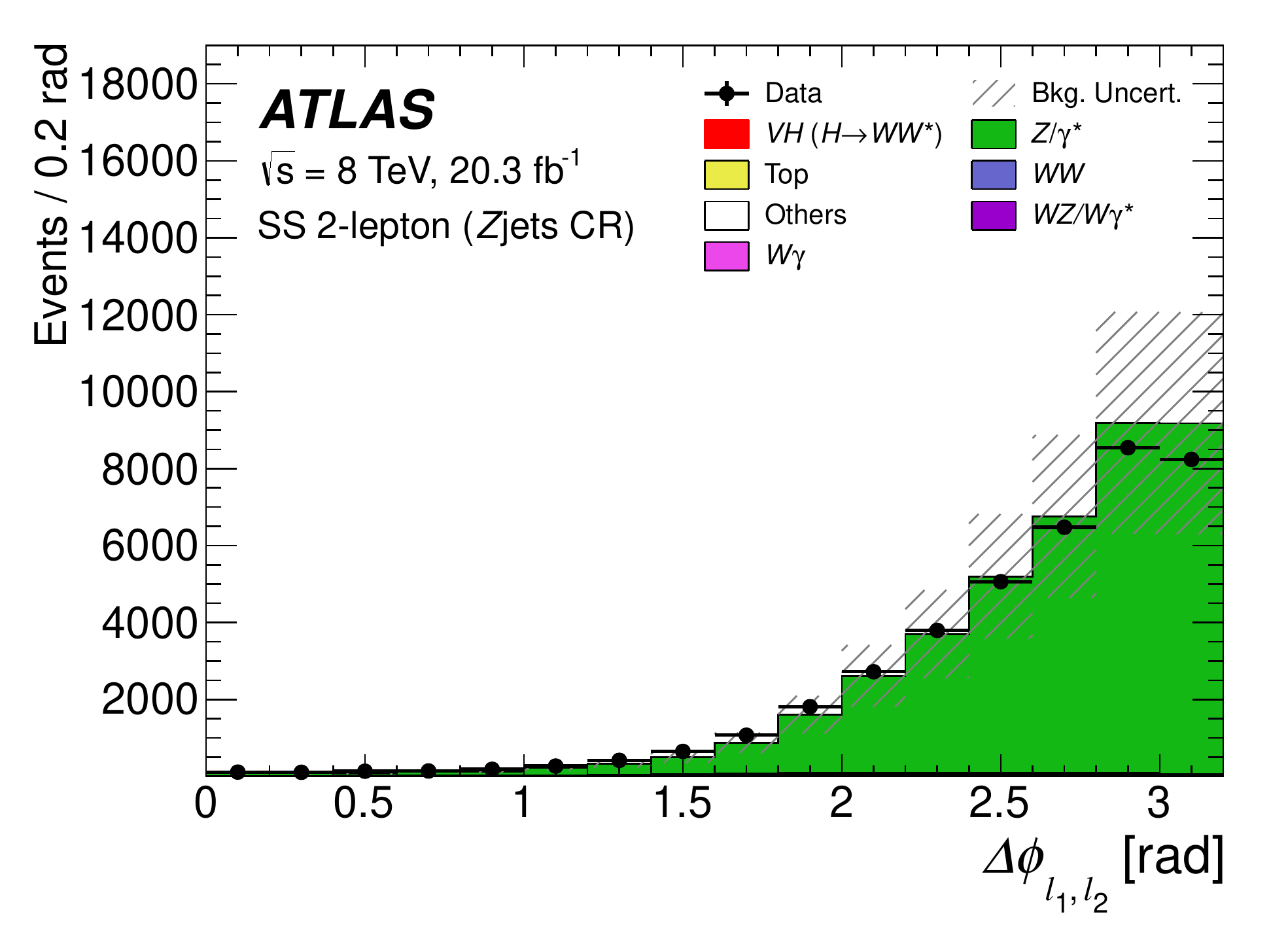}}
\caption{
Distribution of relevant variables in the control regions in the 2$\ell$-SS channel:
(a) azimuthal angle between the two leptons $\Delta \phi_{\ell_1\ell_2}$ in the 2$\ell$-$W\gamma$ CR,
(b) transverse momentum of the leading lepton in the 2$\ell$-$WZ$ CR,
(c) transverse mass $\mT$ in the 2$\ell$-$WW$ CR,
(d) missing transverse momentum $\met$ in the 2$\ell$-SSTop CR and
(e) $\Delta \phi_{\ell_1\ell_2}$, in the 2$\ell$-$Z$jets CR.
Data (points) are compared to the background plus $VH(\hww)$ (\mH=125~\gev) signal expectation 
(stacked filled histograms),
where the background contributions are normalised by applying the normalisation factors shown in table~\ref{tab:NFs}.
The hatched area on the histogram represents total uncertainty, both statistical and systematic (see section~\ref{sec:systematics}), on the total background estimate.
The last bin includes overflows.
}
\label{Figure:CR-SS}
\end{center}
\end{figure}

\clearpage
\newpage

\clearpage
\newpage
\section{Systematic uncertainties                             \label{sec:systematics}         }

The sources of theoretical and experimental systematic uncertainty
on the signal and background are described in this section, and
summarised in table~\ref{table:sys_sign_bck}.

\begin{table}[h!]
\begin{center}
\scalebox{0.75}{
\begin{tabular}{l||cc|ccc|ccc}
\multicolumn{9}{l}{(a) Uncertainties on the $VH(\hww)$ process (\%)}    \\
\hline
Channel &  \multicolumn{2}{c|}{\fourlep}& \multicolumn{3}{c|}{\threelep}   & \multicolumn{3}{c}{\twolep} \\
\hline
Category                         & 2SFOS & 1SFOS & 3SF & 1SFOS & 0SFOS & DFOS & SS2jet & SS1jet \\
\hline
Theoretical uncertainties      &       &       &       &       &       &      &        &        \\
~~~$VH$ acceptance             & 9.2   & 9.3    & 9.9   & 9.9   & 9.9  & 10  & 10    & 9.9    \\
~~~Higgs boson branching fraction  & 4.2   & 4.2   & 4.2   & 4.2    & 4.2  & 4.2  & 4.2   & 4.2    \\
~~~QCD scale                     & 3.1   & 3.0   & 1.2   & 1.0    & 1.0  & 1.3  & 1.0   & 1.0 \\
~~~PDF and $\alpha_{\rm S}$            & 1.0   & 1.1   & 2.1   & 2.2    & 2.2  & 1.9  & 2.3   & 2.2 \\
~~~$VH$ NLO EW corrections       & 1.7   & 1.8   & 1.9   & 1.9    & 1.9  & 1.9  & 1.9   & 1.9 \\
Experimental uncertainties       &       &       &       &        &       &      &       &  \\
~~~Jet                           & 2.0   & 3.1   & 2.5   & 2.5    & 2.9  & 3.2  & 8.9   & 5.8 \\
~~~$\MET$ soft term              & 0.2   & 0.3   & --    & --     & --   & 0.3  & 0.6   & 0.2 \\
~~~Electron                      & 2.6   & 2.8   & 1.6   & 2.2    & 2.2  & 1.5  & 2.1   & 1.7 \\
~~~Muon                          & 2.6   & 2.4   & 2.2   & 1.8    & 1.7  & 0.8  & 1.8   & 1.9 \\
~~~Trigger efficiency            & 0.2   & --    & 0.4   & 0.3    & 0.3  & 0.5  & 0.6   & 0.5 \\
~~~$b$-tagging efficiency          & 0.9   & 0.9   & 0.9   & 0.8    & 0.8  & 2.9  & 3.5   & 2.4 \\
~~~Pile-up                       & 1.9   & 0.7   & 2.0   & 1.4    & 0.8  & 1.7  & 1.0   & 2.4 \\
~~~Luminosity                    & 2.8   & 2.8   & 2.8   & 2.8    & 2.8  & 2.8  & 2.8   & 2.8 \\
\hline
\multicolumn{9}{l}{}        \\
\multicolumn{9}{l}{(b) Uncertainties on the total background (\%)}  \\
\hline
Theoretical uncertainties     &       &       &       &       &       &      &  &  \\
~~~  QCD scale                  & 0.2   & 0.1   & 1.0   & 0.9   & --    & 3.7  & 13     & 2.3 \\
~~~  PDF and $\alpha_{\rm S}$         & 0.2   & 2.4   & 0.3   & 0.3   & 1.6   & 1.4  & 0.5    & 0.6 \\
~~~  $VVV$ $K$-factor               & 2.8   & 8.1   & 1.1   & 1.9   & 0.5   & --   & --     & 0.3 \\
~~~  MC modelling               & 5.3   & 4.3   & 7.0   & 6.6   & --    & 4.1  & 0.8    & 1.4 \\

Experimental uncertainties    &       &       &       &  &  &  &  &  \\
~~~Jet                          & 3.1   & 2.4   & 3.2   & 1.8   & 4.1   & 7.2  & 5.0    & 3.4 \\
~~~$\MET$ soft term             & 2.3   & 0.6   & 1.8   & 1.9   & 0.5   & 1.1  & 0.2    & 0.7 \\
~~~Electron                     & 1.0   & 1.4   & 1.0   & 0.4   & 1.1   & 0.7  & 1.1    & 0.8 \\
~~~Muon                         & 1.1   & 1.2   & 0.4   & 0.7   & 0.2   & 0.2  & 0.4    & 0.8 \\
~~~Trigger efficiency           & --    & 0.2   & 0.2   & --    & --    & 0.1  & --     & -- \\
~~~$b$-tagging efficiency         & 0.6   & 0.8   & 0.6   & 0.8   & 2.6   & 0.7  & 1.4    & 0.3 \\
~~~Fake factor                  & --    & --    & --    & --    & --    & 2.8  & 10     & 10 \\
~~~Charge mis-assignment        & --    & --    & --    & --    & 1.4   & --   & 0.7    & 0.8 \\
~~~Photon conversion rate       & --    & --    & --    & --    & --    & --   & 1.1    & 0.9 \\
~~~Pile-up                      & 1.2   & 1.1   & 1.4   & 0.3   & 1.2   & 0.9  & 1.0    & 1.0 \\
~~~Luminosity                   & 0.4   & 0.8   & 0.1   & 0.2   & 0.7   & --   & 0.7    & 0.3 \\
MC statistics                   & 5.3   & 8.0   & 3.8   & 3.2   & 5.5   & 3.1  & 7.3    & 3.9 \\
CR statistics                   & 8.1   & 6.6   & 4.2   & 3.9   & 8.8   & 2.5  & 2.8    & 3.5 \\

\hline
\end{tabular}
}
\end{center}
\caption{Theoretical and experimental uncertainties, in \%, on the predictions of the (a) signal and (b) total background for each category. 
Fake factor refers to the data-driven estimates of the $W$+jets and multijet backgrounds in the $2\ell$ channels.
The  dash symbol (--) indicates
that the corresponding uncertainties either do not apply or are negligible.
The values are obtained through the fit and given for the 8~\tev\ data sample.
Similar values are obtained for the 7~\tev\ data
sample. 
}
\label{table:sys_sign_bck}
\end{table}
The table shows the uncertainties on the estimated event yield after the fit in 8~\tev\ data samples. Similar values are obtained for the 7~\tev\ data analysis.

Systematic uncertainties have an impact on the estimates of the signal and background
event yields in the SRs and CRs. 
The experimental uncertainties are applied to both the SRs and CRs.
The extrapolation parameter from a SR to a CR is defined as the ratio of MC estimates in the SR and CR.
The theoretical uncertainties are computed on the extrapolation parameters, and applied to SRs.
Correlations between SRs and CRs are taken into account in the fit,
and cancellation effects are expected for uncertainties correlated between SRs and CRs. 
The different uncertainty sources are treated  according to their correlations across the data taking periods,
among different analyses, between signal and background sources. 
Whenever an effect is expected to affect coherently the event yields in two event samples (SRs and/or CRs), for instance 
the muon reconstruction efficiency for both signal and background, a 100\% correlation is assumed and the yields are varied 
coherently in the fit through the introduction of a single parameter (nuisance parameter).
Alternatively when an uncertainty source is not expected to affect the event yields coherently, for instance trigger and luminosity uncertainties across different data taking years, 
different nuisance parameters are introduced to represent uncorrelated effects.

The ``MC statistics''  and  ``CR  statistics''  uncertainties  in table~\ref{table:sys_sign_bck} (b) arise from limited  
simulated  events  in  the  SRs  and  CRs and  from  the  number  of  data events  populating  the  CRs,  respectively.   

\subsection{Theoretical uncertainties}
The theoretical uncertainties on the total Higgs boson production cross section and branching fraction are evaluated by following the recommendation of the LHC Higgs cross-section working group~\cite{Dittmaier:2011ti,Dittmaier:2012vm,Heinemeyer:2013tqa}. Uncertainties concerning QCD renormalisation and factorisation scales, which are hereafter collectively referred to as QCD scales, PDF, the value of $\alpha_{\rm S}$ and branching fraction are estimated.

The main uncertainty on the $VH$($\hww$) process, shown in the ``$VH$ Acceptance'' row in table~\ref{table:sys_sign_bck}(a),
accounts for the uncertainties in the acceptance of the signal
processes and it is evaluated with MINLO-$VH$j {\sc
  Powheg-Box1.0}~\cite{Luisoni:2013cuh} simulation.
The leading contributions are the missing next to leading order QCD
contributions in $qq\rightarrow VH$  simulated with {\sc Pythia8} and
the parton shower uncertainty. The first is evaluated by comparing the
{\sc Pythia8} prediction with the MINLO-$VH$j  interfaced with {\sc
  Pythia8} and amounts to 7\%. The second is computed  by comparing
 MINLO-$VH$j interfaced with {\sc Pythia8}
with MINLO-$VH$j interfaced with {\sc Herwig} and accounts for a further 7\%.

The cuts on the number of jets are found to be the  main
source of such uncertainties. The uncertainty on the acceptance of the $gg\rightarrow ZH$
process is estimated to be  5\%  in the \fourlep\ channel, the only channel where this process is relevant.

Uncertainty on the Higgs boson branching fraction to $WW^{\ast}$, which is particularly important for the $VH$ and VBF production modes, amounts to 4\%.
The QCD scale uncertainty is 1\% for $WH$ production and 3\% for $ZH$ production.
This uncertainty is larger for $ZH$ production due to the $gg\rightarrow ZH$ contribution. 
The ``VH NLO EW corrections'' 
refers to additional uncertainties on the corrections~\cite{Denner:2011rn} to the NLO differential cross section, 
applied as a function of the
\pt\ of the associated weak bosons in the LO $WH$ and $ZH$ production modes generated by {\sc Pythia8}.
 The size of this uncertainty is 2\%.

The uncertainties on ggF production are due to the inclusive cross section dependence from QCD scales, the  PDF choice
and $\alpha_{\rm S}$, they range  from 7\% to 8\%.
In the $\twolep$-DFOS channel, an additional uncertainty due to the
jet multiplicity cut is computed with {\sc MCFM}
\cite{Campbell:2010ff} and amounts to about 12\%.

The uncertainties from the QCD scales of backgrounds are estimated
by varying the scales up and down independently by a factor of two.
{\sc MCFM} is used to estimate this uncertainty in 
the $\fourlep$, $\threelep$ and $\twolep$-SS channels.
In the \fourlep\ channel the uncertainty on $ZZ^{\ast}$ is 4\% in the SRs and CR.
However, due to the cancellation between the SRs and CR, the resultant effect becomes negligible.
In the 3$\ell$-3SF and 3$\ell$-1SFOS SRs the QCD scale uncertainty on $WZ/W\gamma^{\ast}$  
is determined in each bin of the ``BDT Score'' and ranges between 3\% and 6\%.
In  the $\twolep$-SS channel, due to cancellations, the QCD scale uncertainties on $WZ/W\gamma^{\ast}$ and $W\gamma$
are found to be negligible with the exception of the 2$\ell$-SS2jet SR, in which 
 100\% uncertainty is assigned to $W\gamma$. 
The QCD scale uncertainty on the same-sign $WW$+2jet, estimated using  {\sc VBF@NLO}~\cite{Campanario:2008yg},  is of the order of 40\%.
In the $\twolep$-DFOS channel, QCD scale uncertainties on top-quark and $WW$+2jet production with at least two QCD couplings, referred to as QCD $WW$ in the following, are estimated as 9\% and 17\% 
by using {\sc MC@NLO} and {\sc MadGraph}, respectively. 

The PDF uncertainties on backgrounds are calculated
by following the PDF4LHC recipe~\cite{Botje:2011sn}
using the envelope of predictions from MSTW2008, CT10 and NNPDF2.3 PDF sets
with the exception of top-quark production in the $\twolep$-DFOS channel, 
which is evaluated with the same technique used in the ggF-enriched $n_{j} \ge 2$ category in ref.~\cite{HWWllpaper}.
The uncertainties range from 1\% to 6\%, depending on the background process and the 
categories. 
An uncertainty of 33\% on the NLO $K$-factor for the triboson process is evaluated by using {\sc VBF@NLO};
in the 3$\ell$-0SFOS SR this uncertainty is estimated in bins of $\Delta R_{\ell_0,\ell_1}$ and ranges from 1\% to 6\%.

The ``MC modelling'' row in table~\ref{table:sys_sign_bck}(b) takes
the yield variation observed between predictions of different MC generators.
In the \threelep\ and $\twolep$-SS channels, the uncertainties on the $WZ/W\gamma^{\ast}$ event yield
due to the modelling of the underlying event, parton shower and matching of the matrix element to the
parton shower are evaluated by comparing the predictions of    {\sc Powheg-Box+Pythia} and {\sc MC@NLO+Herwig}.
In the $\twolep$-DFOS channel, an  uncertainty  due to  underlying event and parton shower modelling
is assigned to top-quark production and $\Ztau$ by comparing the expectations from {\sc Powheg-Box+Pythia} and {\sc Powheg-Box+Herwig},
and {\sc Alpgen+Pythia} and {\sc Alpgen+Herwig}, respectively.    
The QCD $WW$ event yield in the $\twolep$-DFOS channel is estimated using {\sc Sherpa}. The uncertainty from the 
underlying event, parton shower modelling and matrix element implementation is evaluated through a comparison with {\sc MadGraph+Pythia}.
The uncertainty on the main background in the \fourlep\ channel, $ZZ^{\ast}$, is dominated by the statistical component;
a systematic uncertainty from the different models, underlying event and parton shower is assigned through a comparison between
{\sc Powheg-Box+Pythia} and {\sc Sherpa}.

\subsection{Experimental uncertainties}
One of the dominant experimental uncertainties, labelled ``Jet'' in table~\ref{table:sys_sign_bck}, 
derives from the propagation of the jet energy scale calibration and resolution uncertainties. 
They were derived from a combination of simulation, test-beam data, and in situ measurements~\cite{Aad:2014bia}.
Additional uncertainties due to differences between quark and gluon jets, and between light- and heavy-flavour jets,
as well as the effect of pile-up interactions are included.
For jets used in this analysis, the jet energy scale uncertainty
ranges from 1\% to 7\%, depending on \pt\ and \eta.
The relative uncertainty on the  jet energy resolution
ranges from 2\% to 40\%, with the largest value of
the resolution and relative uncertainty occurring at the
\pt\ threshold of the jet selection.

The ``Muon'' and ``Electron''  uncertainties include those from lepton reconstruction, identification and isolation, 
 as well as lepton energy and momentum measurements. The ``Trigger efficiency'' uncertainty in table~\ref{table:sys_sign_bck} 
refers to the uncertainty on the lepton trigger efficiencies. The uncertainties on the lepton and trigger 
efficiencies are of the order of 1\% or smaller. 

The changes in jet,  electron and muon energy scale uncertainties due to
varying them by their systematic uncertainties are propagated
to the \met\ evaluation and included in the   ``Jet'',  ``Electron'' and  ``Muon'' rows in table~\ref{table:sys_sign_bck}.
An additional ``$\MET$ soft term'' uncertainty is associated with the contribution of calorimeter energy deposits 
not assigned to any reconstructed objects in the $\MET$ 
reconstruction~\cite{ATLAS-CONF-2012-101,ATLAS-CONF-2010-020,ATLAS-CONF-2014-019}.

The ``$b$-tagging efficiency'' row refers to the uncertainties on the efficiency of tagging of $b$-jets and 
 include contributions from $b$-jet identification and charm and light-flavour jet rejection factors~\cite{CONF-2014-046,CONF-2014-004}.  The uncertainties related to $b$-jet identification 
range from $<$1\% to 8\%. The uncertainties
on the misidentification rate for light-quark jets depend
on \pt\ and \eta, and have a range of 9--19\%. The uncertainties
on $c$-jets reconstructed as $b$-jets range between 6\%
and 14\% depending on \pt.

The uncertainty labelled as ``Fake factor'' is associated with the 
 data-driven estimates of the $W$+jets and multijet backgrounds in the $\twolep$ channels; it ranges between 35\% and 45\% depending on the sample
and on categories.
A ``Charge mis-assignment'' systematic uncertainty is estimated to account for the mismodelling of the charge 
flip effect by 
comparing the number of lepton pairs with same charge and opposite charge under the $Z$ boson mass peak in data and MC simulation,
resulting in a 16\% relative uncertainty. The uncertainty is assigned to $WZ/W\gamma^{\ast}$ in the 3$\ell$-0SFOS SR, 
and top-quark production, $Z$+jets and $WW$ in the $\twolep$-SS channel.

The uncertainty labelled as ``Photon conversion rate''  is assigned to $W\gamma$ in the $\twolep$-SS channel,
and is evaluated by comparing the yield in  data and in  MC simulation for events with
two muons and one electron with no hit on the innermost pixel detector layer. It is relevant only for the 
$\twolep$-SS channel and has a size of 6.5\% for $W\gamma$.

The ``Pile-up'' field in table~\ref{table:sys_sign_bck} includes 
the uncertainty on the weights applied to all simulated events to match the distribution of the number of  pile-up interactions to that of data.
The uncertainty on the integrated luminosity for the 2012 data is $\pm$ 2.8\%
 and it is derived following the same methodology as that detailed in ref.~\cite{Aad:2013ucp}.
For the 2011 data the uncertainty on the integrated luminosity is $\pm$ 1.8\%~\cite{Aad:2013ucp}. 
The  backgrounds normalised with CR data are not affected by the luminosity uncertainty.

The dominant systematic uncertainties on the $VH(\hww)$ process in the
$\fourlep$ and $\threelep$ channels are due to uncertainties on lepton reconstruction and on the
jet energy scale and resolution. 
In the $\twolep$ channels, the jet energy scale and resolution uncertainties are the most important.

\section{Results                              \label{sec:results}              }
The data collected at $\sqrt{s}=8 \tev$ and 7~\tev\ are analysed separately and then combined in all channels
in order to search for the Higgs boson in $WH$ and $ZH$ production  using $\hww$ decays.
The analyses are optimised for a Higgs boson mass  of $\mH = 125\gev$. 
For this mass, the selected Higgs bosons decay mainly as 
$H \to WW^{\ast} \to \ell \nu + X$, but a small contamination from the $VH$ production of Higgs bosons, from the decay chain $H \to \tau \tau \to \ell \nu \ell \nu$, is present. 
This contribution is treated as background and normalised to the SM expectation for the $VH$ production cross section $\sigma_{VH}$, and the $H \to \tau \tau$ branching fraction.

This section is subdivided in five sub-sections. Section \ref{sec:yields} shows the event yields in each category of the 8~\tev\ and 7~\tev\ data samples.
Furthermore, distributions of 
some of the relevant variables are shown for the 8~\tev\ data sample. Section \ref{sec:stat_formalism} summarises the statistical treatment used for the signal extraction. 
Section   \ref{sec:characterization} quantifies the agreement with the background only hypothesis in each analysis category and evaluates the observed and expected significance in each of them.
Section  \ref{sec:interpretation} reports the significance for $WH$, $ZH$ and combined $VH$ production, moreover it shows the measurement of their cross sections  divided by their SM expectations ($\mu$).  The $VH$ categories are then combined with the categories of the ggF and VBF analysis
using $H\to WW^{\ast} \to \ell \nu \ell \nu$ decays described in ref.~\cite{HWWllpaper}.
 In addition  the ggF and VBF yields are extracted in a consistent manner together with the $VH$ yields.
Finally, section  \ref{sec:couplings} shows the measurement of the couplings to vector bosons and to fermions obtained from the combination of the analyses sensitive to the three production modes using $H \to WW^*$ decays.    
\subsection{Event yields and distributions} \label{sec:yields}
The number of events in each category is summarised in
table~\ref{table:SR8} and in table~\ref{table:SR7} for the 8~\tev\ and
7~\tev\  data sample, respectively.
In the  \twolep-DFOS and \twolep-SS1jet categories the numbers of observed
events are slightly larger than the expectation.
The lepton flavour composition of the events in the \twolep-SS channel is shown in
table~\ref{table:SR_SSBD}, in which the high event yield of the \twolep-SS1jet SR mainly
originates from  the $\mu\mu$ and $\mu e$ channels. The behaviour of the
observed data yield as a function of the data period was studied. The
events were uniformly distributed according to the acquired luminosity as
expected, excluding temporary failures of the detector subsystems as an explanation
of the excess. Moreover it was checked that known detector defects were not
increasing the rate of particular background sources.
Finally, the kinematic distributions of the events were analysed in
order to look for striking features pointing to some particular
missing background contribution. Because no particular problem was
found, the data excesses were attributed 
to statistical fluctuations. 

Distributions of some of the relevant variables after the event selection
are presented for 8~\tev\ data in figure~\ref{Figure:SR_4lep} 
for the 4$\ell$ analyses, in figure~\ref{Figure:SR_3lep} for the 3$\ell$ analyses and in figure~\ref{Figure:SR_twolep}
for the 2$\ell$ analyses.
Figures~\ref{Figure:SR_4lep}(a) and~\ref{Figure:SR_4lep}(c)
show the opening angle between the leptons in
the frame where the \pt\ of the Higgs boson is zero,
$\Delta \phi_{\ell_{0}\ell_{1}}^{\rm boost}$, in 4$\ell$-1SFOS events.
Figures~\ref{Figure:SR_4lep}(b) and~\ref{Figure:SR_4lep}(d) 
show the $\Delta \phi_{\ell_{0}\ell_{1}}^{\rm boost}$ in 
4$\ell$-2SFOS events. 
Figures~\ref{Figure:SR_3lep}(a) and~\ref{Figure:SR_3lep}(b) present the ``BDT Score'' distribution 
in  the 3$\ell$-3SF and 3$\ell$-1SFOS SRs, respectively, and figure~\ref{Figure:SR_3lep}(c) shows the distribution of $\Delta R$ between the leptons 
from the Higgs boson candidate, $\Delta R_{\ell_{0}\ell_{1}}$, in the
3$\ell$-0SFOS SR.
Figures~\ref{Figure:SR_4lep}(a) and~\ref{Figure:SR_4lep}(b) are obtained by applying 
the selections in table~\ref{tab:event_selection} only down to the cuts on the $\pt^{\mathrm{miss}}$, in order to maximise the number of events, 
while the full selection is applied 
to produce the distributions in figures~\ref{Figure:SR_4lep}(c) and~\ref{Figure:SR_4lep}(d).
The distributions in figures~\ref{Figure:SR_3lep}(a)--\ref{Figure:SR_3lep}(c) are shown with all the selections applied except for the one on the displayed variable.
Figure~\ref{Figure:SR_twolep}(a) presents the transverse mass, $\mT$,
in the 2$\ell$-DFOS SR, while figure~\ref{Figure:SR_twolep}(b) and  figure~\ref{Figure:SR_twolep}(c) show  
the smallest opening angle in the transverse plane between a lepton and a jet, $\Delta \phi_{\ell_{i},j}^{\mathrm{min}}$, in the
 2$\ell$-SS1jet SR  and  the 2$\ell$-SS2jet SR, respectively.
The distributions in figure~\ref{Figure:SR_twolep} are shown with all the selections applied except for the one on the displayed variable.
No data populates figure~\ref{Figure:SR_4lep}(d). 
In all distributions, good agreement 
between data and the MC prediction is observed.

\begin{sidewaystable}[h!]
\begin{center}
\scalebox{0.86}{
\begin{tabular}{l|| D{,}{\pm}{-1} D{,}{\pm}{-1} | D{,}{\pm}{-1} D{,}{\pm}{-1} D{,}{\pm}{-1} | D{,}{\pm}{-1} D{,}{\pm}{-1} D{,}{\pm}{-1} }
 \multicolumn{9}{l}{8~\tev\ data sample}                       \\  
 \hline
Process &  \multicolumn{2}{c|}{\fourlep} & \multicolumn{3}{c|}{\threelep}& \multicolumn{3}{c}{\twolep}    \\
\hline
Category &  \multicolumn{1}{c}{2SFOS} & \multicolumn{1}{c|}{1SFOS} & \multicolumn{1}{c}{3SF} & \multicolumn{1}{c}{1SFOS} & \multicolumn{1}{c|}{0SFOS} & \multicolumn{1}{c}{DFOS} & \multicolumn{1}{c}{SS2jet} & \multicolumn{1}{c}{SS1jet} \\
\hline
Higgs boson &&&&&&&\\
$~VH~(H\rightarrow WW^{\ast})$   &0.203 , 0.030   &0.228 , 0.034   &0.73 , 0.10   &1.61  , 0.18  &1.43 , 0.16   &2.15  , 0.30            &1.04 , 0.18     &2.04  , 0.30 \\ 
$~VH~(H\rightarrow \tau\tau)$    &0.0084 , 0.0032 &0.012 , 0.004 &0.057 , 0.011 &0.152 , 0.023 &0.248 , 0.035 &\multicolumn{1}{c}{---} &0.036 , 0.008 &0.27 , 0.04 \\ 
~ggF                 &\multicolumn{1}{c}{---} &\multicolumn{1}{c|}{---} &0.076,0.015&0.085,0.018 &\multicolumn{1}{c|}{---}  &2.4,0.5              &\multicolumn{1}{c}{---}  & \multicolumn{1}{c}{---}          \\ 
~VBF                 &\multicolumn{1}{c}{---} &\multicolumn{1}{c|}{---} &\multicolumn{1}{c}{---} &\multicolumn{1}{c}{---} &\multicolumn{1}{c|}{---} & 0.180,0.025            & \multicolumn{1}{c}{---} & \multicolumn{1}{c}{---}  \\
~ttH                 &\multicolumn{1}{c}{---} &\multicolumn{1}{c|}{---} &\multicolumn{1}{c}{---} &\multicolumn{1}{c}{---} &\multicolumn{1}{c|}{---} &\multicolumn{1}{c}{---} & \multicolumn{1}{c}{---} & \multicolumn{1}{c}{---}  \\    
\hline

Background    &&&&&&& \\
$~~V$                  &\multicolumn{1}{c}{---}     &\multicolumn{1}{c|}{---}      &0.22,0.16 &1.9,0.6 &0.37,0.15&  14,4& 8 , 4 &15 , 5  \\ 
$~~VV$                 &1.17,0.20&0.31,0.06 &19,3&28,4&4.7,0.6&10.1,1.6& 11.2 , 2.1& 26 , 4   \\
$~~VVV$                &0.12,0.04&0.10,0.04 &0.8,0.3 &2.2,0.7 &2.93,0.29&\multicolumn{1}{c}{---}         & \multicolumn{1}{c}{---} & 0.47 , 0.05 \\
$~~$Top                &0.014,0.011&\multicolumn{1}{c|}{---}      &0.91,0.26 &2.4,0.6 &3.7,0.9&24,4& 0.75 , 0.19 & 1.3 , 0.5   \\
$~~$Others             &\multicolumn{1}{c}{---}     &\multicolumn{1}{c|}{---}   &\multicolumn{1}{c}{---}      &\multicolumn{1}{c}{---}      &\multicolumn{1}{c|}{---}    & 2.3,0.9 & 0.71 , 0.30 & 0.60 , 0.24  \\
$~~$Total              &1.30,0.23&0.41,0.09&22,4&34,6&11.7,1.8&50,5& 21 , 5 & 44 , 6   \\
\hline
Observed events         & \multicolumn{1}{c}{0} & \multicolumn{1}{c|}{3} & \multicolumn{1}{c}{22} & \multicolumn{1}{c}{38} & \multicolumn{1}{c|}{14} & \multicolumn{1}{c}{63} & \multicolumn{1}{c}{25} & \multicolumn{1}{c}{62} \\
\hline
 \end{tabular}
}
\caption{Number of observed and predicted events in the SRs and their composition in the 8~\tev\ data sample. Background
processes that contribute less than 1$\%$ of the total background, and Higgs boson production modes that contribute less than 1$\%$ of the $VH(\hww)$ process, are not included in the table. The uncertainties on event yields include both the statistical and systematic components (see section~\ref{sec:systematics}).}
\label{table:SR8}
\end{center}
\end{sidewaystable}

\begin{sidewaystable}
\begin{center}
\scalebox{0.90}{
\begin{tabular}{l|| D{,}{\pm}{-1} D{,}{\pm}{-1} | D{,}{\pm}{-1} D{,}{\pm}{-1} D{,}{\pm}{-1} | D{,}{\pm}{-1} }
\multicolumn{7}{l}{7~\tev\ data sample}                       \\ 
\hline
Process &  \multicolumn{2}{c|}{\fourlep} & \multicolumn{3}{c|}{\threelep} & \multicolumn{1}{c}{\twolep} \\
\hline
Category &  \multicolumn{1}{c}{2SFOS} & \multicolumn{1}{c|}{1SFOS} & \multicolumn{1}{c}{3SF} & \multicolumn{1}{c}{1SFOS} & \multicolumn{1}{c|}{0SFOS} & \multicolumn{1}{c}{DFOS} \\
\hline
Higgs boson &&&&&&\\
$~V(H\rightarrow WW^{\ast})$          &0.0226 , 0.0033  &0.0208 , 0.0031 &0.129 , 0.013 &0.325 , 0.034  &0.291 , 0.031 &0.28,0.05 \\ 
\hline
$~V(H\rightarrow \tau\tau)$    &0.0031 , 0.0012  &0.0014, 0.0008 &0.0163 , 0.0035 &0.041, 0.006 &0.067, 0.010 &0.0075 , 0.0032 \\ 
~ggF                  &\multicolumn{1}{c}{---}     &\multicolumn{1}{c|}{---}    &0.0045,0.0015     &0.0045,0.0019  &0.0048,0.0027   &0.32,0.09 \\ 
~VBF                  &\multicolumn{1}{c}{---}     &\multicolumn{1}{c|}{---} &\multicolumn{1}{c}{---}   &\multicolumn{1}{c}{---} &\multicolumn{1}{c|}{---}   &0.021,0.004 \\
~$\ttbar$H                  &\multicolumn{1}{c}{---}     &\multicolumn{1}{c|}{---} &\multicolumn{1}{c}{---}   &0.006,0.004 & 0.0041,0.0032   &\multicolumn{1}{c}{---} \\ 

\hline \\
\hline
Background    &&&&&&\\ 
$~~V$                 &\multicolumn{1}{c}{---}               &\multicolumn{1}{c|}{---}                 &0.36,0.30   &0.59,0.34   &0.36,0.22   &3.4,1.3\\ 
$~~VV$                &0.37,0.14     &0.031,0.013     &4.1,0.6   &5.7,1.0     &1.3,0.2   &0.89,0.27 \\ 
$~~VVV$               &0.014,0.005 &0.0095,0.0033 &0.082,0.028 &0.21,0.07 &0.338,0.031 &\multicolumn{1}{c}{---}           \\ 
$~~$Top               &0.006,0.004 &\multicolumn{1}{c|}{---}                 &0.12,0.14   &0.4,0.3   &0.44,0.29   &3.2,0.8 \\ 
$~~$Others            &\multicolumn{1}{c}{---} &\multicolumn{1}{c|}{---} &\multicolumn{1}{c}{---} &\multicolumn{1}{c}{---}&\multicolumn{1}{c|}{---} &\multicolumn{1}{c}{---}\\ 
$~~$Total             &0.39,0.15     &0.041,0.016     &4.6,1.1     &7.0,1.9     &2.5,0.7   &7.5,1.7   \\ 
\hline
Observed events       & \multicolumn{1}{c}{1} & \multicolumn{1}{c|}{0} & \multicolumn{1}{c}{5} & \multicolumn{1}{c}{6} & \multicolumn{1}{c|}{2} & \multicolumn{1}{c}{7} \\ 
\hline
\end{tabular}
}
\caption{Number of observed and predicted events in the SRs and their
  composition in the 7~\tev\ data sample. Background
processes that contribute less than 1$\%$ of the total background, and Higgs boson production modes that contribute less than 1$\%$ of the $VH(\hww)$ process, are not included in the table. The uncertainties on event yields include both the statistical and systematic components (see section~\ref{sec:systematics}).}
\label{table:SR7}
\end{center}
\end{sidewaystable}

\begin{sidewaystable}[h!]
\begin{center}
\scalebox{0.82}{
\begin{tabular}{l|| D{,}{\pm}{-1} D{,}{\pm}{-1} | D{,}{\pm}{-1} D{,}{\pm}{-1} | D{,}{\pm}{-1} D{,}{\pm}{-1} | D{,}{\pm}{-1} D{,}{\pm}{-1} }
 \multicolumn{9}{l}{8~\tev\ data sample}                       \\
 \hline
Process &  \multicolumn{2}{c|}{SS \twolep\ ($ee$)} &\multicolumn{2}{c|}{SS \twolep\ ($e\mu$)}&\multicolumn{2}{c|}{SS \twolep\ ($\mu e$)}&\multicolumn{2}{c}{SS \twolep\ ($\mu\mu$)}    \\
\hline
Category & \multicolumn{1}{c}{SS2jet} & \multicolumn{1}{c|}{SS1jet} & \multicolumn{1}{c}{SS2jet} & \multicolumn{1}{c|}{SS1jet} & \multicolumn{1}{c}{SS2jet} & \multicolumn{1}{c|}{SS1jet} & \multicolumn{1}{c}{SS2jet} & \multicolumn{1}{c}{SS1jet} \\
\hline
Higgs boson &&&&&&&\\
$~VH~(H\rightarrow WW^{\ast})$      & 0.15 , 0.07    & 0.39 , 0.08   & 0.36 , 0.08   & 0.50 , 0.12   & 0.24 , 0.08   & 0.69 , 0.14   & 0.28 , 0.06   & 0.46 , 0.11 \\

$~VH~(H\rightarrow\tau\tau)$ & 0.009 , 0.004 & 0.063 , 0.013   & 0.012 , 0.004 & 0.081 , 0.018 & 0.0040 , 0.0020 & 0.080 , 0.014 & 0.012 , 0.004 & 0.046 , 0.010 \\
$~$ggF                      & 0.0028 , 0.0021 & 0.0011 , 0.0011  &\multicolumn{1}{c}{---} &\multicolumn{1}{c|}{---} &\multicolumn{1}{c}{---} &\multicolumn{1}{c|}{---} &\multicolumn{1}{c}{---} &\multicolumn{1}{c}{---} \\
~VBF                        &\multicolumn{1}{c}{---} &\multicolumn{1}{c|}{---} &\multicolumn{1}{c}{---} &\multicolumn{1}{c|}{---} &\multicolumn{1}{c}{---} &\multicolumn{1}{c|}{---} &\multicolumn{1}{c}{---} &\multicolumn{1}{c}{---} \\
~$\ttbar$H                        &\multicolumn{1}{c}{---} &\multicolumn{1}{c|}{---} &\multicolumn{1}{c}{---} &\multicolumn{1}{c|}{---} &\multicolumn{1}{c}{---} &\multicolumn{1}{c|}{---} &\multicolumn{1}{c}{---} &\multicolumn{1}{c}{---} \\
\hline
Background    &&&&&&& \\
$~~V$                       & 2.9 , 2.8       & 5.0 , 2.0       & 2.5 , 1.0       & 6.6 , 2.3     & 1.4 , 0.6     & 2.1 , 0.7   & 1.1 , 0.8     & 1.1 , 0.8 \\
$~~VV$                      & 2.8 , 0.9     & 6.3 , 1.3       & 4.7 , 1.3       & 10.9 , 1.8    & 2.3 , 0.7     & 5.6 , 1.1     & 1.3 , 0.4     & 3.5 , 0.8 \\
$~~VVV$                     &\multicolumn{1}{c}{---} &\multicolumn{1}{c|}{---} &\multicolumn{1}{c}{---} &\multicolumn{1}{c|}{---} &\multicolumn{1}{c}{---} & 0.13 , 0.02 &\multicolumn{1}{c}{---} & 0.11 , 0.03 \\
$~~$Top                     & 0.12 , 0.08   & 0.49 , 0.14     & 0.36 , 0.15     & 0.48 , 0.23   & 0.24 , 0.08   & 0.39 , 0.13 & 0.035 , 0.032   &\multicolumn{1}{c}{---} \\
$~~$Others                  & 0.07 , 0.04   &\multicolumn{1}{c|}{---}                  & 0.26 , 0.12     & 0.44 , 0.17   & 0.32 , 0.15     &\multicolumn{1}{c|}{---}                & 0.061 , 0.031   & 0.056 , 0.029 \\
Total                       & 5.9 , 3.1       & 11.9 , 2.4      & 7.9 , 1.7       & 18.6 , 3.0    & 4.3 , 0.9     & 8.2 , 1.3     & 2.6 , 0.9      & 4.8 , 1.2 \\
\hline
Observed events             & \multicolumn{1}{c}{4} & \multicolumn{1}{c|}{12} & \multicolumn{1}{c}{8} & \multicolumn{1}{c|}{25} & \multicolumn{1}{c}{7} & \multicolumn{1}{c|}{14} & \multicolumn{1}{c}{6} & \multicolumn{1}{c}{11} \\
\hline
\end{tabular}
}
\caption{Number of observed and predicted events in the \twolep-SS channel in the 8~\tev\ data sample with different lepton flavour combinations: $ee$, $e\mu$, $\mu e$ and $\mu\mu$.
Background processes that contribute less than 1$\%$ of the total background, and Higgs boson production modes that contribute less than 1$\%$ of the $VH(\hww)$ process, are not included in the table. The uncertainties on event yields include both the statistical and systematic components (see section~\ref{sec:systematics})}
\label{table:SR_SSBD}
\end{center}
\end{sidewaystable}

\begin{figure}[h!]
\begin{center}
\subfigure[]{\includegraphics[width = 0.48\textwidth]{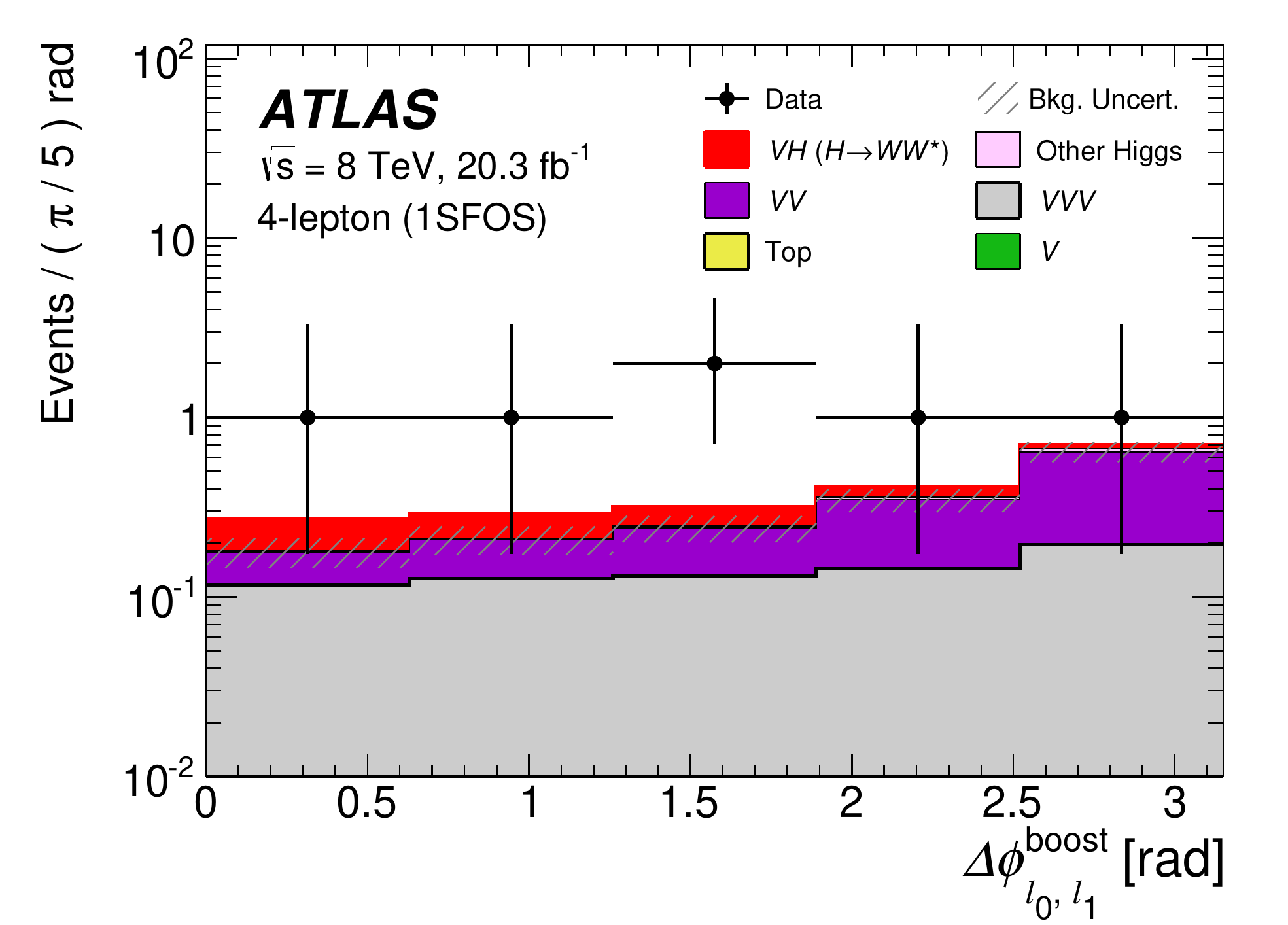}}
\subfigure[]{\includegraphics[width = 0.48\textwidth]{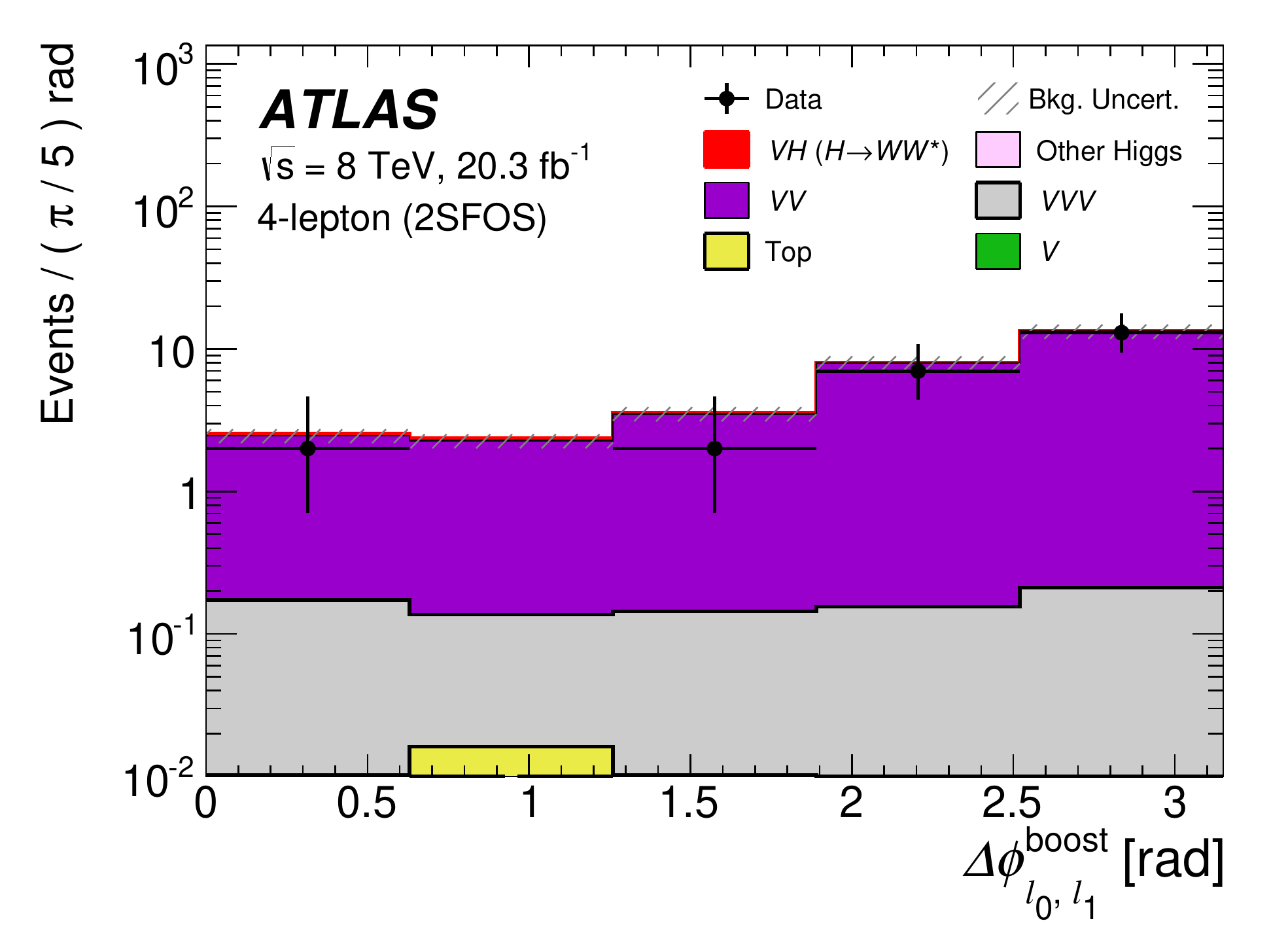}}
\subfigure[]{\includegraphics[width = 0.48\textwidth]{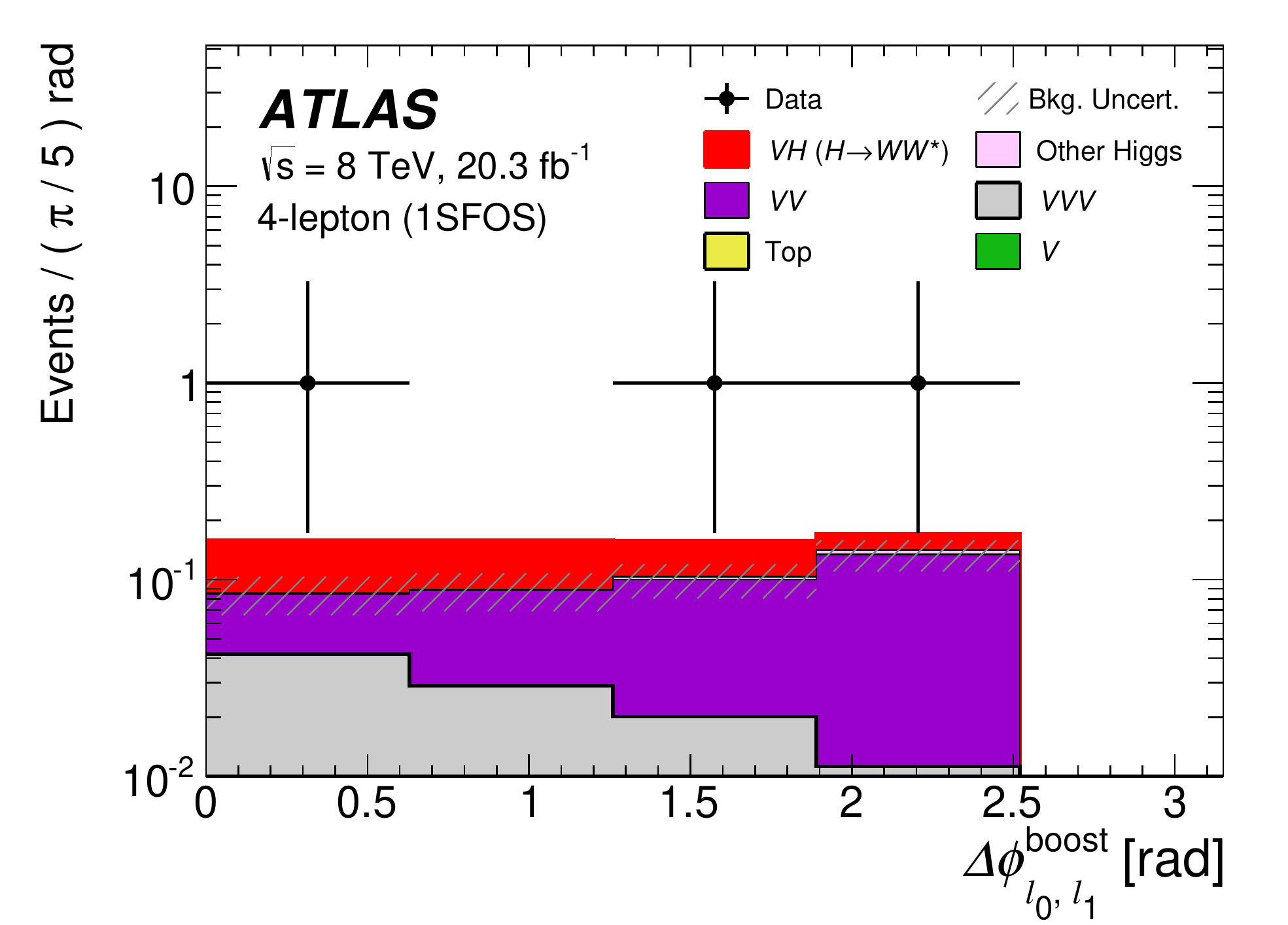}}
\subfigure[]{\includegraphics[width = 0.48\textwidth]{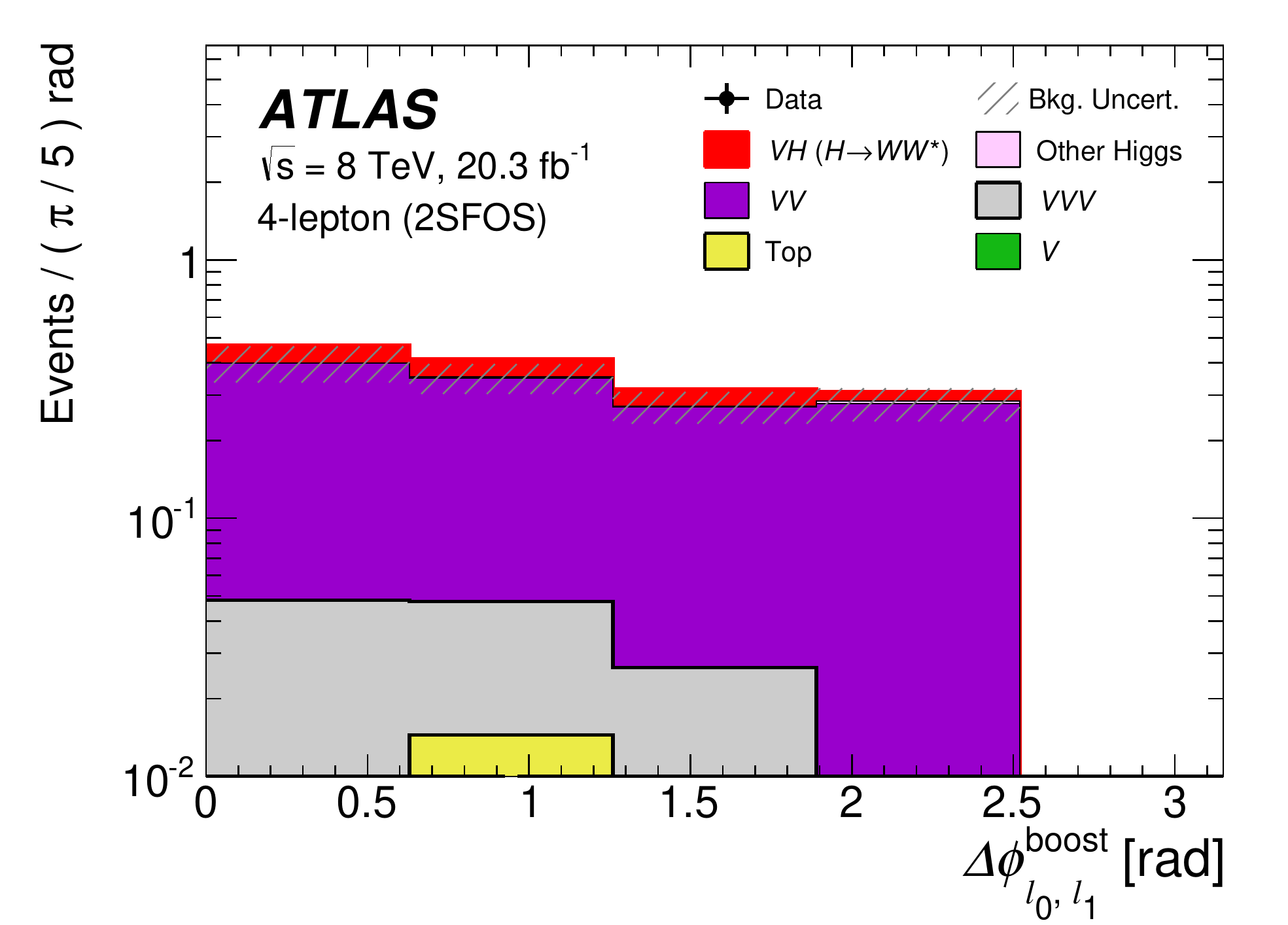}}
\caption{Distributions of the angular separation in $\phi$ between the opposite-sign lepton pair
from the Higgs boson decay candidate in the frame where the \pt\ of the Higgs boson is zero, $\Delta \phi_{\ell_{0}\ell_{1}}^{\rm boost}$, in 
the 4$\ell$ analyses using the 8~\tev\ data sample:
(a) and (c) with 4$\ell$-1SFOS events, and (b) and (d) with 4$\ell$-2SFOS events.
Figures (a) and (b) are 
obtained by applying the cuts in table~\ref{tab:event_selection} down to the track missing \pt\ ($\pt^{\mathrm{miss}}$) selection, 
removing the other selections in order to increase the otherwise very limited number of events,
while the distributions in (c) and (d) have all the selections applied. 
Data (points) are compared to the background plus the $VH(\hww)$ (\mH=125~\gev) signal expectation (stacked filled histograms), 
where the background components are normalised by applying the normalisation factors shown in table~\ref{tab:NFs}.
The hatched area on the histogram represents the total uncertainty on the background estimate including the statistical and systematic uncertainties added in quadrature.
}
\label{Figure:SR_4lep}
\end{center}
\end{figure}

\begin{figure}[h!]
\begin{center}
\subfigure[]{\includegraphics[width = 0.48\textwidth]{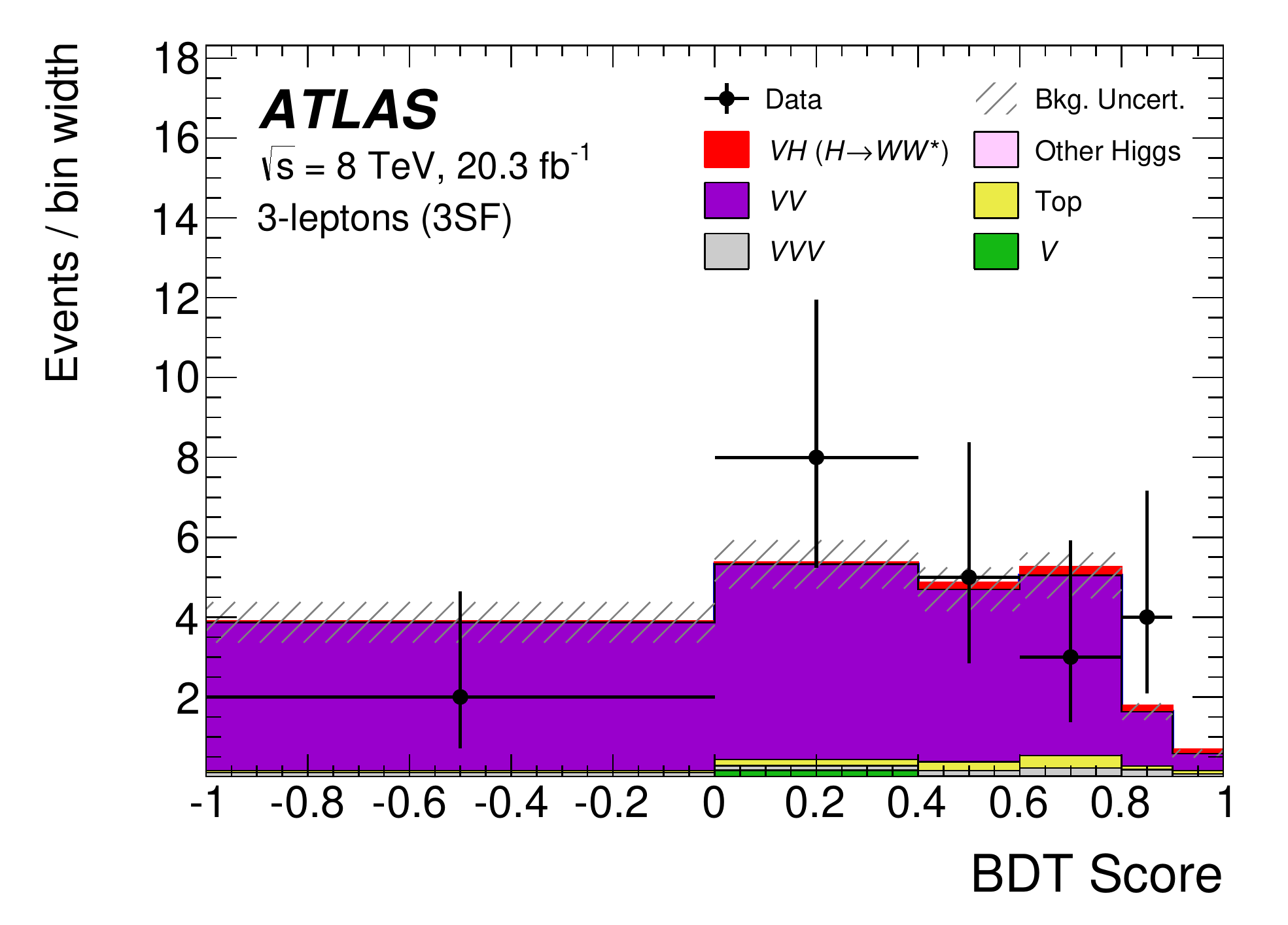}}
\subfigure[]{\includegraphics[width = 0.48\textwidth]{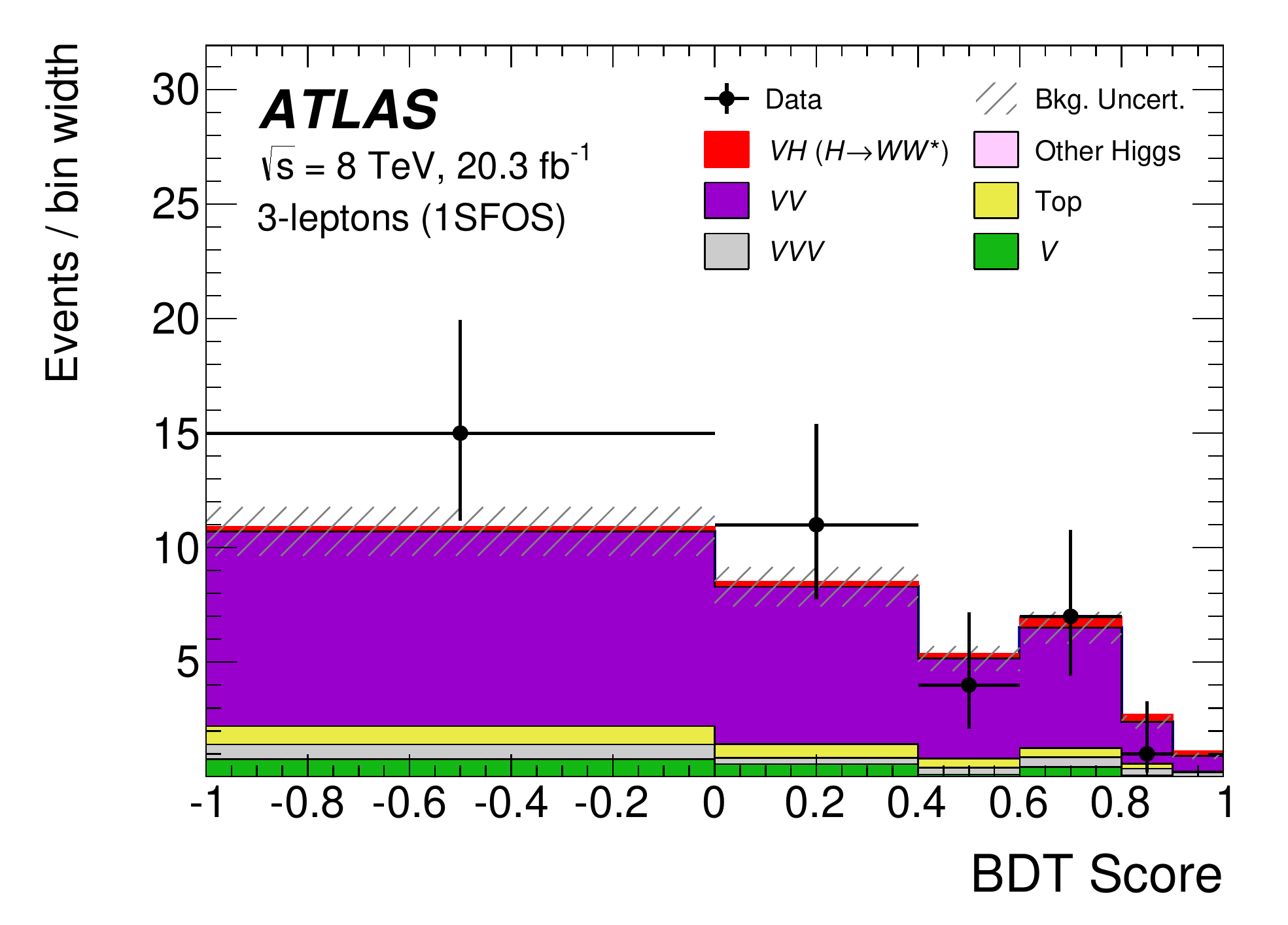}}
\subfigure[]{\includegraphics[width = 0.48\textwidth]{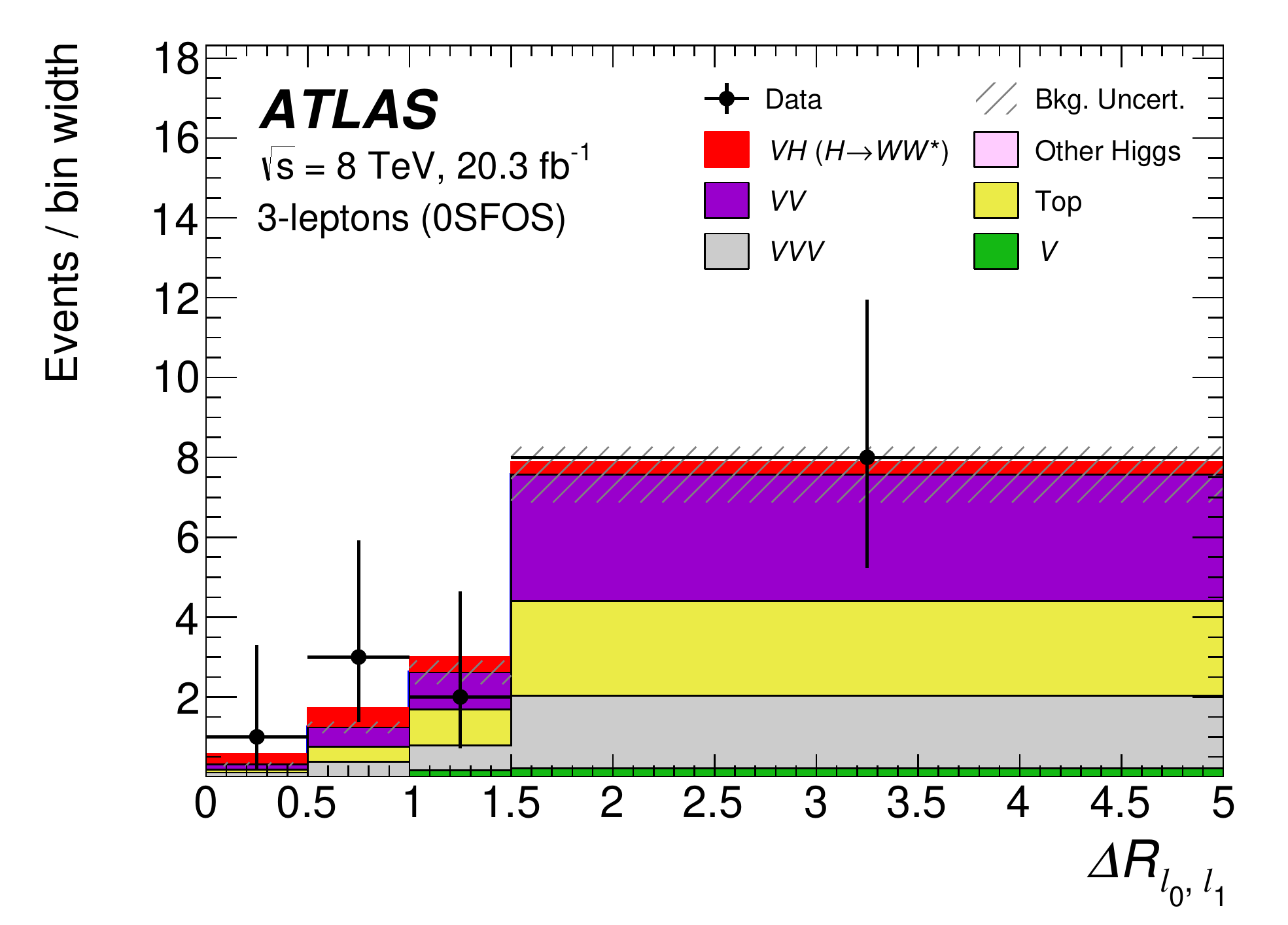}}
\caption{Distributions of relevant quantities for the 
3$\ell$ analyses using the 8~\tev\ data sample: 
(a) ``BDT Score'' in the 3$\ell$-3SF and (b) in the 3$\ell$-1SFOS SRs, and 
(c) the angular separation in $R$ of the two opposite-sign leptons with smaller $\Delta R$ distance, $\Delta R_{\ell_{0}\ell_{1}}$, in the 3$\ell$-0SFOS SR.
The distributions are shown with all
the selections applied except for the one on the displayed variable.
Data (points) are compared to the background plus the $VH(\hww)$ (\mH=125~\gev) signal expectation (stacked filled histograms), 
where the background components are normalised by applying the normalisation factors shown in table~\ref{tab:NFs}.
The hatched area on the histogram represents the total uncertainty on the background estimate including the statistical and systematic uncertainties added in quadrature.
}
\label{Figure:SR_3lep}
\end{center}
\end{figure}

\begin{figure}[h!]
\begin{center}
\subfigure[]{\includegraphics[width = 0.48\textwidth]{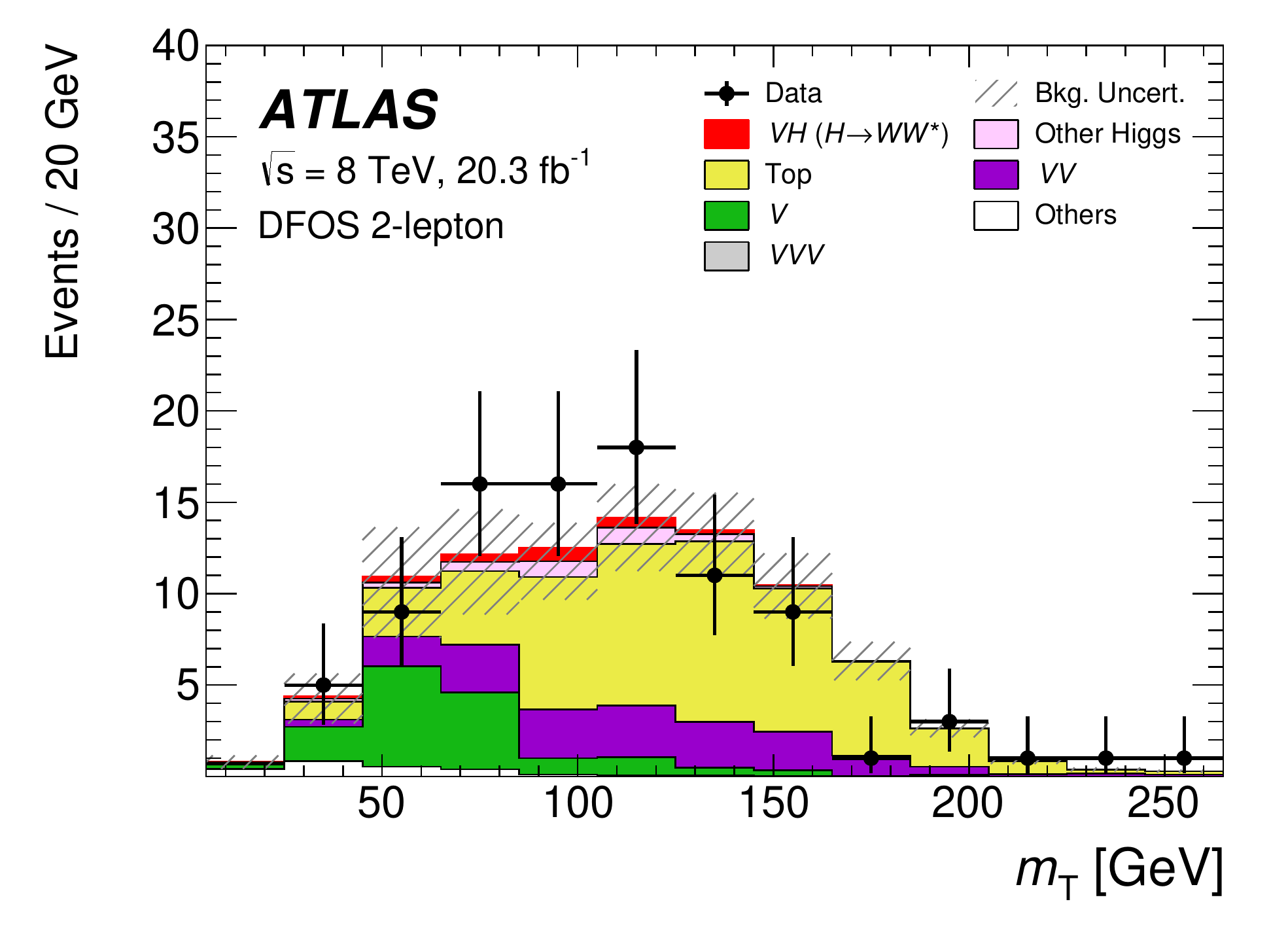}}\\
\subfigure[]{\includegraphics[width = 0.48\textwidth]{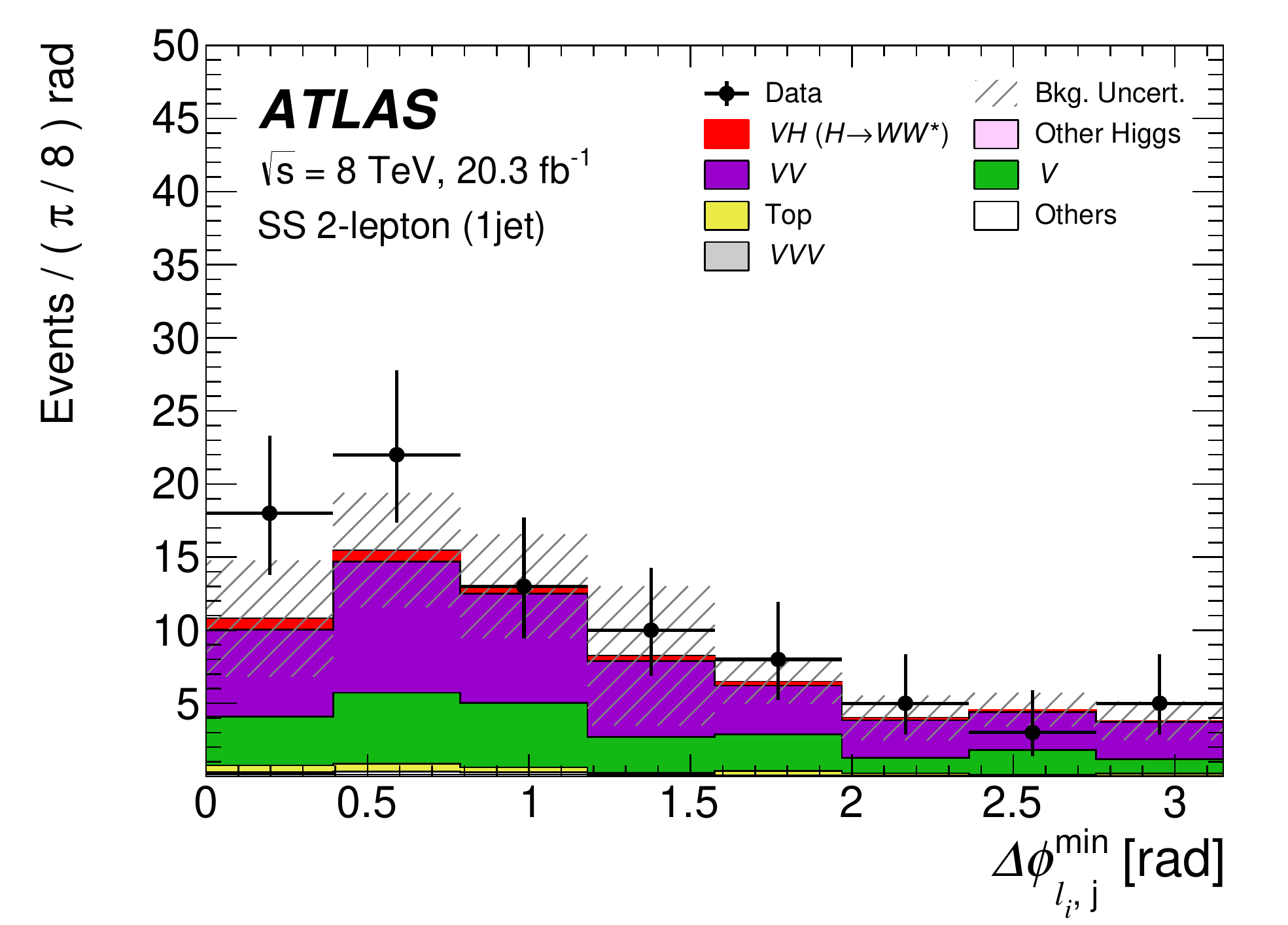}}
\subfigure[]{\includegraphics[width = 0.48\textwidth]{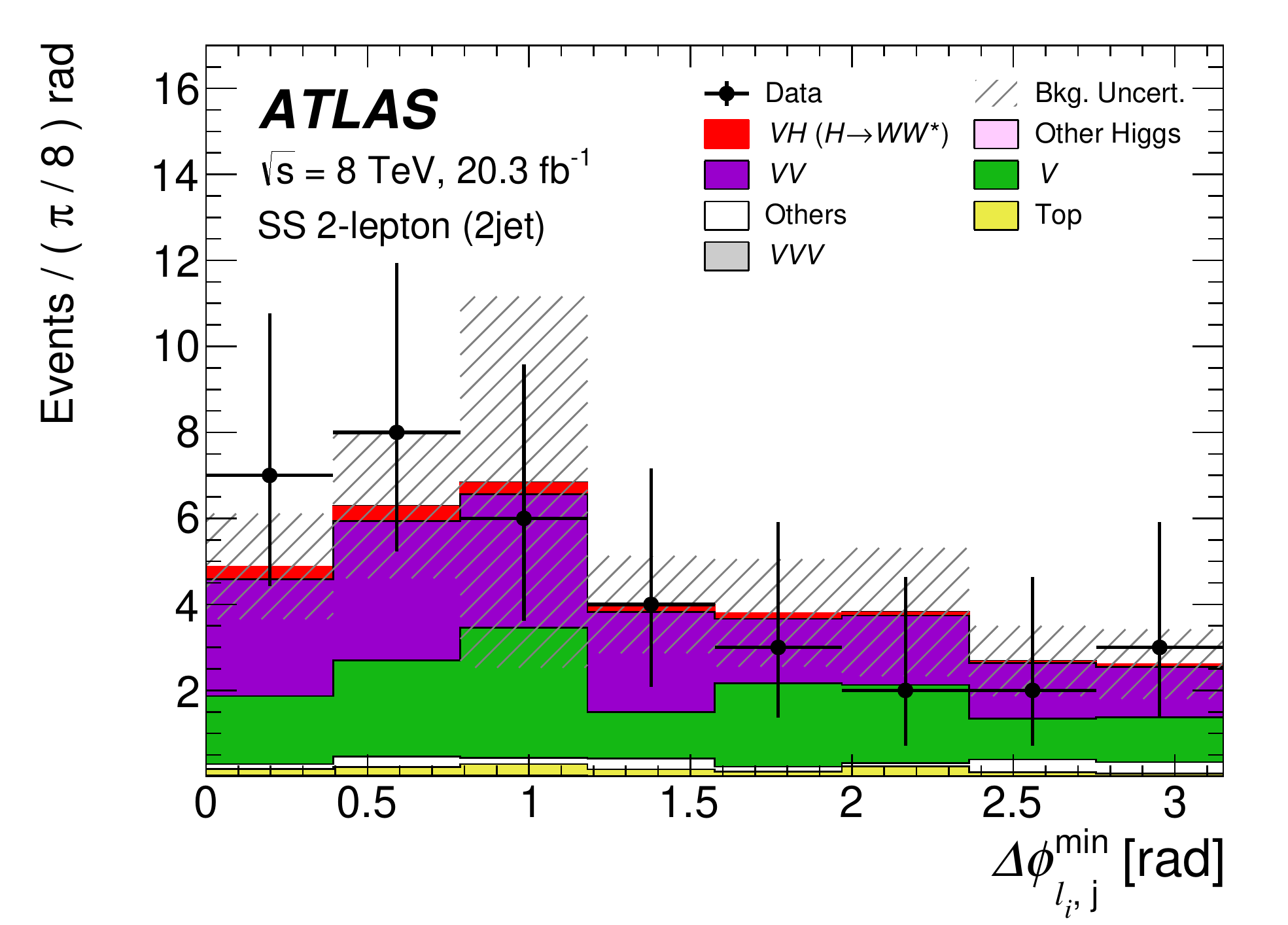}}
\caption{Distributions of relevant quantities for the  2$\ell$ analyses using the 8~\tev\ data sample: 
(a) transverse mass $\mT$ in the 2$\ell$-DFOS SR,
the smallest azimuthal opening angle between a lepton and a jet, $\phi_{\ell_{i}j}^{\mathrm{min}}$, (b) in the 2$\ell$-SS1jet and (c) 2$\ell$-SS2jet SRs.
The distributions are shown with all
the selections applied except for the one on the displayed variable.
Data (points) are compared to the background plus the $VH(\hww)$ (\mH=125~\gev) signal expectation (stacked filled histograms), 
where the background components are normalised by applying the normalisation factors shown in table~\ref{tab:NFs}.
The hatched area on the histogram represents the total uncertainty on
the background estimate including the statistical and systematic
uncertainties added in quadrature.
}
\label{Figure:SR_twolep}
\end{center}
\end{figure}

\subsection{Statistical method \label{sec:stat_formalism} }
The signal extraction is performed using the profile likelihood ratio
method~\cite{asymptotics}, which consists of maximising a binned likelihood function \makebox{$\mathcal{L}(\mu,
\boldsymbol{\theta}\mid\boldsymbol{n})$}. 
The likelihood is the product of  Poisson distributions for each SR and CR. 
The mean
values of the distributions are the sum of the  expected yields  of signal and background. The
symbol $\boldsymbol{n}$ represents the observed events in each SR
and CR. The signal and background expectations are functions of the 
signal-strength parameter, $\mu$, and a set of nuisance parameters, $\boldsymbol{\theta}$.
The signal strength $\mu$ multiplies the SM predicted signal event yield
in all categories, while background normalisation factors, included as
nuisance parameters, represent corrections for  background
sources normalised to data. 
Signal and background predictions are affected by systematic uncertainties
that are described by nuisance parameters. 
The normalisation factors are left free in the fit, while the constraints on the systematic uncertainties are chosen to be log-normal distributions.

The test statistic $q_{\mu}$  is defined as
\begin{equation}
q_{\mu} = -2{\rm
  ln}\frac{\mathcal{L}(\mu,\boldsymbol{\hat{\theta}_{\mu}})}{\mathcal{L}_{\rm
    max}} = -2{\rm
  ln \Lambda}.
\label{eq:q_mu}
\end{equation}
The symbol $\boldsymbol{\hat{\theta}_{\mu}}$  indicates the nuisance parameter values at the maximum of the
likelihood function for a given $\mu$. 
The denominator is the maximum value of $\mathcal{L}$
obtained with both $\mu$ and $\theta$ floating.
When the denominator is maximised, $\mu$ takes the value of $\hat{\mu}$.
The $p_0$ value is computed for the test statistic $q_0$ evaluated at $\mu=0$ in eq.~(\ref{eq:q_mu}), and is defined to be the
probability to obtain a value of $q_0$ larger than the observed value
under the background-only hypothesis. There are no bounds on
$\hat{\mu}$, although $q_0$ is defined to be negative if $\hat{\mu}
\le 0$. 
The equivalent formulation, expressed in terms of the number of standard deviations $\sigma$, is
referred to as the local significance $Z_0$.  
The signal acceptance for all production
modes and decays are computed 
assuming $\mH=125.36\gev$, which is the mass measured in $H \to \gamma \gamma$ and $H \to 4 \ell$ decays by ATLAS~\cite{Aad:2014aba}.
The acceptance for this mass is obtained  interpolating  between the values computed at $\mH = 125$ and $130 \gev$.  
\clearpage
\subsection{Characterisation of the excess and $VH$ signal region splitting                       \label{sec:characterization}
}

Table~\ref{tab:split} shows the expected sensitivity to the SM Higgs boson with mass $\mH=125.36\gev$, 
the observed signal significance $Z_0$ for \hww\ decays and the measured $\mu$ value using the categories described in section~\ref{sec:selection}. 
The 3$\ell$-3SF and 3$\ell$-1SFOS SRs are further split in the likelihood function according to the value of the ``BDT Score'', 
while the 3$\ell$-0SFOS SR is split into intervals of $\Delta R_{\ell_0,\ell_1}$, as discussed in section~\ref{subs:3l}. 
The intervals are shown in figures~\ref{Figure:SR_3lep}(a)--\ref{Figure:SR_3lep}(c). 
Each of the 2$\ell$-SS2jet and 2$\ell$-SS1jet SRs is further split into four sub-categories according to the flavour of the leading and sub-leading leptons. 
For the 2$\ell$-DFOS a single SR is considered. 
The numbers in table~\ref{tab:split} are computed by adding the contributions from the ggF and VBF production to the signal component, and the relative strengths of $VH$, 
ggF and VBF production are fixed to the SM values and constrained with their theoretical uncertainties.

\begin{table*}[h!]
{\tiny
\begin{tabular*}{1\textwidth}{lp{0.002\textwidth}l}
\dbline
\begin{tabular*}{0.370\textwidth}{llll}
& \multicolumn{3}{c}{Signal significance $Z_0$}
\\
\clineskip
\cline{2-4}
\clineskip
Category
& Exp.
& Obs.
& Obs.
\\
& $\, \, Z_0$
& $\, \, Z_0$
& $\, \, Z_0$
\\
\sgline
\balkenscale{480}{-0.1}
\fourlep               &$0.41$ &$1.9$ &{\myr\Balkenx{0}{1.91}{0}{0}{0}}\\
\quad 2SFOS            &$0.19$  &$0$ &{\myb\Balkenx{0}{0}{0}{0}{0}}\\
\quad 1SFOS            &$0.36$ &$2.5$ &{\myb\Balkenx{0}{2.48}{0}{0}{0}}\\
\sgline
\threelep              &$0.79$ &$0.66$ &{\myr\Balkenx{0}{0.66}{0}{0}{0}}\\
\quad 1SFOS and 3SF  &$0.41$ &$0$ &{\myb\Balkenx{0}{0}{0}{0}{0}}\\
\quad  0SFOS     &$0.68$ &$1.2$ &{\myb\Balkenx{0}{1.18}{0}{0}{0}}\\
\sgline                                         
\twolep          &$0.59$ &$2.1$ &{\myr\Balkenx{0}{2.10}{0}{0}{0}}\\
\quad DFOS             &$0.54$ &$1.2$ &{\myb\Balkenx{0}{1.18}{0}{0}{0}}\\
\quad SS2jet           &$0.17$ &$1.4$ &{\myb\Balkenx{0}{1.40}{0}{0}{0}}\\
\quad SS1jet           &$0.27$ &$2.3$ &{\myb\Balkenx{0}{2.29}{0}{0}{0}}\\
\sgline
&&&\hspace{-7.85pt}
\begin{bchart}[step=1,max=3.,width=0.080\textwidth,scale=1.188]\end{bchart}
\\
\end{tabular*}
&
\begin{tabular*}{0.570\textwidth}{lp{0.002\textwidth} ll ll}
& \multicolumn{5}{c}{Observed signal strength $\sigmu$}\\
\clineskip
\cline{1-6}
\clineskip
 \multicolumn{1}{c}{{\bf $\sigmu$}}
& \multicolumn{2}{c}{Tot.\,err.}
& \multicolumn{2}{c}{Syst.\,err.}

& \multicolumn{1}{c}{\quad  $\sigmu$}
\\
& \multicolumn{1}{c}{$+$} & \multicolumn{1}{c}{$-$}
& \multicolumn{1}{c}{$+$} & \multicolumn{1}{c}{$-$}
&\\
\sgline
\balkenscale{150}{-8.0}

{\bf $4.9$} & $4.6$ &$3.1$ &$1.1$ &$0.40$ &{\myr\Balkenx{4.9}{4.45}{3.07}{1.08}{0.40}} \\
{\bf  $-5.9$} & $6.8$ &$4.1$ &$0.33$ &$0.72$ &{\myb\Balkenx{-5.8}{6.79}{4.04}{0.33}{0.72}} \\
{\bf $9.6$} & $8.1$  &  $5.4$ & $2.1$ & $0.64$ &{\myb\Balkenx{9.6}{7.87}{5.33}{2.08}{0.65}} \\
\sgline
{\bf $0.72$} & $1.3$ &$1.1$ &$0.40$ &$0.29$ &{\myr\Balkenx{0.72}{1.29}{1.01}{0.40}{0.29}} \\
{\bf $-2.9$} & $2.7$ &$2.1$ &$1.2$ &$0.92$ &{\myb\Balkenx{-2.87}{2.45}{1.90}{1.25}{0.92}} \\
{\bf $1.7$} & $1.9$ &$1.4$  &$0.51$ &$0.29$ &{\myb\Balkenx{1.7}{1.84}{1.41}{0.51}{0.29}} \\
\sgline
{\bf $3.7$} & $1.9$ &$1.5$  &$1.1$ &$1.1$ &{\myr\Balkenx{3.7}{1.45}{1.37}{1.15}{1.136}} \\
{\bf $2.2$} & $2.0$ &$1.9$  &$1.0$ &$1.1$ &{\myb\Balkenx{2.20}{1.66}{1.54}{1.04}{1.08}} \\
{\bf $7.6$} & $6.0$ &$5.4$  &$3.2$ &$3.2$ &{\myb\Balkenx{7.67}{5.06}{4.37}{3.2}{3.2}} \\
{\bf $8.4$} & $4.3$ &$3.8$  &$2.3$ &$2.0$ &{\myb\Balkenx{8.396}{3.57}{3.23}{2.34}{1.96}} \\
\sgline
            &       &       &       &       &
\hspace{-20.0pt}
\begin{bchart}[step=2,min=-10,max=17,width=0.255\textwidth,scale=1.04]\end{bchart}
$\nq$
\end{tabular*}
\\
\dbline
\end{tabular*}

}
\caption{
The signal significance $Z_0$, and the \hww\ signal strength $\mu$ evaluated
in the signal regions, combining the 8 TeV and 7
TeV data.  The expected
  (exp.) and observed (obs.) values are shown. The two plots represent
  the observed significance and the observed  $\mu$. In the $\mu$ plot the statistical uncertainty
  (stat.) is represented by the thick line, the total uncertainty (tot.) by the
  thin line. The first entry in each group (in red) indicates the combination of more than one category. All values are computed for a Higgs boson mass of $125.36\GeV$.
}
\label{tab:split}
\end{table*}

\subsection{Signal significance extraction and determination of signal strengths                             \label{sec:interpretation}
}

The $VH$-targeted categories are then combined with the categories of the ggF and VBF analysis
using $H\to WW^{\ast} \to \ell \nu \ell \nu$ decays described in ref.~\cite{HWWllpaper}.
The combination is again performed by building a likelihood
function that includes the SRs and CRs of the ggF, VBF and $VH$ analyses.
The experimental and theoretical uncertainties affecting the same sources
are correlated among different production modes. 
The $\mu$ values for each production mode ($\mu_{\rm ggF}$, $\mu_{\rm
  VBF}$, $\mu_{VH}$) are correlated in all categories and fitted together while the
background NFs are uncorrelated among the different
analyses as they cover different phase-space regions. Therefore, when
extracting $\mu_{VH}$, $\mu_{WH}$ and $\mu_{ZH}$, the ggF and VBF
productions are treated as background and their yields determined by
the global fit.

\noindent The fit results for the $\mu$ values for the $WH$, $ZH$ and $VH$
production
are:

\[
\mu_{WH} = 2.1^{+1.5}_{-1.3}{\,{(\rm stat.)}}^{+1.2}_{-0.8}{\,{(\rm sys.)}} , \, \,
\mu_{ZH} = 5.1^{+3.8}_{-3.0}{\,{(\rm
    stat.)}}^{+1.9}_{-0.9}{\,{(\rm sys.)}} , \, \, \\
\]
\[
\mu_{\small{VH}} = 3.0^{+1.3}_{-1.1}{\, {(\rm stat.)}}^{+1.0}_{-0.7}{\,{(\rm sys.)}}.
\]

The uncertainties on the $\mu_{VH}$ value are shown in
table \ref{table:sys_mu_VH} (see section \ref{sec:systematics} for
their description). The derivative of $\mu_{VH}$ with $m_{H}$ has
been evaluated to be -5.8 \%/\gev\ at $m_H = 125.36$~\gev.

\begin{table}[h!]
\begin{center}
\scalebox{0.75}{
\begin{tabular}{l||cc}
\multicolumn{2}{l}{ Uncertainties on the signal strength $\mu_{VH}$ (\%)}    \\
\hline
Signal theoretical uncertainties  &     \multicolumn{2}{c}{$\Delta \mu_{VH}/\mu_{VH}$} \\ 
                                               & + & - \\
~~~$VH$ acceptance             & 11 & 7 \\
~~~Higgs boson branching fraction & 7 & 4 \\ 
~~~QCD scale                     & 1.6 & 0.7 \\ 
~~~PDF and $\alpha_{\rm S}$            & 3.2 & 1.5  \\ 
~~~$VH$ NLO EW corrections         & 2.5 & 1.2 \\
\hline
Background theoretical uncertainties  &      \\ 
~~~  QCD scale                  & 10 & 9 \\ 
~~~  PDF and $\alpha_{\rm S}$         & 2.3 & 2.0 \\ 
~~~  $VVV$ $K$-factor               & 3.0 & 3.0 \\
~~~  MC modelling               & 7.5 & 6.9  \\ 
\hline
\\
\hline
Experimental uncertainties       &      \\
~~~Jet                           & 14 & 9  \\ 
~~~$\MET$ soft term              & 3.4 & 2.3 \\ 
~~~Electron                      & 4.8 & 2.9   \\ 
~~~Muon                          & 4.8 & 3.2   \\ 
~~~Trigger efficiency            & 1.7 & 0.9  \\ 
~~~$b$-tagging efficiency          & 4.7 & 3.2  \\ 
~~~Fake factor                  & 14 & 12 \\
~~~Charge mis-assignment        & 1.1 & 1.0 \\
~~~Photon conversion rate       & 0.8 & 0.7 \\
~~~Pile-up                       & 3.0 & 1.9  \\ 
~~~Luminosity                    & 5.4 & 3.3 \\
MC statistics & 8 & 8 \\ 
CR statistics              & 18 & 15  \\
ggF SR statistics        & 5.5 & 4.4 \\
VBF SR statistics        & 1.9 & 1.5 \\
ggF+VBF CR statistics        & 10 & 9 \\
\hline
\end{tabular}
}
\end{center}
\caption{Percentage theoretical and experimental uncertainties on the
  observed $VH$ signal strength $\mu_{VH}$. The contributions from signal-related and
  background-related theoretical uncertainties are specified. The ``VH
  acceptance'' is evaluated using both the $qq \to (W/Z)H$ and the
  $gg \to ZH$ production. The
  statistical uncertainty due to the ggF and VBF subtraction
  measured in the categories of the ggF and VBF analysis are indicated with ``ggF SR
  statistics'' and ``VBF SR statistics'', for the contribution from
  the signal regions, and ``ggF+VBF CR statistics'' for the
  contribution from the control regions. The row ``MC statistics''
  shows the uncertainty due to the statistics of the simulated samples.
The values are obtained from the combination of the 8~\tev\ and 7~\tev\ data samples. 
}
\label{table:sys_mu_VH}
\end{table}
Figure \ref{fig:muvh_muzh} shows the value of the test statistic as a function of
$\mu_{WH}$ and $\mu_{ZH}$; as shown, the correlation between
the two parameters is weak.
\begin{figure}
\begin{center}
\includegraphics[width=0.6\textwidth]{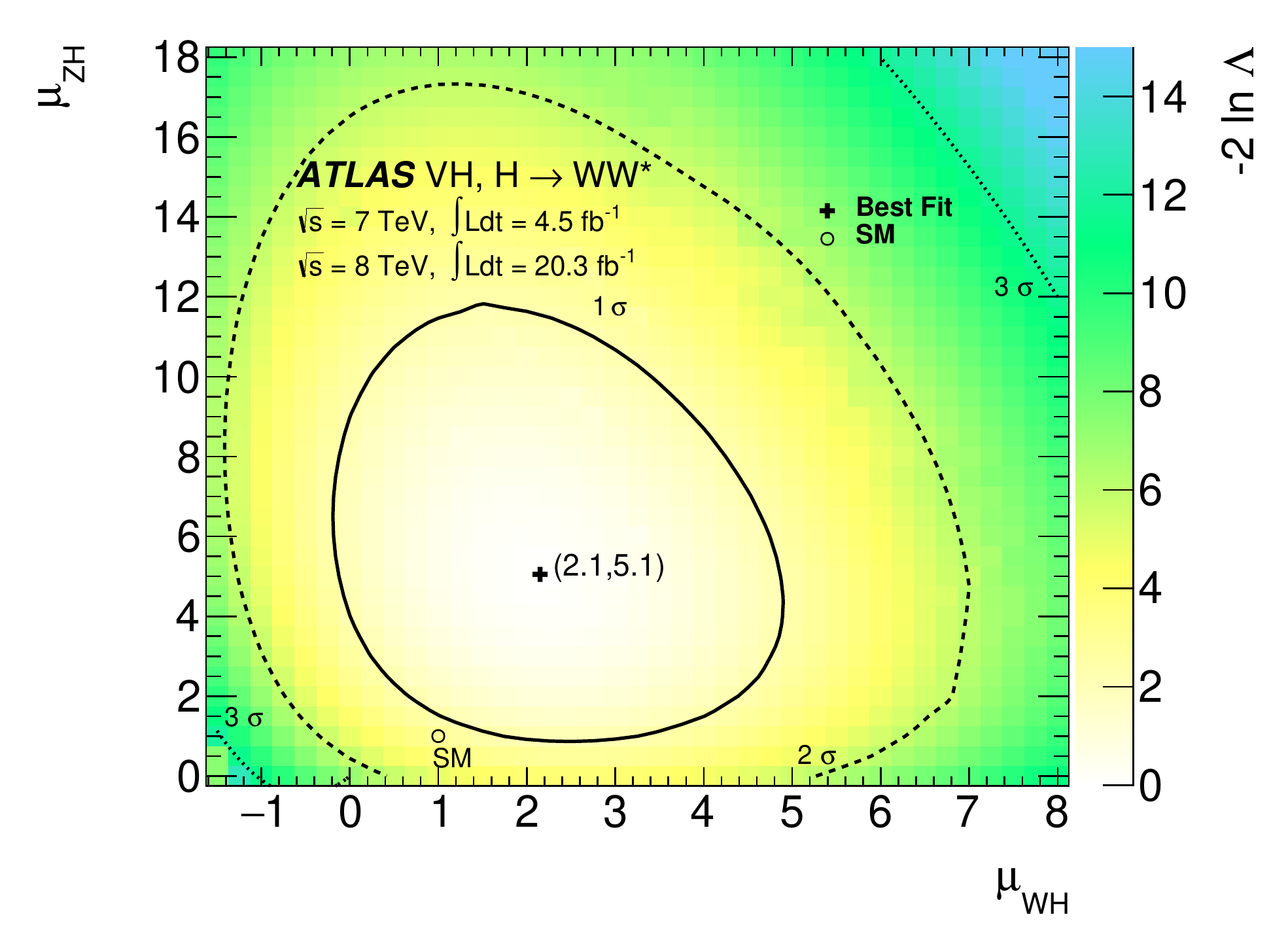}
\end{center}
\caption{The value of the test statistic as a function of $\mu_{WH}$ and
  $\mu_{ZH}$, for $\mH=125.36\gev$. The contours 
correspond to the
  values of $(\mu_{WH},\mu_{ZH})$ associated with 
the  68\%,
  90\% and 95\% confidence levels. 
  The black cross indicates the best fit to the data and the open circle represents the SM expectation ($\mu_{WH}$, $\mu_{ZH}$)=(1,1).} \label{fig:muvh_muzh}
\end{figure}

 Table~\ref{tab:summary} summarises the signal strengths for each 
production mode and their combination at a value of 
$\mH = 125.36\gev$, together with the observed and the expected
$Z_0$. The combined signal strength is $\mu = 1.16^{+0.16}_{-0.15}
{\rm (stat.)}^{+0.18}_{-0.15} {\rm (sys.)}$,
and the significance of the excess respect to the background only
hypothesis is 6.5  $\sigma$, while the expected significance in the
presence of a Higgs boson decaying to $WW^{*}$ is 5.9 $\sigma$.  
Figure \ref{fig:mu_s} shows the value of the test statistic
as a function of the signal strength of each production mode ($\mu_{\rm
  ggF}, \mu_{\rm VBF}, \mu_{VH}$) and as a
function of the combined signal strength $\mu_{HWW}$. 

\begin{figure}
\begin{center}
\begin{tabular}{cccc}
\includegraphics[width=0.45\textwidth]{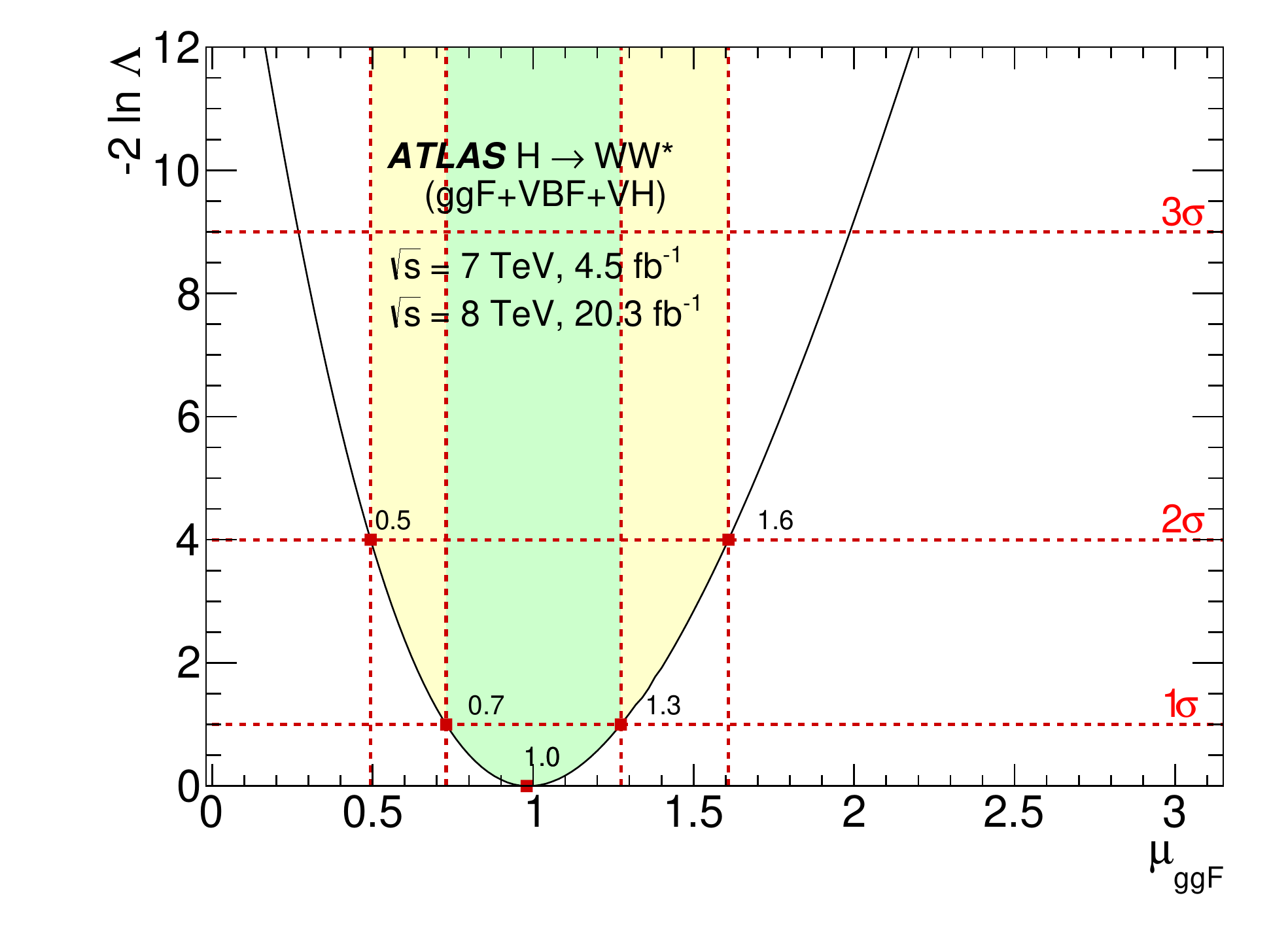} &
\includegraphics[width=0.45\textwidth]{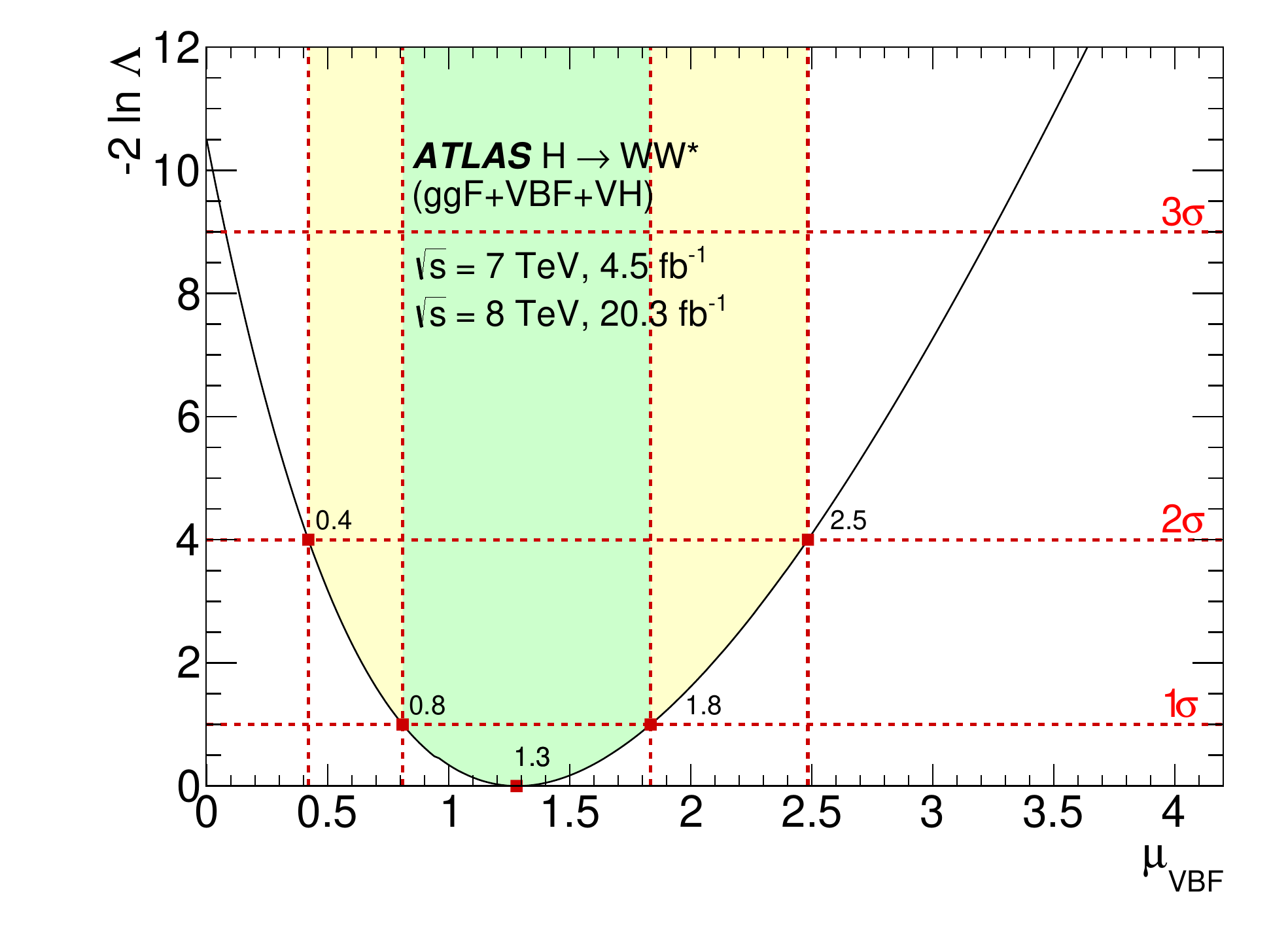} \\
(a) & (b) \\
\includegraphics[width=0.45\textwidth]{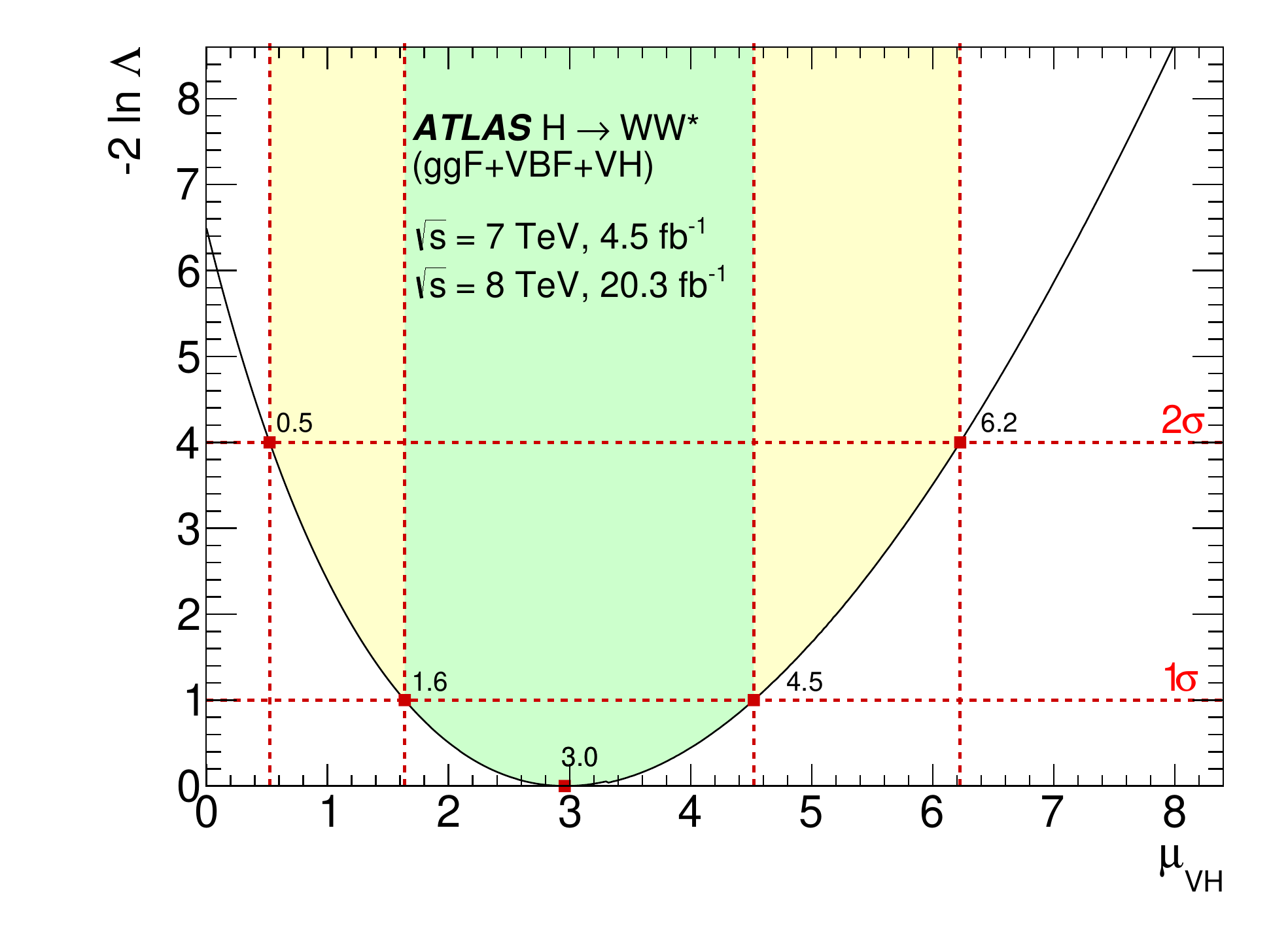} &
\includegraphics[width=0.45\textwidth]{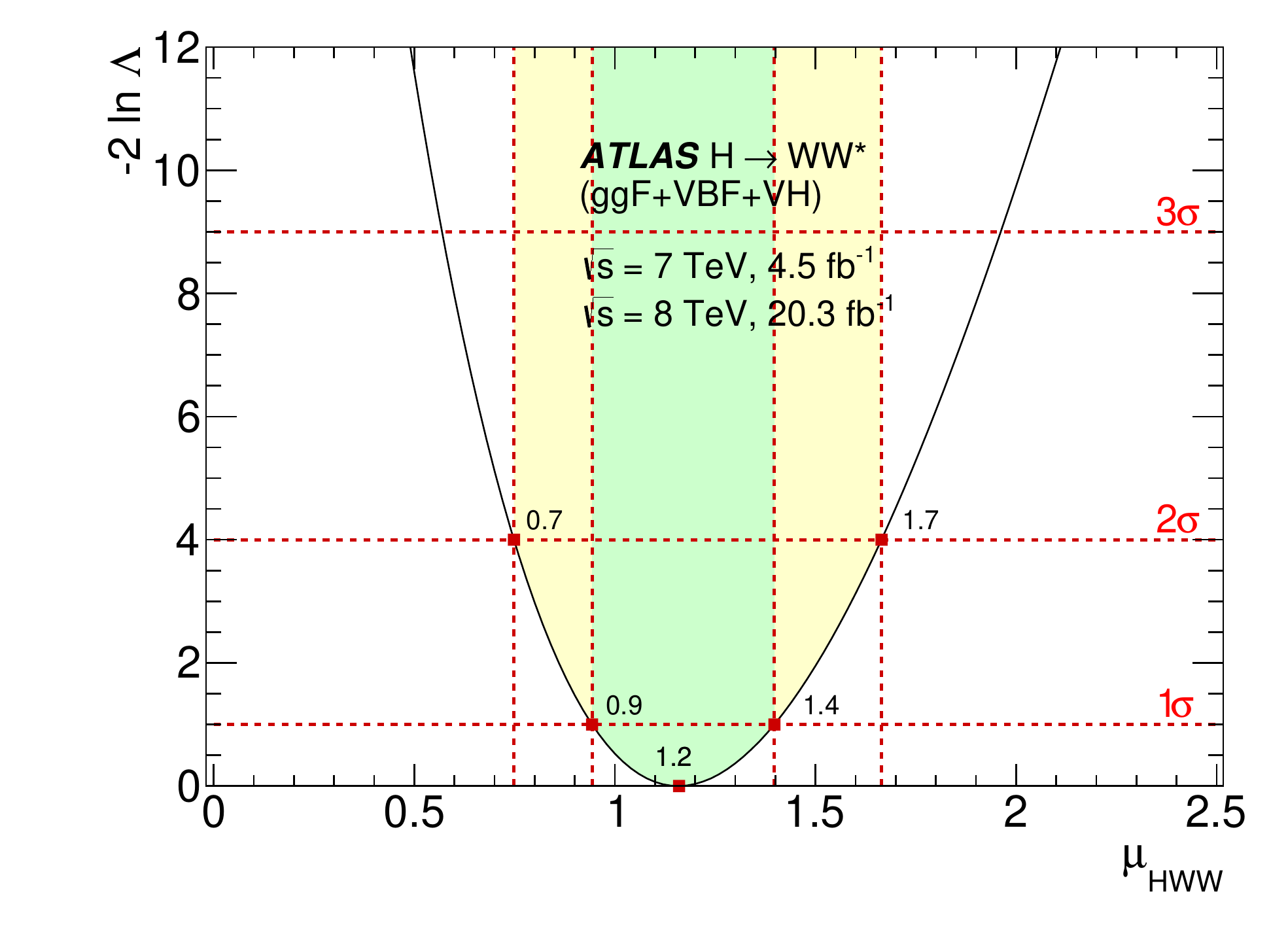} \\
(c) & (d) \\
\end{tabular}
\end{center}
\caption{The 
value of the test statistic as a function of the $\mu$ value from the
different production modes (a) ggF, (b) VBF, (c) $VH$ and (d) all
combined. All values are extracted from the combined fit.
The best fit values are represented by the markers at the likelihood minima, with the $\pm 1 \sigma$ and $\pm 2 \sigma$  uncertainties given by the green and yellow shaded bands.
}\label{fig:mu_s}
\end{figure}

The obtained values are all compatible with the SM expectation within 
1.4 standard deviations. Figures~\ref{fig:correlation}(a) and \ref{fig:correlation}(b)  show 
the two-dimensional dependence of the likelihood on $\mu_{VH}$ and $\mu_{\rm ggF}$, and on $\mu_{VH}$ and $\mu_{\rm VBF}$. 
The $\mu$ values not displayed are kept as free unconstrained parameters in the fit. 
The correlation between the parameters shown in figure~\ref{fig:correlation} is small.
The central values obtained for $\mu_{\rm ggF}$ and $\mu_{\rm VBF}$ are slightly
different from those reported in ref.~\cite{HWWllpaper}. 
The shift represents a few percent of the quoted errors 
and is pulled up by the presence of a small contamination by $VH(\hww)$ events
in the 1-jet and 2-jets categories of the ggF and VBF analysis.

\begin{table*}[hb!]
{\tiny
\begin{tabular*}{1\textwidth}{lp{0.002\textwidth}l}
\dbline
\begin{tabular*}{0.370\textwidth}{llll}
& \multicolumn{3}{c}{Signal significance $Z_0$}
\\
\clineskip
\cline{2-4}
\clineskip
Category
& Exp.
& Obs.
& Obs.
\\
& $\, \, Z_0$
& $\, \, Z_0$
& $\, \, Z_0$ 
\\
\sgline
\balkenscale{250}{-0.1}
\quad ggF                   &$4.4$  &$4.2$ &{\myb\Balkenx{0}{4.2}{0}{0}{0}}\\
\quad VBF                   &$2.6$ &$3.2$ &{\myb\Balkenx{0}{3.2}{0}{0}{0}}\\
\sgline
\quad $VH$    &$0.93$ &$2.5$ &{\myr\Balkenx{0}{2.5}{0}{0}{0}}\\ 
\quad $WH$  only     &$0.77$ &$1.4$ &{\myb\Balkenx{0}{1.4}{0}{0}{0}}\\
\quad $ZH$  \, only  &$0.30$ &$2.0$&{\myb\Balkenx{0}{2.0}{0}{0}{0}} \\
\sgline
\quad  ggF+VBF+$VH$  &$5.9$ &$6.5$ &{\myr\Balkenx{0}{6.5}{0}{0}{0}}\\
\sgline                                         
&&&\hspace{-8.25pt}
\begin{bchart}[step=1,max=7.,width=0.080\textwidth,scale=1.43]\end{bchart}
\\
\end{tabular*}
&
\begin{tabular*}{0.570\textwidth}{lp{0.002\textwidth} ll ll}
& \multicolumn{5}{c}{Observed signal strength $\sigmu$}\\
\clineskip
\cline{1-6}
\clineskip
 \multicolumn{1}{c}{{\bf $\sigmu$}}
& \multicolumn{2}{c}{Tot.\,err.}
& \multicolumn{2}{c}{Syst.\,err.}

& \multicolumn{1}{c}{\quad $\sigmu$}
\\
& \multicolumn{1}{c}{$+$} & \multicolumn{1}{c}{$-$}
& \multicolumn{1}{c}{$+$} & \multicolumn{1}{c}{$-$}
&\\
\sgline
\balkenscale{440}{+0.5}
{\bf $0.98$} & $0.29$ &$0.26$  &$0.22$ &$0.18$ &{\myb\Balkenx{0.98}{0.19}{0.19}{0.22}{0.18}} \\
{\bf $1.28$} & $0.55$  &  $0.47$ & $0.32$ & $0.25$ &{\myb\Balkenx{1.28}{0.45}{0.40}{0.32}{0.25}} \\
\sgline
{\bf $3.0$} & $1.6$ & $1.3$  & $0.95$ & $0.65$ &{\myr\Balkenx{3.0}{1.3}{1.26}{0.95}{0.65}} \\
{\bf $2.1$} & $1.9$ & $1.6$  & $1.2$ & $0.79$ &{\myb\Balkenx{2.1}{1.47}{1.39}{1.2}{0.79}} \\
{\bf $5.1$} & $4.3$ & $3.1$  & $1.9$ & $0.89$ &{\myb\Balkenx{5.1}{3.86}{2.97}{1.9}{0.89}}  \\
\sgline
{\bf $1.16$} & $0.24$ &$0.21$  &$0.18$ &$0.15$ &{\myr\Balkenx{1.16}{0.16}{0.15}{0.18}{0.15}} \\%
\sgline
           &       &       &       &       &
\hspace{-15.50pt}
\begin{bchart}[step=1,min=0,max=10,width=0.203\textwidth,scale=1.435]\end{bchart}
$\nq$
\end{tabular*}
\\
\dbline
\end{tabular*}
}
\caption{
The signal significance $Z_0$, and the signal strength $\mu$ evaluated
 for the different production modes: ggF, VBF and $VH$ for
 $\mH = 125.36$ GeV, for the 8 TeV and 7 TeV data combined. 
The two plots represent
  the observed significance and the observed  $\mu$. In the $\mu$ plot the statistical uncertainty
  (stat.) is represented by the thick line, the total uncertainty (tot.) by the
  thin line. Combinations of different categories (in red) are shown
  too. All values are computed for a Higgs boson mass of
  $125.36\GeV$.
} \label{tab:summary}
\end{table*}

\clearpage
\begin{figure}[h]
\subfigure[]{\includegraphics[width=0.49\textwidth]{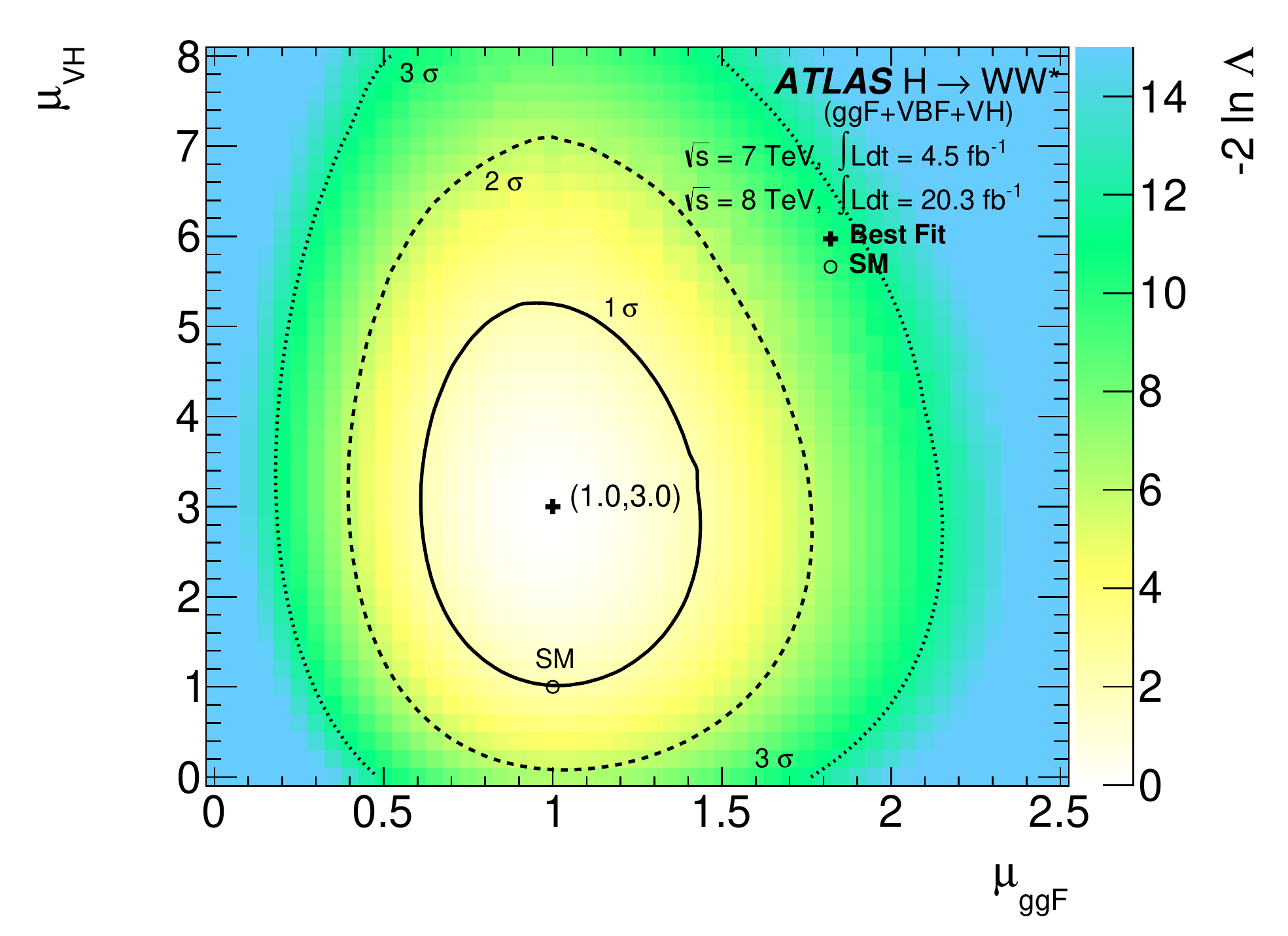}}
\subfigure[]{\includegraphics[width=0.49\textwidth]{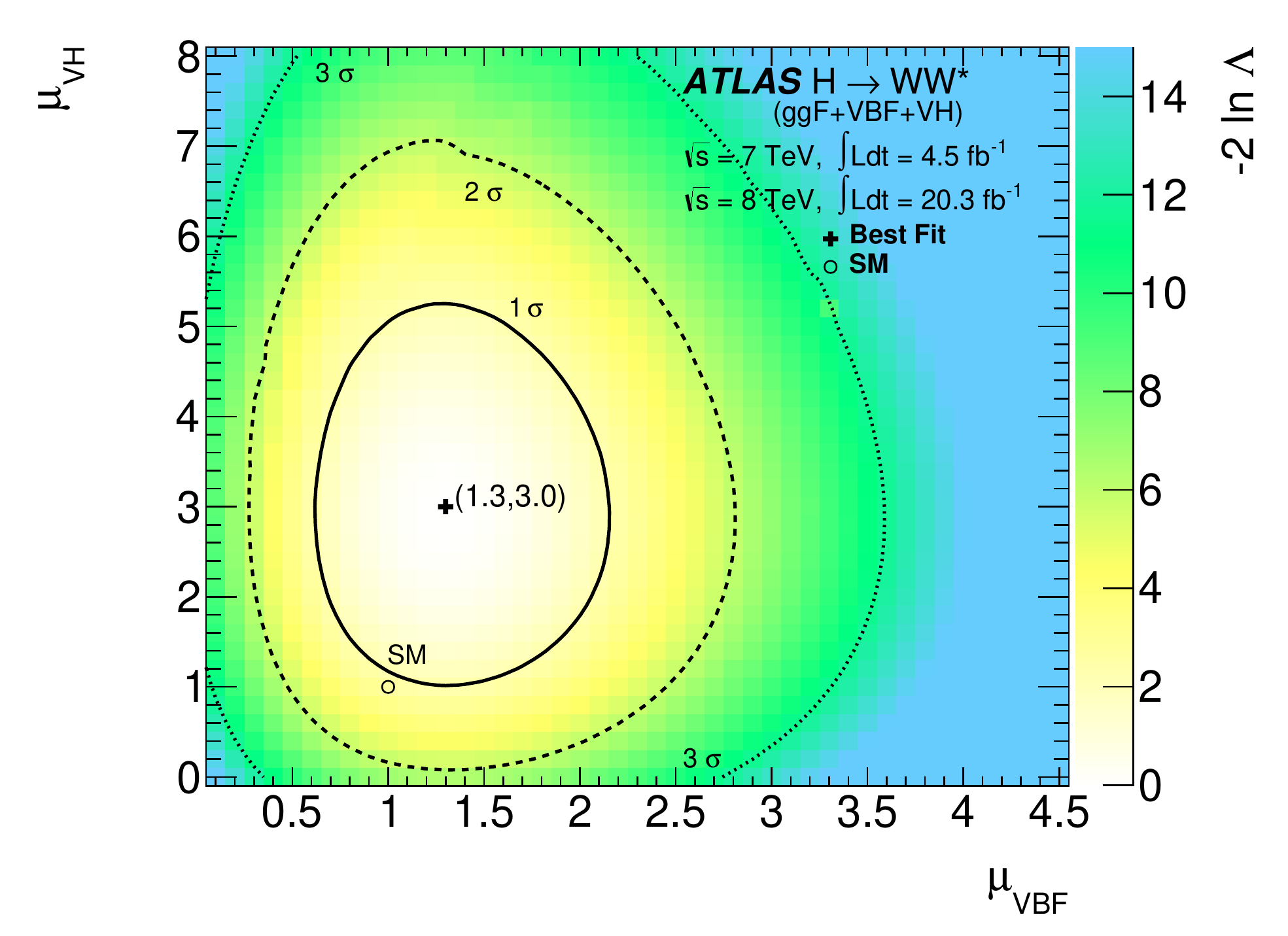}}
\caption{Likelihood as a function of the production mode signal
  strengths in two-dimensional planes of (a)  $\mu_{VH}$  vs $\mu_{\rm ggF}$ and
(b) $\mu_{VH}$ vs $\mu_{\rm VBF}$.  
  The black cross indicates the best fit to the data and the open circle represents the SM expectation (1,1).
}
\label{fig:correlation}
\end{figure}
\subsection{Measurement of the couplings to vector bosons and fermions}  
\label{sec:couplings}         

The values of $\mu_{\rm ggF}$, $\mu_{\rm VBF}$ and $\mu_{VH}$ can be used to test the compatibility of the bosonic
and fermionic couplings of the Higgs boson with the SM prediction using
the formalism developed in ref.~\cite{Heinemeyer:2013tqa}.
Assuming the validity of the SU(2) custodial symmetry and a universal
scaling of the fermion couplings relative to the SM prediction, two
parameters are defined: the scale factor for the SM coupling to the vector bosons ($\kappa_V$) 
and the scale factor for the coupling to the fermions ($\kappa_F$).
Loop-induced processes are assumed to scale as in the
SM. The $H \to \tau \tau$ contribution is treated
as signal and its yield is parameterised as a function of $\kappa_V$ and
$\kappa_F$.
The total width of the Higgs boson can be expressed as the sum of the different 
partial widths, each one rescaled by the square of the appropriate scaling
factor. Neglecting the small contribution from $\Gamma(H \to\gamma\gamma)$ and rarer decay modes,
the $\hww$ decay branching fraction is expressed as:

\begin{equation*}
{\small
{\mathrm{Br}}(\hww) = \frac{\kappa_V^2 \Gamma_{\rm SM} (\hww)}{\kappa_F^2\Gamma_{\rm SM} (H \to
  f\overline{f}) +
  \kappa_F^2\Gamma_{\rm SM}(H \to gg) +  \kappa_V^2\Gamma_{\rm SM}(H\to VV)},
}
\end{equation*}

\noindent where $\Gamma_{\rm SM}(H\to f\overline{f})$,  $\Gamma_{\rm SM} (H\to gg)$ and $\Gamma_{\rm SM} (H\to VV)$ are the SM partial decay
widths to fermions, gluons and weak bosons, respectively.

The ggF ($gg \to H$) process depends directly on the fermion scale factor $\kappa_F^2$ through the top and bottom quark loops, while the VBF ($qq \to Hqq$) and $VH(qq \to VH)$ production cross sections are proportional to
$\kappa_V^2$, as expressed by the following relations:

\begin{equation*}
{\small
\sigma(gg \to H) = \kappa_F^2\sigma_{\rm SM}(gg \to H), \quad \sigma(qq \to H qq) =
\kappa_V^2\sigma_{\rm SM}(qq \to Hqq),
}
\end{equation*}
\begin{equation*}
\sigma(qq \to WH,ZH) = \kappa_V^2\sigma_{\rm SM}(qq \to WH,ZH).
\end{equation*}
where the $\sigma$ without subscript indicates the $(\kappa_V$, $\kappa_F)$-dependent cross sections
and $\sigma_{\rm SM}$ represents the SM cross sections.
The $gg \to ZH$ production cross sections are more complex functions of both $\kappa_V$ and $\kappa_F$~\cite{CombinationNote}:
\begin{equation*}
\sigma(gg \to ZH)_{\,8\tev} = (0.37\times \kappa_F^2 - 1.64\times \kappa_F\times \kappa_V + 2.27\times \kappa_V^2)\,\sigma_{\rm SM}(gg \to ZH)_{\,8\tev},
\end{equation*}
\begin{equation*}
\sigma(gg \to ZH)_{\,7\tev} = (0.35\times \kappa_F^2 - 1.58\times \kappa_F\times \kappa_V + 2.24\times \kappa_V^2)\,\sigma_{\rm SM}(gg \to ZH)_{\,7\tev},
\end{equation*}
The signal event yield is
expressed as $ \sigma \cdot {\rm Br}(\hww)$ using the narrow-width 
approximation.  
Only the relative sign between $\kappa_V$ and $\kappa_F$ is observable and hence
in the following only $\kappa_V>0$ is considered, without loss of
generality.

Sensitivity to the sign results from negative interference, in the $gg \to ZH$ process, between the
box diagram in which both the $Z$ and $H$ bosons are produced directly
from the heavy-quark loop and the triangle diagram in which only the
$Z^{*}$ is produced and subsequently radiates a Higgs boson~\cite{Englert:2013vua}.
Because the relative weights of such processes depend on the
$\sqrt{s}$ of the interaction, different coefficients appear in the
expression for 8 and 7~\tev.  

\begin{figure}[t!]
\begin{center}
\includegraphics[width=\textwidth]{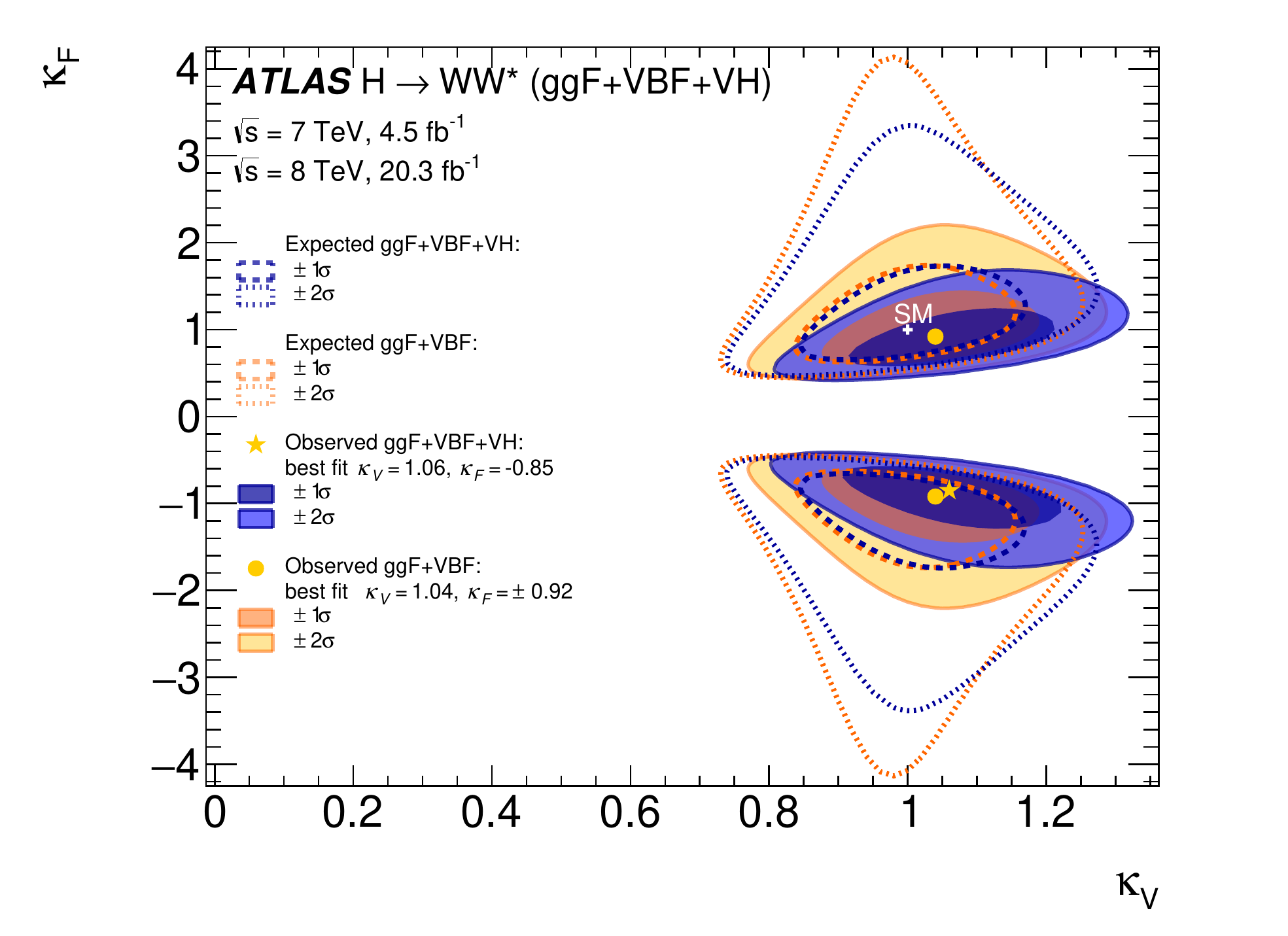}%
\end{center}
\caption{
The likelihood scan as a function of $\kappa_V$ and $\kappa_F$ both with and without the $VH(\hww)$
contribution.~Both the expected and observed contours corresponding to the 68\%,
and 95\% C.L. are shown.
The yellow star and circles indicate the best fit values to the data, and the white cross represents the SM expectation ($\kappa_V$, $\kappa_F$)=(1,1).
}
\label{fig:couplings}
\end{figure}

The likelihood dependence on $\kappa_V$ and $\kappa_F$ is shown in figure~\ref{fig:couplings}.
The product $\sigma(gg \to H)\cdot {\rm Br}(\hww)$, which is measured with good accuracy, does not depend on $|\kappa_F|$ in the limit $|\kappa_F| \gg \kappa_V$. 
This explains the 
low sensitivity to high values of $\kappa_F$.

On the other hand $\mu_{\rm VBF}$ and $\mu_{VH}$, as measured for the $\hww$ decay, should vanish in the limit 
$|\kappa_F| \gg \kappa_V$ due to the increased value of the Higgs boson total width and the consequent reduction of the $\hww$ branching fraction. 
The observation of significant excesses in the VBF and $VH$ production modes therefore leads to an exclusion of the $|\kappa_F| \gg \kappa_V$ region.

The fit to the data results in two local minima and, although the negative $\kappa_F$ solution is preferred to the positive solution at $0.5\sigma$, the observed results are compatible with the SM expectation, and the best fit values are:
\begin{eqnarray*}
|\kappa_F| = 0.85^{+0.26}_{-0.20}, \quad |\kappa_V| = 1.06^{+0.10}_{-0.10},
\end{eqnarray*}
and their correlation is $\rho = 0.54$.

\section{Conclusions                              \label{sec:conclusions}              }
A search for the Standard Model Higgs boson produced in association with a
$\Wboson$ or $\Zboson$ boson and decaying into $WW^{\ast}$
is presented. 
Associated $WH$ production is studied in the final states in which the
three $W$ bosons decay to leptons or where one $W$ boson decays to
hadrons while the others decay leptonically.
The two-lepton and four-lepton final states are used to search for $ZH$ production.
The dataset corresponds to integrated luminosities of
\lumieleven\ifb\ and \lumi$\ifb$
recorded by the
ATLAS experiment with LHC proton--proton collisions 
at $\sqrt{s} = $ 7~\tev\ and 8~\tev, respectively.
For the Higgs boson mass of 125.36~\gev, the observed (expected) deviation from the
background-only hypothesis, which includes the Standard Model expectation for $H \to \tau \tau$,  corresponds to a significance of 2.5 (0.9) 
standard deviations. The ratio of the measured signal yield to its
Standard Model expectation for the $VH$ production is found to be
$\mu_{\rm VH} =3.0^{+1.3}_{-1.1}{\, {(\rm
    stat.)}}^{+1.0}_{-0.7}{\,{(\rm sys.)}}$. A combination
with the gluon fusion and vector boson fusion analyses using the
$H \to WW^{*} \to \ell \nu \ell \nu$ decay is also 
presented. Including $VH$ production the observed 
significance for a Higgs boson decaying to $WW^{*}$ is 6.5 $\sigma$ with
an expectation of 5.9 $\sigma$ for a Standard Model Higgs boson of
mass $m_H = 125.36$~\gev. The combined signal strength is $\mu
= 1.16^{+0.16}_{-0.15} {\rm (stat.)}^{+0.18}_{-0.15} {\rm (sys.)}$. The 
data were analysed using a model where  all
Higgs boson couplings to the vector bosons are scaled by a common factor
$\kappa_V$ and those to the fermions  by a factor $\kappa_F$. They are
measured as $|\kappa_V| = 1.06^{+0.10}_{-0.10}$ and $|\kappa_F|=0.85^{+0.26}_{-0.20}$.
 
\clearpage
\section*{Acknowledgements}

We thank CERN for the very successful operation of the LHC, as well as the
support staff from our institutions without whom ATLAS could not be
operated efficiently.

We acknowledge the support of ANPCyT, Argentina; YerPhI, Armenia; ARC,
Australia; BMWFW and FWF, Austria; ANAS, Azerbaijan; SSTC, Belarus; CNPq and FAPESP,
Brazil; NSERC, NRC and CFI, Canada; CERN; CONICYT, Chile; CAS, MOST and NSFC,
China; COLCIENCIAS, Colombia; MSMT CR, MPO CR and VSC CR, Czech Republic;
DNRF, DNSRC and Lundbeck Foundation, Denmark; EPLANET, ERC and NSRF, European Union;
IN2P3-CNRS, CEA-DSM/IRFU, France; GNSF, Georgia; BMBF, DFG, HGF, MPG and AvH
Foundation, Germany; GSRT and NSRF, Greece; RGC, Hong Kong SAR, China; ISF, MINERVA, GIF, I-CORE and Benoziyo Center, Israel; INFN, Italy; MEXT and JSPS, Japan; CNRST, Morocco; FOM and NWO, Netherlands; BRF and RCN, Norway; MNiSW and NCN, Poland; GRICES and FCT, Portugal; MNE/IFA, Romania; MES of Russia and NRC KI, Russian Federation; JINR; MSTD,
Serbia; MSSR, Slovakia; ARRS and MIZ\v{S}, Slovenia; DST/NRF, South Africa;
MINECO, Spain; SRC and Wallenberg Foundation, Sweden; SER, SNSF and Cantons of
Bern and Geneva, Switzerland; NSC, Taiwan; TAEK, Turkey; STFC, the Royal
Society and Leverhulme Trust, United Kingdom; DOE and NSF, United States of
America.

The crucial computing support from all WLCG partners is acknowledged
gratefully, in particular from CERN and the ATLAS Tier-1 facilities at
TRIUMF (Canada), NDGF (Denmark, Norway, Sweden), CC-IN2P3 (France),
KIT/GridKA (Germany), INFN-CNAF (Italy), NL-T1 (Netherlands), PIC (Spain),
ASGC (Taiwan), RAL (UK) and BNL (USA) and in the Tier-2 facilities
worldwide.

\bibliography{main}
\bibliographystyle{JHEP}

\newpage 
\begin{flushleft}
{\Large The ATLAS Collaboration}

\bigskip

G.~Aad$^{\rm 85}$,
B.~Abbott$^{\rm 113}$,
J.~Abdallah$^{\rm 151}$,
O.~Abdinov$^{\rm 11}$,
R.~Aben$^{\rm 107}$,
M.~Abolins$^{\rm 90}$,
O.S.~AbouZeid$^{\rm 158}$,
H.~Abramowicz$^{\rm 153}$,
H.~Abreu$^{\rm 152}$,
R.~Abreu$^{\rm 30}$
% ...
\bigskip
\\
% ...
$^{11}$ Institute of Physics, Azerbaijan Academy of Sciences, Baku, Azerbaijan\\
% ...
$^{30}$ CERN, Geneva, Switzerland\\
% ...
$^{85}$ CPPM, Aix-Marseille Universit{\'e} and CNRS/IN2P3, Marseille, France\\
% ...
$^{90}$ Department of Physics and Astronomy, Michigan State University, East Lansing MI, United States of America\\
% ...
$^{107}$ Nikhef National Institute for Subatomic Physics and University of Amsterdam, Amsterdam, Netherlands\\
% ...
$^{113}$ Homer L. Dodge Department of Physics and Astronomy, University of Oklahoma, Norman OK, United States of America\\
% ...
$^{151}$ Institute of Physics, Academia Sinica, Taipei, Taiwan\\
$^{152}$ Department of Physics, Technion: Israel Institute of Technology, Haifa, Israel\\
$^{153}$ Raymond and Beverly Sackler School of Physics and Astronomy, Tel Aviv University, Tel Aviv, Israel\\
% ...
$^{158}$ Department of Physics, University of Toronto, Toronto ON, Canada\\
% ...
\end{flushleft}

\end{document}